\documentclass{aa}
\usepackage{txfonts}
\usepackage{graphicx}
\usepackage{epsfig}
\usepackage{natbib}
\bibpunct{(}{)}{;}{a}{}{,} 
\newcommand{\refeq}[1]{Eq.\,(\ref{#1})}
\newcommand{\refeqs}[2]{Eqs.\,(\ref{#1}-\ref{#2})}
\newcommand{\reffig}[1]{Fig.\,\ref{#1}}
\newcommand{\reffigs}[2]{Figs.\,\ref{#1} and~\ref{#2}}
\newcommand{\reftab}[1]{Tab.\,\ref{#1}}
\newcommand{\refsec}[1]{Sect.\,\ref{#1}}
\newcommand{\refobs}[1]{\textbf{obs.~\#{#1}}}
\newcommand{\refobsPulseAsymmetry}{\refobs{1}}
\newcommand{\refobsPulseEnergy}{\refobs{2}}
\newcommand{\refobsLags}{\refobs{3}}
\newcommand{\refobsHardToSoftEvolution}{\refobs{4}}
\newcommand{\refobsHIC}{\refobs{5}}
\newcommand{\refobsHFC}{\refobs{6}}
\newcommand{\refobsHardnessDuration}{\refobs{7}}
\newcommand{\figAEGSmpezc}{\ref{fig:ABC_4BATSEchannels},\ref{fig:ABC_Ep_alpha},\ref{fig:ABC_WE_lags},\ref{fig:ABC_HIC_HFC},\ref{fig:HELC},\ref{fig:HESP},\ref{fig:HESPslopep}} % REFERENCE CASE
\newcommand{\figAEGSmpezv}{\ref{fig:AB_zeta_4BATSEchannels},\ref{fig:HELC},\ref{fig:HESP},\ref{fig:HESPslopep}}
\newcommand{\figAEGSmpgzc}{\ref{fig:duration},\ref{fig:HESPslopep}}
\newcommand{\figAEGSmpgzv}{\ref{fig:duration},\ref{fig:HESPslopep}}
\newcommand{\figBEGSmpazc}{\ref{fig:B_Epeak_p},\ref{fig:B_HIC_p},\ref{fig:B_p_WE_lags}}
\newcommand{\figBEGSmpczc}{\ref{fig:B_Epeak_p},\ref{fig:B_HIC_p},\ref{fig:B_p_WE_lags}}
\newcommand{\figBEGSmpizc}{\ref{fig:B_Epeak_p},\ref{fig:B_HIC_p},\ref{fig:B_p_WE_lags}}
\newcommand{\figBEGSmpkzc}{} 
\newcommand{\figBEGSmpmzc}{}
\newcommand{\figBEGSmpezc}{\ref{fig:ABC_4BATSEchannels},\ref{fig:ABC_Ep_alpha},\ref{fig:ABC_WE_lags},\ref{fig:ABC_HIC_HFC},\ref{fig:B_Epeak_p},\ref{fig:B_HIC_p},\ref{fig:B_p_WE_lags},\ref{fig:B_Dynamics},\ref{fig:HELC},\ref{fig:HESP},\ref{fig:HESPslopep}} % REFERENCE CASE
\newcommand{\figBEGSmpezv}{\ref{fig:B_zeta_Spectrum},\ref{fig:AB_zeta_4BATSEchannels},\ref{fig:B_zeta_Epeak_WE_lags},\ref{fig:B_zeta_HIC},\ref{fig:HELC},\ref{fig:HESP},\ref{fig:HESPslopep}}
\newcommand{\figBEGSmpgzc}{\ref{fig:B_Epeak_p},\ref{fig:B_HIC_p},\ref{fig:B_p_WE_lags},\ref{fig:duration},\ref{fig:HESPslopep}}
\newcommand{\figBEGSmpgzv}{\ref{fig:duration},\ref{fig:HESPslopep}}
\newcommand{\figBEGShpezc}{\ref{fig:B_Gamma_4BATSEchannels_WE_lags},\ref{fig:B_Dynamics},\ref{fig:HELC},\ref{fig:HESP}}
\newcommand{\figBEGShpezv}{\ref{fig:B_Gamma_4BATSEchannels_WE_lags},\ref{fig:HELC},\ref{fig:HESP}}
\newcommand{\figBMGSmpezc}{\ref{fig:B_Mdot_4BATSEchannels_WE_lags},\ref{fig:B_Dynamics},\ref{fig:HELC},\ref{fig:HESP}}
\newcommand{\figBMGSmpezv}{\ref{fig:B_Mdot_4BATSEchannels_WE_lags},\ref{fig:HELC},\ref{fig:HESP}}
\newcommand{\figCEGSmpezc}{\ref{fig:ABC_4BATSEchannels},\ref{fig:ABC_Ep_alpha},\ref{fig:ABC_WE_lags},\ref{fig:ABC_HIC_HFC}} % REFERENCE CASE

\begin{document}
\title{Spectral evolution in gamma-ray bursts: predictions of the internal shock model and comparison to observations}
\author{\v Zeljka Bo\v snjak\inst{1,2,3} and Fr\'{e}d\'{e}ric Daigne\inst{2} }
\institute{AIM (UMR 7158 CEA/DSM-CNRS-Universit\'e Paris Diderot) Irfu/Service d'Astrophysique, Saclay, F-91191 Gif-sur-Yvette Cedex, France
\and
UPMC-CNRS, UMR7095, Institut d'Astrophysique de Paris, F-75014, Paris, France
\and Department of Physics, University of Rijeka, 51000 Rijeka, Croatia\\
\email{zeljka.bosnjak@cea.fr ; daigne@iap.fr}
}
\abstract
{Several trends have been identified in the prompt gamma-ray burst (GRB) emission:
e.g.  hard-to-soft evolution, pulse width evolution with energy, time lags, hardness-intensity{$/$}-fluence correlations. 
Recently \textit{Fermi} has significantly extended  the spectral coverage of GRB observations
and improved the characterization of this
 spectral evolution.}
{We want to study how 
internal shocks
can reproduce these observations.
 In this model the emission comes from the synchrotron radiation of shock accelerated electrons, and the spectral evolution is governed by the evolution of the physical conditions in the shocked regions.}
{We present a comprehensive set of simulations of a single pulse and investigate the impact of the model parameters, related to the shock microphysics 
and to the initial conditions 
in the  ejecta.}
{We find a general qualitative agreement between the model and the various observations used for the comparison. All these properties or relations are governed by the evolution of the peak energy and photon indices of the spectrum.
In addition, we identify 
the conditions for
 a quantitative agreement.
We find that the best agreement is obtained for (i) steep electron slopes ($p\ga 2.7$),  (ii) microphysics parameters varying with shock conditions so that more electrons are accelerated in stronger shocks, (iii) steep variations of the initial Lorentz factor in the ejecta. When simulating short GRBs by contracting all timescales, all other parameters being unchanged, we show that the hardness-duration correlation is reproduced, as well as the evolution with duration of the pulse properties. 
Finally, we investigate the signature at high energy
of these different scenarios
 and find distinct properties -- delayed onset, longer emission, and flat spectrum in some cases -- suggesting that internal shocks could have a significant contribution to the prompt LAT emission.}
{Spectral evolution is an important property of GRBs that is not easily reproduced in most models for the prompt emission. 
We find that the main observed features can be accounted for in a quantitative way 
within the internal shock model.
However
the current uncertainties on  shock acceleration in the mildly relativistic regime and relativistic ejection by compact sources prevent us from deciding if one or several of the proposed scenario are viable. It may be possible by combining observations over the whole spectral range of \textit{Fermi} to identify in the future specific signatures imprinted by this uncertain underlying physics.}
\authorrunning{Bo\v snjak \& Daigne}
\titlerunning{Spectral evolution in GRBs: internal shock predictions}
%\titlerunning{Spectral evolution in GRBs: internal shock predictions and comparison to observations}
\date{Received: 22.07.2013 -- Accepted: 05.04.2014}
\maketitle

\section{Introduction}

Since the launch of the \textit{Swift} (2004) \citep{gehrels:04} and \textit{Fermi} (2008) satellites, there is a significantly growing sample
of gamma-ray bursts (GRBs) with a known
redshift and a well characterized gamma-ray prompt emission \citep[see e.g. the recent review by][]{gehrels:13}. 
The high-energy domain ($>100$ MeV) is currently explored by \textit{Fermi}-LAT \citep{atwood:09}. The sample of detected bursts is still small but has allowed the identification
of several important 
spectral and temporal properties
\citep{omodei:09,zhang:11,LATcatalog:13}, that are summarized in \refsec{sec:LAT}. 
In the soft gamma-ray range, the GRB sample is much larger and not limited to the brightest bursts.
Thanks to its large spectral range (8 keV-40 MeV), 
\textit{Fermi}-GBM \citep{meegan:09}
has significantly improved the description of the GRB  
properties in the keV-MeV range. This effort follows the results already obtained by several past or current missions, especially  
BATSE (Burst And Transient Source Experiment) on board the \textit{Compton Gamma-Ray Observatory} 
\citep{kaneko:06},
\textit{Beppo-SAX}
\citep{guidorzi:11},
and \textit{HETE-2} \citep{lamb:04,sakamoto:05}. 
Based on this large set of observations, our current know\-ledge of the spectral and temporal properties of the GRB prompt soft gamma-ray emission is summarized in \refsec{sec:obs}.

The standard GRB model associates the prompt gamma-ray emission to internal dissipation within an ultra-relativistic outflow 
($\Gamma \ga$ 100) ejected by a new-born compact source \citep[see e.g.][]{piran:99}. The nature of the dissipation mechanism and of the associated radiative process remains to be identified.
 In order to account for the
observed short time scale variability ($\sim$ ms), the internal shock model \citep{rees:94}, where
variations of the bulk Lorentz factor lead to the formation of shock waves within the ejecta, was proposed for the extraction 
of the jet kinetic energy.
The dissipated energy is distributed between
protons, electrons, and magnetic field; the prompt GRB emission 
model is due to the synchrotron radiation of shock accelerated electrons, with an additional component due to inverse Compton scatterings.
Detailed calculations of the expected light curves and spectra  are available \citep{kobayashi:97,daigne:98,bosnjak:09,asano:11} 
and show a good agreement with observations except for a notable exception, the low-energy photon index, which is usually observed to  be larger than the standard fast cooling synchrotron slope $-3/2$. Several
 solutions 
have been proposed, as the role of inverse Compton scatterings in Klein Nishina regime \citep{derishev:01,bosnjak:09,nakar:09,wang:09,daigne:11}
or the magnetic field decay  in the shocked region \citep{derishev:07,zhao:14}. 

Other mechanisms 
 could also play a role. 
Thermal emission is expected at the photosphere when the ejecta becomes transparent for its own radiation. Depending both on the efficiency of the acceleration process and of the non-thermal emission above the photosphere, this emission could be bright  \citep{meszaros:00,daigne:02,hascoet:13}. 
It produces
in principle a narrow quasi-Planckian component \citep{goodman:86,peer:08,beloborodov:11}; however
different possible sub-photospheric dissipation processes  
may affect the spectrum, especially due to the comptonization, so that it appears as non-thermal \citep{thompson:94,rees:05,peer:06,giannios:07,beloborodov:10,vurm:11,toma:11,veres:12,veres:13}. The peak energy is governed by a detailed balance between the emission/absorption and scattering processes \citep{vurm:13} and can reproduce the observed values (\citealt{beloborodov:13}, see however \citealt{zhang:12}). The lateral structure of the jet may also affect the photospheric spectrum \citep{lundman:13,lazzati:13}.

Magnetized ejecta offer a third possibility.
A large initial magnetization may play a major role for the acceleration of the jet to
relativistic speed
\citep[see e.g.][]{begelman:94,daigne:02b,vlahakis:03,komissarov:09,tchekhovskoy:10,komissarov:10,granot:11} and is already invoked for this reason in some scenarios where the emission is due to the photosphere and/or internal shocks. However, if the ejecta is still magnetized at large distance, magnetic reconnection can provide 
 a new dissipation process \citep{spruit:01,drenkhahnspruit:02,lyutikov:03,gianniosspruit:05,zhangyan:11,mckinney:12}. Compared to the
 previous possibilities, this model cannot provide yet detailed predictions for the GRB light curves and spectra 
\citep[see the preliminary calculation of the temporal properties by][]{zhang:13}.

The photospheric emission should be present in all scena\-rios, even if very weak. On the other hand, magnetic reconnection requires a large magnetization at large distance which may prevent internal shock formation and propagation \citep{giannios:08,mimica:10,narayan:11}. Therefore, depending on the magnetization in the emission site, only one of the two mechanisms  should be at work. Recent observations of two components in the soft gamma-ray spectrum of a few bright \textit{Fermi}/GBM bursts, one being quasi-Planckian and the other being non-thermal \citep{ryde:10,guiriec:11,axelsson:12,guiriec:13}, indicate that both the photosphere and either internal shocks or reconnection may be indeed at work in GRBs \citep{hascoet:13}.

Aside from interpreting the light curves and spectra, a successful theoretical model should also reproduce
the observed spectral evolution with time, which is
 is mainly governed by the evolution of the peak energy of the spectrum.
 It can be related either to the physics of the dissipative mechanism in the outflow, or to the curvature of the emitting surface. In the first case, the spectral evolution is due to an intrinsic evolution of the physical conditions in the flow, whereas it is a geometrical effect (delay, Doppler shift) in the second case.
 The spectral evolution in a pulse associated to the curvature effect has been studied by several authors and does not agree with observations \citep{fenimore:96,dermer:04,shen:05,shenoy:13}. Then, the spectral evolution has to be understood from the physics of the dissipative mechanism and may therefore represent an important test to discriminate between the different possible prompt emission models listed above.
 
Regarding the photospheric emission, the spectral evolution has been computed only in the case of non-dissipative photospheres \citep{daigne:02,peer:08}. As mentioned above, this model cannot reproduce the observed spectrum. In the case of a dissipative photosphere, the peak energy of the spectrum is fixed by a complex physics \citep{beloborodov:13}, which makes difficult a prediction of the spectral evolution. It is usually assumed that modulations in the properties at the base of the flow will lead to the observed evolution (see for instance \citet{giannios:07} in the case where the dissipation is associated to magnetic reconnection). However, dissipative photospheric  models require that the dissipation occurs just below the photosphere for the spectrum to be affected. It is not obvious that a change in the central engine leading to a displacement of the photosphere will affect the dissipation process in the same way so that it remains well located. Therefore, it remains to be demonstrated that these models can reproduce the observed spectral evolution.
 In the context of an emission produced above the photosphere, several authors have investigated  the time development of the photon spectrum without specifying the dissipation mechanism and relate the observed spectral evolution 
 to the evolution of the electron/photon injection rate and/or the decaying magnetic field 
 \citep[e.g.][]{liang:97,stern:04,asano:09,asano:11}.
 It is encouraging that a reasonable agreement with observations is found in some cases.
To reach a final conclusion, it is however necessary to carry such a study in the context of a physical model for the dissipation, which gives a prescription for the accelerated electrons and the magnetic field.
It is still out of reach for reconnection models due to the lack of any spectral calculation. It has been done for the internal shock model using a very simple spectral calculation including only synchrotron radiation \citep{daigne:98,daigne:02}. 
Since these early calculations, the observational description of the spectral evolution has improved a lot, as well as the modeling of the emission from internal shocks.
Therefore, we examine here the quantitative prediction of this model for the  spectral evolution in GRBs. 

For the first time the detailed dynamical evolution is combined with the calculation of the radiative processes, and the outcome is confronted to the large set of 
observed properties
summarized in \refsec{sec:obs}
(e.g. hard-to-soft evolution, pulse width evolution with energy, time lags, hardness-intensity/fluence correlation).
In \refsec{sec:internalshocks}, we present our approach, which is based on the model developed in
\citet{bosnjak:09}.
Following \citet{daigne:11} we define three \textit{reference cases},
 which are representative of the different possible spectral shapes in the keV-MeV range, and
 present a detailed comparison of their temporal and spectral properties  
 with observations.
Then
we investigate 
in \refsec{sec:EffectMicrophysics} and \refsec{sec:EffectDynamics}
the effect 
on our results
of different
assumptions for 
the microphysics and dynamics of the relativistic ejecta.
The specific signatures in the \textit{Fermi}-LAT range are presented in \refsec{sec:LAT}. 
We discuss our results in \refsec{sec:discussion} and conclude
in \refsec{sec:conclusions}.

\section{GRB temporal and spectral properties: observations}
\label{sec:obs}

There are several global  trends in GRB spectra and light curves that have been identified in the prompt emission observations by various missions during the past three decades. 
Spectral variations were already observed  
by the KONUS experiment providing time resolved data between 40 and 700  keV \citep{golenetskii:83}.
BATSE provided the largest database 
of high temporal and spectral resolution prompt GRB data and allowed  detailed studies of the correlations between spectral and temporal properties 
\citep[e.g.][]{band:93,ford:95,norris:96,kaneko:06}.
Since GRB peak energies are usually above the higher energy limit of the \textit{Swift}-BAT($\sim 150$ keV), it is difficult to examine the analogous correlations for the sample of \textit{Swift} GRBs. Due to the large number of events with determined redshifts, this
sample is  however of great interest to test spectral and temporal properties in the source frame \citep{krimm:09,ukwatta:10}. 
\textit{Fermi}-GBM data are providing new insights in the temporal and spectral behavior of GRBs, extending the spectral coverage
to high-energies and to energies below the BATSE low-energy threshold 
\citep[e.g.][]{lu:12,bhat:12}.
We list here the commonly observed trends in the spectral and temporal properties to which we are referring in the subsequent sections when comparing the internal shock model with observations.
We distinguish between short and long GRBs when necessary. It should be noted that
generally, GBM data indicate that the temporal and spectral evolution of short GRBs are very similar to long ones, 
but with light curves
contracted in time and with harder spectra due to higher peak energies  \citep{guiriec:10,bhat:12}. 

\noindent\refobsPulseAsymmetry\textbf{~-- Pulse asymmetry.} When the burst has apparent separated pulses in the time histories, a fast rise and an exponential decay of the pulse is often observed 
\citep{fishman:95}.
\citet{nemiroff:94}
showed that the individual long pulses in GRBs are time-asymmetric. 
The most thorough study of pulses in long GRB light curves was provided by \citet{norris:96}
\citep[see also e.g.][]{quilligan:02,hakkila:11}
using 64 ms resolution BATSE data. They found a typical rise-to-decay time ratio $\sim$0.3-0.5 for long pulses independently of the energy band and also showed that short pulses tend to be more symmetric. This trend is confirmed by the analysis of pulses in short GRB light curves by 
\citet{mcbreen:01}.

\noindent\refobsPulseEnergy\textbf{~-- Energy-dependent pulse asymmetry.} \citet{norris:96}
found that the dominant trend in the pulse shapes observed in different energy channels is a faster onset at higher energies and a longer decay at lower energies. The dependence between the energy $E_\mathrm{obs}$ and the pulse width is approximately a power law, $W(E_\mathrm{obs}) \propto E_\mathrm{obs}^{-a}$ with $a\simeq 0.40$, in a sample of long BATSE bursts 
\citep{fenimore:95,norris:96}. The same evolution with $a\simeq 0.40$ is also found in sample of long pulses with large time lags \citep{norris:05}.
\citet{bissaldi:11} found the same trend for \textit{Fermi} bursts with $a\simeq 0.40$ \citep[see also][]{bhat:12}.

%%%%%%%%%%%%%%%%%%%%%%%%%%%%%%%%%%%%%%%%%%%%%%
%%%%%%%%%%%%%%%%%%%%%%%%%%%%%%%%%%%%%%%%%%%%%%
%%%%%%%%%%%%%%%%%%%%%%%%%%%%%%%%%%%%%%%%%%%%%%
%%%
%%% Table with all cases discussed in the paper
%%%
%%%%%%%%%%%%%%%%%%%%%%%%%%%%%%%%%%%%%%%%%%%%%%
%%%%%%%%%%%%%%%%%%%%%%%%%%%%%%%%%%%%%%%%%%%%%%
%%%%%%%%%%%%%%%%%%%%%%%%%%%%%%%%%%%%%%%%%%%%%%
\begin{table*}[!t]\begin{center}
\resizebox{\textwidth}{!}{\begin{minipage}{19.5cm}
\begin{tabular}{l||cccc||ccc||cc||cccc|l}
\textbf{Case} & \multicolumn{4}{c||}{\textbf{Dynamics}} & \multicolumn{3}{c||}{\textbf{Microphysics}} & \multicolumn{2}{c||}{\textbf{Spec. @ max.}} & \multicolumn{4}{c}{\textbf{Spectro-temporal properties}} &  \textbf{Figs.} \\
 & Ejection & $E_\mathrm{kin,iso}$ & $\Gamma(t)$ & $\bar{\Gamma}$ & $\zeta$ & $\epsilon_\mathrm{B}$ & $p$ & $E_\mathrm{p,obs}$ & $\alpha$  & $\tau_\mathrm{r}/\tau_\mathrm{d}$ & $a\left(W(E)\right)$ & $\delta$ (HIC) & $\kappa$ (HIC) &\\
 &  & [erg] & & & & & & [keV] & & &  & & & \\
\hline
\hline
 A & $\mathbf{\dot{E}=\mathrm{cst}}$ & $\mathbf{ 1.00\times 10^{      54}}$ & \textbf{smooth} & $\mathbf{     340}$ & $\mathbf{ 3.00\times 10^{      -3}}$ & $\mathbf{1/3}$                    & $\mathbf{2.5}$ & $\mathbf{     731}$ & $\mathbf{-1.5}$ & $\mathbf{0.38}$ & $\mathbf{ 0.29}$ & $\mathbf{2.28}$ & $\mathbf{2.16}$& \figAEGSmpezc \\
   &                                 &                                      &                 &                     &         $ 3.40\times 10^{      -3}$  &                                   & $        2.7 $ & $             731 $ & $        -1.5 $ & $        0.39 $ & $         0.30 $ & $        2.15 $ & $        1.97 $& \figAEGSmpgzc \\
   &                                 &                                      &                 &                     & varying                              &                                   &                & $             744 $ & $        -1.4 $ & $        0.31 $ & $         0.28 $ & $        2.23 $ & $        1.55 $& \figAEGSmpezv \\
   &                                 &                                      &                 &                     & varying                              &                                   & $        2.7 $ & $             744 $ & $        -1.4 $ & $        0.30 $ & $         0.29 $ & $        2.12 $ & $        1.48 $& \figAEGSmpgzv \\
\hline
\hline
   &                                 &                                      &                 &                     &         $ 4.00\times 10^{      -4}$  &                                   & $        2.1 $ & $             912 $ & $        -1.2 $ & $        0.41 $ & $         0.14 $ & $        /    $ & $        /    $& \figBEGSmpazc \\
   &                                 &                                      &                 &                     &         $ 8.80\times 10^{      -4}$  &                                   & $        2.3 $ & $             666 $ & $        -1.1 $ & $        0.46 $ & $         0.18 $ & $        /    $ & $        /    $& \figBEGSmpczc \\
 B & $\mathbf{\dot{E}=\mathrm{cst}}$ & $\mathbf{ 1.00\times 10^{      54}}$ & \textbf{smooth} & $\mathbf{     340}$ & $\mathbf{ 1.00\times 10^{      -3}}$ & $\mathbf{        10^{      -3} }$ & $\mathbf{2.5}$ & $\mathbf{     642}$ & $\mathbf{-1.1}$ & $\mathbf{0.43}$ & $\mathbf{ 0.23}$ & $\mathbf{/   }$ & $\mathbf{/   }$& \figBEGSmpezc \\
   &                                 &                                      &                 &                     &         $ 1.10\times 10^{      -3}$  &                                   & $        2.7 $ & $             619 $ & $        -1.1 $ & $        0.54 $ & $         0.24 $ & $        0.97 $ & $        0.89 $& \figBEGSmpgzc \\
   &                                 &                                      &                 &                     &         $ 1.15\times 10^{      -3}$  &                                   & $        2.9 $ & $             630 $ & $        -1.1 $ & $        0.54 $ & $         0.27 $ & $        1.23 $ & $        1.05 $& \figBEGSmpizc \\
   &                                 &                                      &                 &                     &         $ 1.20\times 10^{      -3}$  &                                   & $        3.1 $ & $             619 $ & $        -1.1 $ & $        0.54 $ & $         0.27 $ & $        1.31 $ & $        1.07 $& \figBEGSmpkzc \\
   &                                 &                                      &                 &                     &         $ 1.23\times 10^{      -3}$  &                                   & $        3.3 $ & $             619 $ & $        -1.1 $ & $        0.54 $ & $         0.28 $ & $        1.32 $ & $        1.06 $& \figBEGSmpmzc \\
   &                                 &                                      &                 &                     & varying                              &                                   &                & $             679 $ & $        -1.1 $ & $        0.33 $ & $         0.24 $ & $        0.96 $ & $        0.80 $& \figBEGSmpezv \\
   &                                 &                                      &                 &                     & varying                              &                                   & $        2.7 $ & $             679 $ & $        -1.1 $ & $        0.32 $ & $         0.27 $ & $        1.27 $ & $        0.97 $& \figBEGSmpgzv \\
   &                                 &          $ 1.50\times 10^{      54}$ &                 &        ${     360}$ & varying                              &                                   &                & $             691 $ & $        -1.1 $ & $        0.37 $ & $         0.24 $ & $        /    $ & $        /    $& \figBEGSmpezv \\
   &                                 &          $ 1.50\times 10^{      54}$ &                 &        ${     360}$ & varying                              &                                   & $        2.7 $ & $             679 $ & $        -1.1 $ & $        0.36 $ & $         0.26 $ & $        0.92 $ & $        0.78 $& \figBEGSmpgzv \\
   &                                 &          $ 5.85\times 10^{      53}$ &         sharp   &      
   &         $ 2.00\times 10^{      -3}$  &                                   &                & $             744 $ & $        -1.2 $ & $        0.68 $ & $         0.18 $ & $        /    $ & $        /    $& \figBEGShpezc \\
   &                                 &          $ 5.85\times 10^{      53}$ &         sharp   &       
   & varying                              &                                   &                & $             772 $ & $        -1.1 $ & $        0.04 $ & $         0.25 $ & $        /    $ & $        /    $& \figBEGShpezv \\
   & $        \dot{M}=\mathrm{cst} $ &          $ 1.85\times 10^{      54}$ &                 &                     &         $ 6.00\times 10^{      -4}$  &                                   &                & $             679 $ & $        -1.1 $ & $        0.75 $ & $         0.16 $ & $        0.13 $ & $        0.17 $& \figBMGSmpezc \\
   & $        \dot{M}=\mathrm{cst} $ &          $ 1.85\times 10^{      54}$ &                 &                     & varying                              &                                   &                & $             630 $ & $        -1.1 $ & $        0.60 $ & $         0.16 $ & $        /    $ & $        /    $& \figBMGSmpezv \\
\hline
\hline
 C & $\mathbf{\dot{E}=\mathrm{cst}}$ & $\mathbf{ 1.00\times 10^{      53}}$ & \textbf{smooth} & $\mathbf{    1020}$ & $\mathbf{ 1.00\times 10^{      -3}}$ & $\mathbf{        10^{      -1} }$ & $\mathbf{2.5}$ & $\mathbf{     164}$ & $\mathbf{-0.7}$ & $\mathbf{0.55}$ & $\mathbf{ 0.11}$ & $\mathbf{/   }$ & $\mathbf{/   }$& \figCEGSmpezc \\
\end{tabular}
\end{minipage}}
\end{center}
\caption{\textbf{Parameters of all the GRB pulse models discussed in the paper.} The three reference cases defined in \refsec{sec:internalshocks} are listed in bold face. For other models discussed in \refsec{sec:EffectMicrophysics} and \refsec{sec:EffectDynamics}, we list only the input parameters that are modified compared to the reference case. The first columns list the parameters for the dynamics and the microphysics (see text).
In all cases, $\epsilon_\mathrm{e}=1/3$.
The last columns list a few properties of the corresponding simulated GRB pulse : spectral properties at the maximum of the GBM  light curve (peak energy and low-energy photon index), and four indicators of the spectral and temporal properties: ratio of the rise time over the decay time of the pulse (BATSE 2+3 channel), index $a$ for the evolution of the pulse width $W(E)$ with energy ($W(E)\propto E^{-a}$), slopes of the hardness-intensity correlation ($\delta$ is the slope when using the photon flux and $\kappa$ the energy flux), see text. Cases where it is not possible to define the slopes of the HIC are indicated with '$/$'. For reference, typical observed values 
are 
$\tau_\mathrm{r}/\tau_\mathrm{d}\simeq 0.3-0.5$ \citep{norris:96},
$a\simeq 0.3-0.4$ \citep{norris:96,bissaldi:11}, $\delta\simeq 0.4-1.1$ \citep{ryde:02} and $\kappa\simeq 0.3-1.2$ \citep{lu:12} : see \refsec{sec:obs}. The last column lists figures in the paper where some properties of each case are shown.
}
\label{tab:allmodels}
\end{table*}
%%%%%%%%%%%%%%%%%%%%%%%%%%%%%%%%%%%%%%%%%%%%%%
%%%%%%%%%%%%%%%%%%%%%%%%%%%%%%%%%%%%%%%%%%%%%%
%%%%%%%%%%%%%%%%%%%%%%%%%%%%%%%%%%%%%%%%%%%%%%

\noindent\refobsLags\textbf{~-- Time lags.} Time lags are commonly observed in GRB pulses: pulses tend to peak earlier at higher energy in the soft gamma-ray range \citep{norris:96}. 
Time lags were studied for a large sample of BATSE GRBs 
\citep{band:97,norris:02,hakkila:08}.
Short lags ($<$ 350 ms) dominate the BATSE sample, even if a long lag ($>$ 1 s) subpopulation was identified by 
\citet{norris:02}.
\citet{norris:01} and \citet{norris:06}
found negligible lags
  for a sample of BATSE, \textit{Swift} and \textit{HETE-2} short GRBs \citep[see also][]{yi:06}. 
  \citet{guiriec:10}
confirmed negligible spectral lags 
below 1 MeV for three bright short GRBs observed by {\it Fermi}. 

\noindent\refobsHardToSoftEvolution\textbf{~-- Hard-to-soft evolution.} 
\citet{norris:86}
examined a handful of bursts observed by the {\it Solar Maximum Mission} satellite between 50 and 300 keV,
and found that the pulse emission evolved from hard to soft with the hardness maximum preceeding the peak of the intensity.
More detailed studies followed using BATSE data 
\citep{bhat:94,ford:95,band:97}:
 it was found that the spectral peak energy $E_\mathrm{p,obs}$ is rising or slightly preceding the pulse intensity increase, and is softening during the pulse decay (hard-to-soft evolution within a pulse). The later pulses in burst time history were also found to be softer than earlier ones (global hard-to-soft evolution). For \textit{Fermi} GBM bursts, 
 \citet{lu:12}
reported the same  hard-to-soft evolution 
in the variation with time of the spectral peak energy in the majority of GRBs,
but also found cases where the peaking energy is simply tracking the intensity. These bursts show usually more symmetric pulses.
Short \textit{Fermi} GRBs are usually found to follow the 'intensity tracking' pattern.

\noindent\refobsHIC\textbf{~-- Hardness-intensity correlation (HIC).} 
\citet{golenetskii:83}
reported the discovery of a correlation between the instantaneous luminosity and the temperature $kT$ characterizing the photon spectrum, $L \propto (kT)^\gamma$, with
$\gamma \approx$ 1.5-1.7. 
\citet{kargatis:94}
investigated bursts from the SIGNE experiment (50-700 keV) and confirmed the luminosity-temperature correlation in $\sim$50\% of the events with a larger spread for the exponent, $\gamma \approx$ 1.3-2.7. 
\citet{kargatis:95}
performed the spectral analysis of the decay phase in BATSE GRB pulses and found that the spectral peak energy correlates with the instantaneous energy flux $F(t)$ as $F(t) \propto E_\mathrm{p,obs}^{1.7}$ (hardness intensity correlation, or HIC).
\citet{borgonovo:01} 
studied this HIC
using 
the value of $E F_E$ at the peak energy
to represent the intensity, and found it was proportional to 
$ E_\mathrm{p,obs}^{\eta}$. This makes  the correlation less dependent on the observational spectral window. The mean value of $\eta$ was found to be $2.0$ and it corresponds to a mean value of $\gamma$ = 1.9. 
\citet{ryde:00,ryde:02}
studied 
the HIC in a large sample of BATSE bursts using 
the instantaneous photon flux $N(t)$, 
and found $E_\mathrm{p,obs} \propto N(t)^\delta$, with $\delta \sim$ 0.4 - 1.1. More recently, a similar HIC was found in GBM bursts using the energy flux $F(t)$, $E_\mathrm{p,obs}\propto F(t)^\kappa$, with $\kappa\simeq 0.4-1.2$ \citep{lu:12,guiriec:13}. We focus on these two most recent studies of the HIC to compare with model predictions in next sections.

\noindent\refobsHFC\textbf{~-- Hardness-fluence correlation (HFC).} This correlation was discovered by 
\citet{liang:96}.
It describes the exponential decay of the spectral hardness as a function of the cumulative photon fluence $\Phi(t)$, $E_\mathrm{p,obs}(t) \propto$ exp$(-\Phi(t)/\Phi_0)$. 
\citet{crider:99}
 tested a reformulated correlation using energy fluence instead of the photon fluence, and confirmed the decay pattern. 
 \citet{ryde:00}
 used this correlation combined with the HIC correlation to obtain a self-consistent description for the temporal behavior of the instantaneous photon flux and got a good agreement with BATSE data. 
 
\noindent\refobsHardnessDuration\textbf{~-- Hardness-duration correlation.} 
\citet{kouveliotou:93}
reported that short
GRBs are harder than long ones.
Hardness was characterized by the ratio of the total counts in the two BATSE energy channels (usually 100-300 keV and 50-100 keV energy range). 
\citet{ghirlanda:04}
 argued that the hardness of the short events is owing to a harder low energy spectral slope (photon index $\alpha$) in short bursts, rather than to a higher  peak energy $E_\mathrm{p,obs}$. This is not confirmed by
the detailed analysis of three bright GBM short  GRBs by \citet{guiriec:10}, which shows
 that the hardness of short bursts is due both to hard low-energy photon indexes $\alpha$ and high peak energies $E_\mathrm{p,obs}$.
 
\section{Spectral evolution in the internal shock model}
\label{sec:internalshocks}
\subsection{Modelling the emission from internal shocks}
Several steps are necessary to model the prompt GRB emission
from internal shocks.
From the initial conditions in the ultra-relativistic outflow ejected by the central engine, the dynamical evolution must be calculated. This allows to know how many internal shocks will form and propagate within the outflow and to compute the time-evolution of the physical conditions in each of the shocked regions (Lorentz factor, mass and energy density, ...). Then, the distribution of shock-accelerated electrons and the intensity of the shock-amplified magnetic field must be evaluated. This is the most uncertain step and is usually done using a very simple parametrization of the local microphysics. Knowing the distribution of relativistic electrons accelerated at each shock and the magnetic field, it is then possible to compute the emission produced in the comoving frame of each shocked region, taking into account the relevant radiative processes. Finally, the contributions of each emitting region are summed up taking into account relativistic effects (Doppler boosting, relativistic beaming), the curvature of the emitting surface (integration on equal-arriving time surfaces) and cosmological effects (redshift, time dilation). This full procedure allows to predict  light curves and time-evolving spectra for synthetic GRBs and therefore to make a detailed comparison with observations. 

To follow this procedure, we use
 the model described in \citet{bosnjak:09}.  
We assume that the outflow at large distance from the source has a negligible magnetization, which can be achieved either in the standard fireball or in an efficient magnetic acceleration scenario, which is preferred by observations \citep{hascoet:13}. A moderate magnetization ($\sigma\ga 0.01-0.1$) may affect the spectrum of internal shocks, especially at high energy \citep{mimica:12}.
The model parameters are of two types: (i) the initial conditions for the \textit{dynamics} of the relativistic outflow, given by the duration of the relativistic ejection $t_\mathrm{w}$, the initial distribution of the Lorentz factor $\Gamma(t_\mathrm{ej})$, where $0\le t_\mathrm{ej}\le t_\mathrm{w}$ is the time of the ejection, and the initial kinetic power $\dot{E}(t_\mathrm{j})$; (ii) the \textit{microphysics} parameters :  it is assumed that a fraction $\epsilon_\mathrm{B}$ of the dissipated energy in a shocked region is injected into the amplified magnetic field, and that a fraction $\epsilon_\mathrm{e}$ of the energy is injected into a fraction $\zeta$ of the electrons to produce a non-thermal population. The distribution of these accelerated relativistic electrons is a power-law with a slope $-p$. 

We refer to \citet{bosnjak:09} for a detailed description of the model, which is based on a multi-shell approximation for the dynamics and a radiative code that solves simultaneously the time evolution of the electron and photon distributions in the comoving frame of the emitting material, taking into account all the relevant processes : synchrotron radiation and self-absorption, inverse Compton scattering (including Klein-Nishina regime), $\gamma\gamma$ annihilation, and adiabatic cooling. 

Depending on the choice of microphysics parameters, the typical Lorentz factor of the shock-accelerated electrons and the  intensity of the magnetic field can be rather different, even for similar relativistic outflows and dynamical evolutions. Then, the dominant radiative process in the soft gamma-ray range (BATSE or GBM) could be either direct synchrotron radiation or inverse Compton scattering of low-energy synchrotron photon (SSC). \citet{bosnjak:09} have shown that the second case would predict a bright additional component in the GeV, due to the second inverse Compton scatterings. Such a peak does not seem to be observed by \textit{Fermi} and the SSC scenario is  probably ruled out, as discussed for instance by \citet{piran:09}. Therefore we focus here on the scenario where the prompt GRB emission is dominated by synchrotron radiation from shock-accelerated electrons in internal shocks. In this case, there are in principle two components in the spectrum, one, peaking in the soft gamma-ray range, due to synchrotron radiation, and a second one, peaking at high energy, associated to inverse Compton scatterings, which are very likely in the Klein-Nishina regime.
%%%%%%%%%%%%%%%%%%%%%%%%%%%%%%%%%%%%%%%%%%%%%%
%%%%%%%%%%%%%%%%%%%%%%%%%%%%%%%%%%%%%%%%%%%%%%
%%%%%%%%%%%%%%%%%%%%%%%%%%%%%%%%%%%%%%%%%%%%%%
%%%
%%% FIGURE 1 : reference cases A,B,C in four BATSE channels
%%%
%%%%%%%%%%%%%%%%%%%%%%%%%%%%%%%%%%%%%%%%%%%%%%
%%%%%%%%%%%%%%%%%%%%%%%%%%%%%%%%%%%%%%%%%%%%%%
%%%%%%%%%%%%%%%%%%%%%%%%%%%%%%%%%%%%%%%%%%%%%%
\begin{figure*}[!t]
\begin{center}
\begin{tabular}{ccc}
\centering
\includegraphics[width=0.3\textwidth]{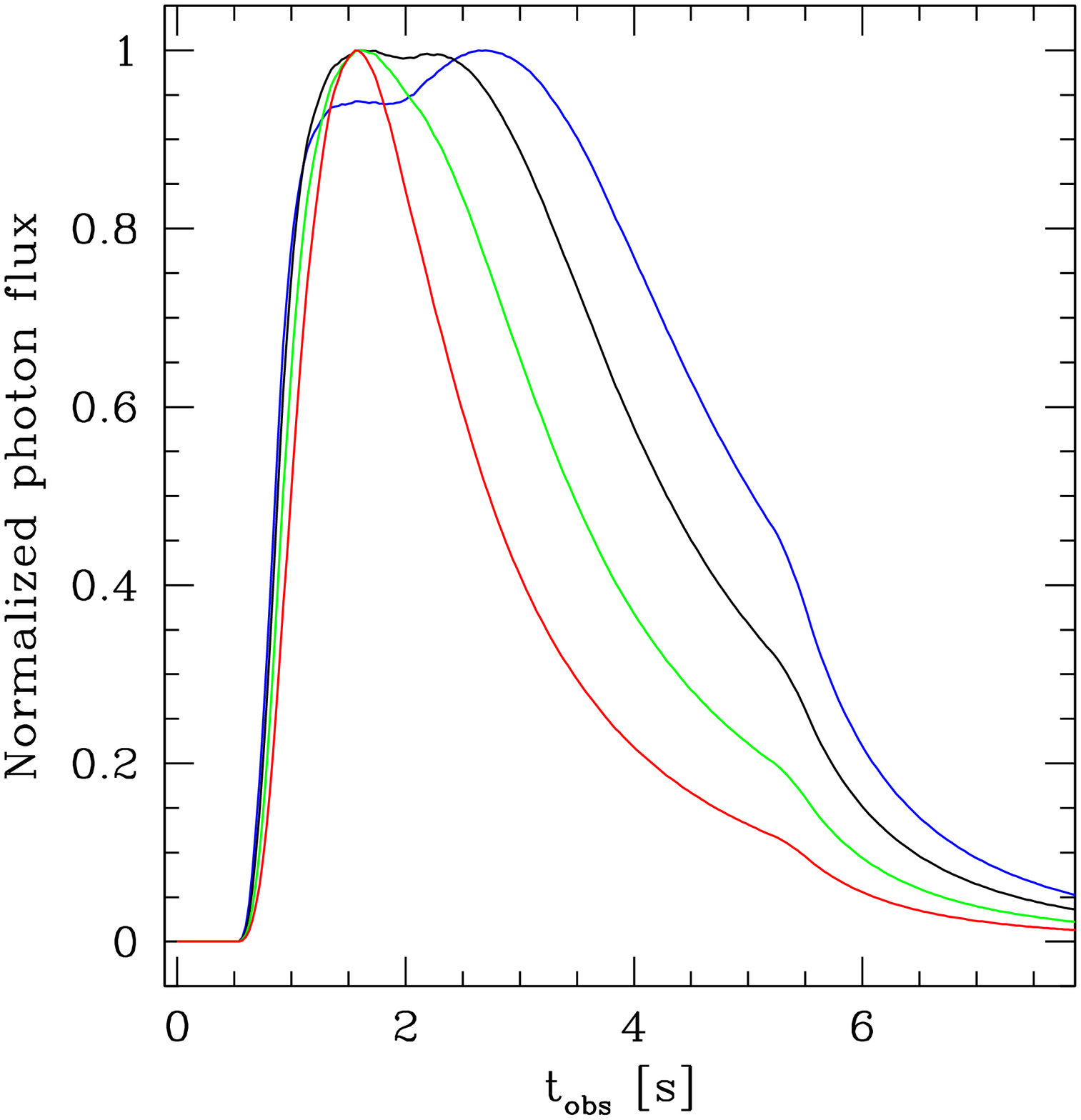}&
\includegraphics[width=0.3\textwidth]{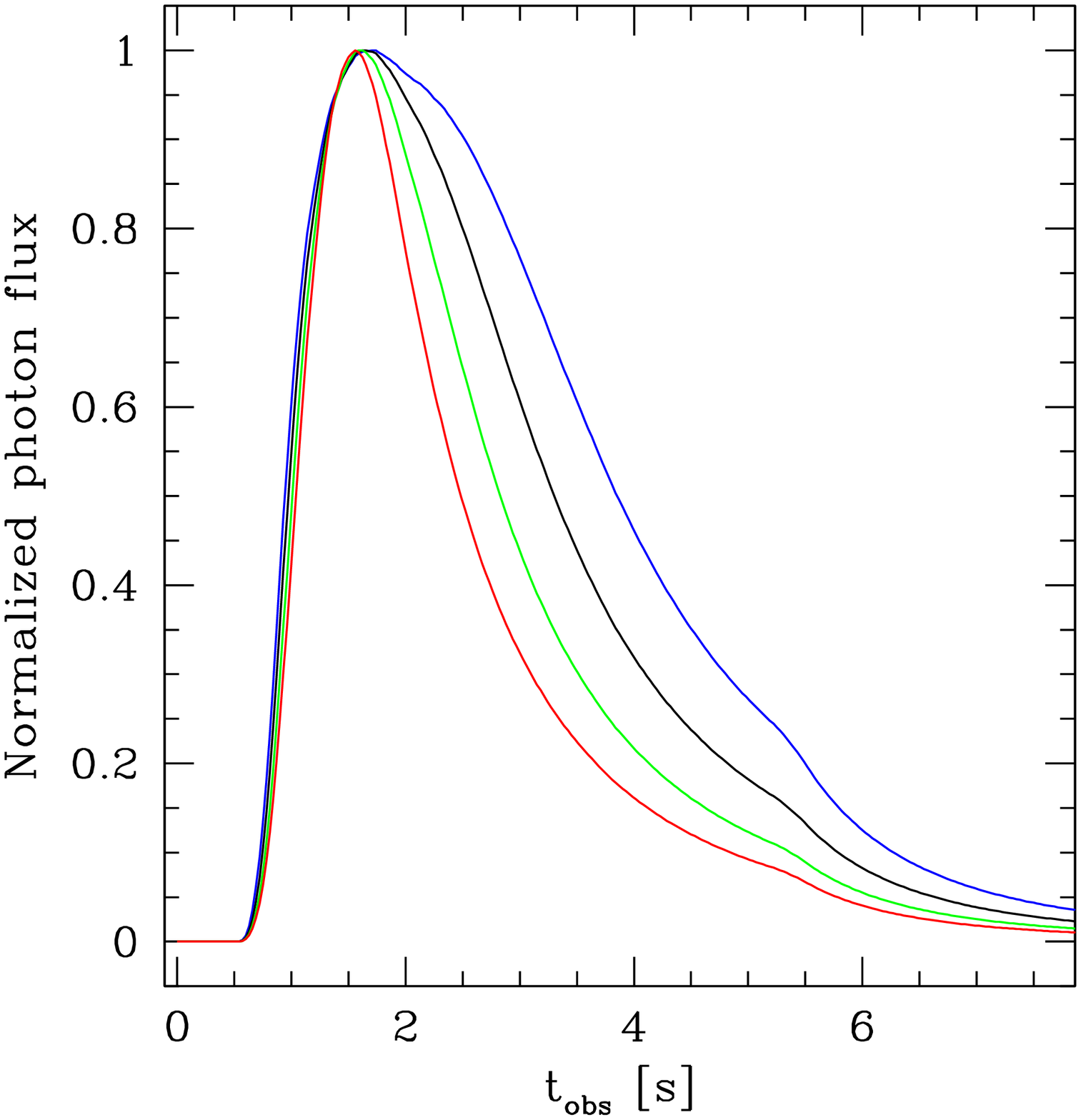}&
\includegraphics[width=0.3\textwidth]{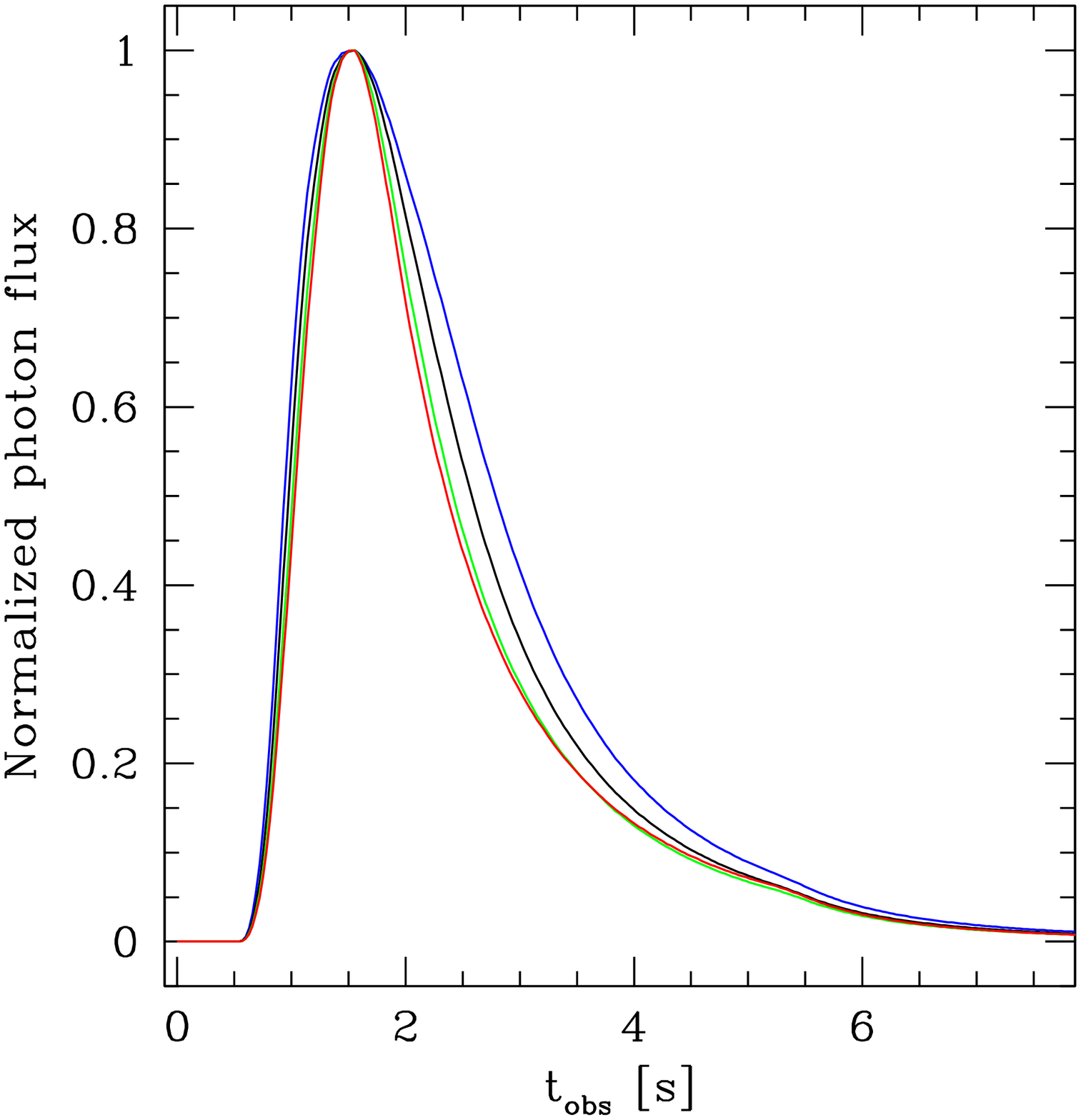}\\
\end{tabular}
\end{center}
%\vspace*{-3ex}

\caption{ \textbf{Three reference cases: normalized light curves in the four BATSE channels.} (blue line: 20-50 keV; black line: 50-100 keV; green line: 100-300 keV; red line: 300-2000 keV) for the cases A, B \& C (see text). 
}
\label{fig:ABC_4BATSEchannels}
\end{figure*}
%%%%%%%%%%%%%%%%%%%%%%%%%%%%%%%%%%%%%%%%%%%%%%
%%%%%%%%%%%%%%%%%%%%%%%%%%%%%%%%%%%%%%%%%%%%%%
%%%%%%%%%%%%%%%%%%%%%%%%%%%%%%%%%%%%%%%%%%%%%%

\subsection{Three reference cases}

Even in the synchrotron scenario, the spectral shape of the prompt emission still depends strongly on the assumptions for the microphysics parameters. Assuming that electrons are radiatively efficient (synchrotron fast cooling, \citealt{sari:98}), which is required both by the variability timescale and the energetics of the prompt emission, there are three main possibilities, that have been described by \citet{daigne:11} and illustrated with three reference cases. Each of these cases corresponds to an example of a single pulse burst that should be seen as a building block for more complex light curves. The initial Lorentz factor is rising during the ejection, according to the simple law
\begin{equation}
\Gamma(t_\mathrm{ej})  = \left\lbrace\begin{array}{cl}
\frac{\Gamma_\mathrm{max}+\Gamma_\mathrm{min}}{2}-\frac{\Gamma_\mathrm{max}-\Gamma_\mathrm{min}}{2} \cos{\left(\pi\frac{t_\mathrm{ej}}{0.4 t_\mathrm{w}}\right)} & \mathrm{for} \, 0 \le \frac{t_\mathrm{ej}}{t_\mathrm{w}}\le 0.4 \\
\Gamma_\mathrm{max} & \mathrm{for}\, 0.4 \le \frac{t_\mathrm{ej}}{t_\mathrm{w}}\le 1
\end{array}\right.\, , 
\label{eq:SmoothGamma}
\end{equation}
with $\Gamma_\mathrm{min}=100$  (resp. $300$) and $\Gamma_\mathrm{max}=400$  (resp. $1200$) in cases A and B (resp. case C): 
 see Fig.~1 in \citet{bosnjak:09}.
The duration of the ejection is $t_\mathrm{w}=2\, \mathrm{s}$ and the injected kinetic power is constant and equals $\dot{E}=10^{54}\, \mathrm{erg.s^{-1}}$ (resp. $5\times 10^{52}\, \mathrm{erg.s^{-1}}$) in cases A and B (resp. in case C).
 Then, the collision of the `slow' and `rapid' parts in the ejecta lead to the formation of two internal shock waves, a short lived `forward' shock and a `reverse' shock that crosses most of the ejecta and dominates the emission. The three reference cases differ mainly by the microphysics: in all three cases, $\epsilon_\mathrm{e}=1/3$ and $p=2.5$, but the two other microphysics parameters $\epsilon_\mathrm{B}$ and $\zeta$ are different. 
 
\noindent\textit{Case A. Pure fast cooling synchrotron case, with $\epsilon_\mathrm{B}=\epsilon_\mathrm{e}=1/3$ and $\zeta=0.003$.} For $\epsilon_\mathrm{B}\ga \epsilon_\mathrm{e}$, inverse Compton scatterings are very inefficient. Then the radiated spectrum does not show an additional component at high energy and is very close to the standard fast cooling synchrotron spectrum with a low-energy photon index $\alpha=-3/2$.  

\noindent\textit{Case B. Fast cooling synchrotron case affected by inverse Compton scattering in Klein Nishina regime, with $\epsilon_\mathrm{B}=10^{-3}$ and $\zeta=0.001$.} For $\epsilon_\mathrm{B}\ll \epsilon_\mathrm{e}$, inverse Compton scattering becomes efficient and leads to a second spectral component at high energy, which
 remains however weak due to the Klein-Nishina regime. In addition, as the scatterings are more efficient for low-energy photons in this regime, the low-energy photon index of the synchrotron component is modified and becomes larger than the standard fast cooling value: $-3/2\la \alpha \la -1$ \citep[see also a discussion of this case for GRB 080916C in][]{wang:09}. 

\noindent\textit{Case C. Marginally fast cooling case, with $\epsilon_\mathrm{B}=0.1$ and $\zeta=0.001$.} In some conditions, it is possible that the cooling frequency $\nu_\mathrm{c}$ becomes very close to the synchrotron frequency $\nu_\mathrm{m}$. In such a situation where $\nu_\mathrm{c}\la \nu_\mathrm{m}$, radiation is still efficient (fast cooling), but the intermediate region of the spectrum where $\alpha=-3/2$ disappears, so that large photon indices (possibly as large
as $\alpha=-2/3$ usually expected in the slow cooling regime) can be measured. It would require some fine tuning for this marginally fast cooling regime to be reached in most GRBs but the necessary conditions may be encountered in some parts of a burst prompt phase, especially when the flux is weaker and the emission softer. Recently, \citet{beniamini:13} found that the conditions for the marginally fast cooling may indeed be found in a noticeable region of the parameter space.

\begin{figure}[t!]
\begin{center}
\includegraphics[width=0.45\textwidth]{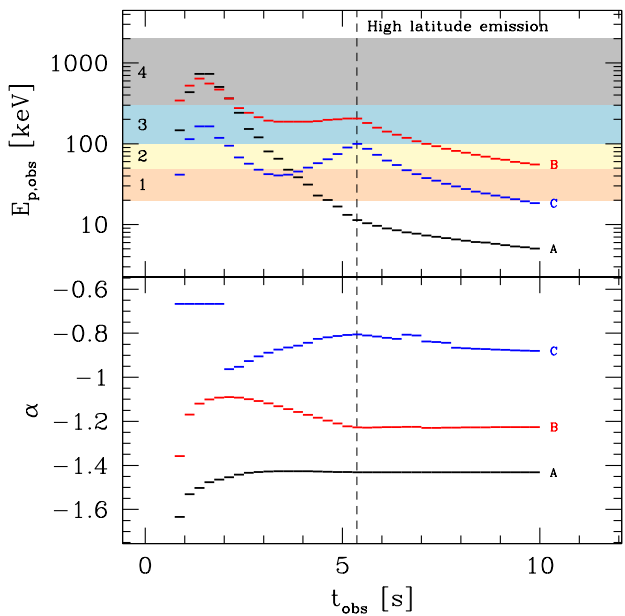}\\
\end{center}
\caption{ \textbf{Three reference cases: spectral evolution.} 
The time evolution of the observed peak energy $E_\mathrm{p,obs}$ (top) and low-energy photon index $\alpha$ (bottom) is plotted for the three reference cases: A (black), B (red) and C (blue). In the top pannel, the energy range of the four BATSE channels is also indicated. }
\label{fig:ABC_Ep_alpha}
\end{figure}

The corresponding
GBM and LAT light curves 
 for these three  
cases are plotted in Figs.~8, 9 and 10 in \citet{daigne:11}, together with the predicted time-evolution of the peak-energy $E_\mathrm{p,obs}$ and the low-energy photon index $\alpha$ of the synchrotron component (GBM range), assuming a
 source redshift $z=1$. 
We have plotted for each case the normalized light curves in the four BATSE channels in \reffig{fig:ABC_4BATSEchannels} and 
 the spectral evolution 
 in \reffig{fig:ABC_Ep_alpha}.
 In the three cases, spectral evolution is found, with a global trend for the peak-energy to follow the intensity, and with a hard-to-soft evolution during the pulse decay.

%%%%%%%%%%%%%%%%%%%%%%%%%%%%%%%%%%%%%%%%%%%%%%
%%%%%%%%%%%%%%%%%%%%%%%%%%%%%%%%%%%%%%%%%%%%%%
%%%%%%%%%%%%%%%%%%%%%%%%%%%%%%%%%%%%%%%%%%%%%%
%%%
%%% FIGURE 2 : reference cases A,B,C - spectral evolution
%%%
%%%%%%%%%%%%%%%%%%%%%%%%%%%%%%%%%%%%%%%%%%%%%%
%%%%%%%%%%%%%%%%%%%%%%%%%%%%%%%%%%%%%%%%%%%%%%
%%%%%%%%%%%%%%%%%%%%%%%%%%%%%%%%%%%%%%%%%%%%%%
%%%%%%%%%%%%%%%%%%%%%%%%%%%%%%%%%%%%%%%%%%%%%%
%%%%%%%%%%%%%%%%%%%%%%%%%%%%%%%%%%%%%%%%%%%%%%
%%%%%%%%%%%%%%%%%%%%%%%%%%%%%%%%%%%%%%%%%%%%%%

\subsection{Origin of the temporal and spectral evolution}
Three possible time scales can in principle govern the observed evolution: the radiative timescale for the cooling of shock accelerated electrons, the hydrodynamical timescale associated to the propagation of internal shocks,
and the time scale associated to the curvature effect or high-latitude emission. The radiative time scale has to be the shortest to allow for the observed short time scale variability \citep{rees:94,sari:96,kobayashi:97} and can not be the main driver of the observed spectral evolution. The curvature effect leads to a strong temporal and spectral evolution which is not observed during most of the prompt emission \citep{fenimore:96,dermer:04,shenoy:13}, but is most probably responsible for the early steep decay found in the X-ray afterglow  by \textit{Swift}/XRT
\citep[see][for a recent discussion in the context of different GRB prompt emission models]{hascoet:12}, as demonstrated in several studies
 \citep[see e.g.][]{kumar:00,liang:06,butler:07,zhang:07,qin:08,genet:09,willingale:10}.
Therefore, the temporal and spectral evolution of GRB pulses in the internal shock model has to be mainly governed by the hydrodynamical timescale \citep{daigne:03}.

When an internal shock 
is propagating within the outflow, 
the corresponding 
 bolometric luminosity  
 is given by
\begin{equation}
L_\mathrm{bol}  =  f_\mathrm{rad} \epsilon_\mathrm{e} 4\pi R^2 \Gamma_*^2 \rho_* \epsilon_* c\, ,
\end{equation}
where $R$ is the shock radius,
$\Gamma_*$ is the Lorentz factor of the shocked material, $\rho_*$ and $\epsilon_*$ are the mass 
 and the specific internal energy density in the shocked region (comoving frame), and $f_\mathrm{rad}$ the radiative efficiency, which is close to $1$ for synchrotron radiation in fast cooling regime. For typical values representative of cases A, B and C  close to their maximum, we have
\begin{eqnarray}
L_\mathrm{bol}
& \simeq & 2.7\times 10^{51}\,  f_\mathrm{rad} 
\left(\frac{\epsilon_\mathrm{e}}{1/3}\right)
\left(\frac{R}{3\times 10^{14}\, \mathrm{cm}}\right)^2 
\nonumber\\
& &
\times\left(\frac{\Gamma_*}{300}\right)^2 
\left(\frac{\rho_*}{10^{-14}\, \mathrm{g.cm^{-3}}}\right) 
\left(\frac{\epsilon_* /c^2}{0.1}\right)\,
\mathrm{erg.s^{-1}}
\, ,
\label{eq:lbol}
\end{eqnarray}
\reffig{fig:B_Dynamics} shows  
the time evolution of $\Gamma_*$, $\rho_*$ and $\epsilon_*$ in 
cases A and B. Case C would show a similar behavior.
In these three examples,
the bolometric luminosity is initially rising when the shock forms (increase of $\Gamma_*$ and $\epsilon_*$), reaches a maximum, and then decreases again due to the radial expansion (decrease of $\epsilon_*$ and~$\rho_*$): see \citet{bosnjak:09} for a more detailed discussion.
 This leads to the pulse shape of the light curve. In more realistic cases, a large number of shock waves will form and propagate in the flow and several of them can contribute at the same time, leading to a complex, multi-pulses, light curve.

%%%%%%%%%%%%%%%%%%%%%%%%%%%%%%%%%%%%%%%%%%%%%%
%%%%%%%%%%%%%%%%%%%%%%%%%%%%%%%%%%%%%%%%%%%%%%
%%%%%%%%%%%%%%%%%%%%%%%%%%%%%%%%%%%%%%%%%%%%%%
%%%
%%% FIGURE 3 : reference cases A,B,C : pulse width and time lags
%%%
%%%%%%%%%%%%%%%%%%%%%%%%%%%%%%%%%%%%%%%%%%%%%%
%%%%%%%%%%%%%%%%%%%%%%%%%%%%%%%%%%%%%%%%%%%%%%
%%%%%%%%%%%%%%%%%%%%%%%%%%%%%%%%%%%%%%%%%%%%%%
\begin{figure}[!t]
\begin{center}
\includegraphics[width=0.45\textwidth]{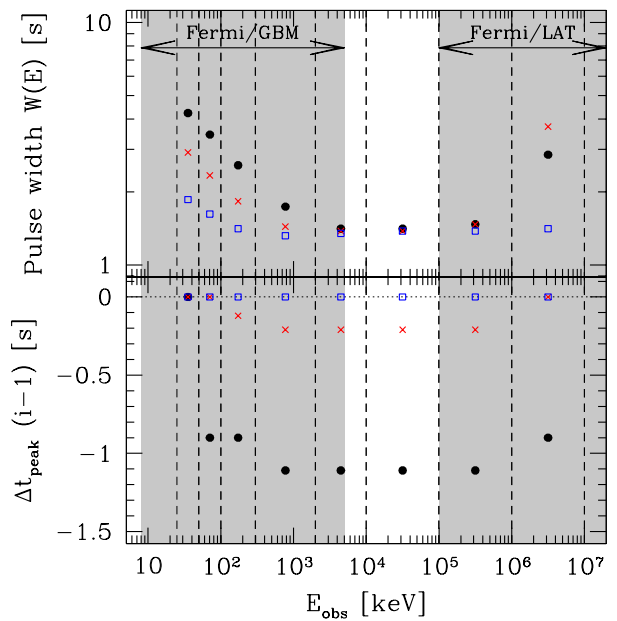}
\end{center}
%\vspace*{-3ex}

\caption{\textbf{Three reference cases: dependence of the pulse shape on energy.}  \textit{Top:} pulse width as a function of  energy. \textit{Bottom:} position of the pulse maximum in channel $i$ with respect to the lowest considered energy band, 20-50 keV (BATSE channel 1). Dashed lines show the energy bands for which the light curves were calculated. 
The first 4 energy bands correspond to the same four BATSE channels used in \reffig{fig:ABC_4BATSEchannels}. The shaded regions indicate the spectral coverage of  \textit{Fermi} GBM and LAT detectors. Black dots correspond to reference case A, red crosses to case B, and blue square symbols to case C. 
}
\label{fig:ABC_WE_lags}
\end{figure}
%%%%%%%%%%%%%%%%%%%%%%%%%%%%%%%%%%%%%%%%%%%%%%
%%%%%%%%%%%%%%%%%%%%%%%%%%%%%%%%%%%%%%%%%%%%%%
%%%%%%%%%%%%%%%%%%%%%%%%%%%%%%%%%%%%%%%%%%%%%%

In addition to the evolution of the bolometric luminosity, the spectral evolution affect the details of the pulse shape in a given energy band.
When the synchrotron peak is not too much affected by inverse Compton scattering in Klein-Nishina regime (see below), the observed peak energy can be expressed as $E_\mathrm{p,obs}\simeq \Gamma_* h \nu'_\mathrm{m}$, where 
$\nu'_\mathrm{m}$ is the peak of the synchrotron spectrum in the comoving frame, leading to
\begin{equation}
E_\mathrm{p,obs}\simeq \frac{380\, \mathrm{keV}}{1+z}\left(\frac{\Gamma_*}{300}\right)\left(\frac{\Gamma_\mathrm{m}}{10^4}\right)^2\left(\frac{B}{3000\, \mathrm{G}}\right)\, ,
\end{equation}
where 
$B$ is the intensity of the magnetic field (comoving frame) and $\Gamma_\mathrm{m}$ the minimum Lorentz factor of the power-law distribution of accelerated electrons. These two last quantities can be estimated from the microphysics parameters, leading to
\begin{eqnarray}
E_\mathrm{p,obs} & \simeq & \frac{1.4\, \mathrm{MeV}}{1+z} 
\left(\frac{\Gamma_*}{300}\right)\left(\frac{\rho_*}{10^{-14}\, \mathrm{g.cm^{-3}}}\right)^{1/2}\left(\frac{\epsilon_*/c^2}{0.1}\right)^{5/2}\nonumber\\
& &\times \left(\frac{\epsilon_\mathrm{B}}{1/3}\right)^{1/2} \left(\frac{(p-2)/(p-1)}{1/3}\right)^2 
\left(\frac{\epsilon_\mathrm{e}}{1/3}\right)^2
\left(\frac{\zeta}{0.001}\right)^{-2}
\, ,
\label{eq:Epeak}
\end{eqnarray}
using typical values for $\Gamma_*$, $\rho_*$ and $\epsilon_*$
at the time corresponding to the
maximum of the 
light curve in case A or B.
The spectrum around $E_\mathrm{p,obs}$ is very close to the standard fast cooling synchrotron spectrum (i.e. $\alpha=-3/2$) in case A and shows a larger
 low-energy photon index in case B ($\alpha\simeq -1.1$) due to inverse Compton scattering in Klein Nishina regime. As shown in \citet{daigne:11}, this effect is expected for large values of both the Compton parameter given by
\begin{equation}
Y_\mathrm{Th}=\frac{p-2}{p-1}\frac{\epsilon_\mathrm{e}}{\epsilon_\mathrm{B}}
\label{eq:Yth}
\end{equation}
and the parameter $w_\mathrm{m}$ defined by
\begin{equation}
w_\mathrm{m} = \Gamma_\mathrm{m} \frac{h \nu'_\mathrm{m}}{m_\mathrm{e}c^2}\, .
\label{eq:wm}
\end{equation}
Note that the Compton parameter $Y_\mathrm{Th}$ defined in \refeq{eq:Yth} is computed assuming the Thomson regime for the scatterings. The effective Compton parameter in the simulations presented here is always smaller due to the Klein-Nishina corrections. 
Close to the maximum of the pulse light curve, $Y_\mathrm{Th}\simeq 110$ (resp. $0.33$) and $w_\mathrm{m}\simeq 60$ (resp. $40$) for case B (resp. case A).

From \refeqs{eq:Epeak}{eq:wm}, the evolution of the physical conditions ($\Gamma_*$, $\rho_*$ and $\epsilon_*$) in the shocked region during the propagation of the shock wave leads to an evolution of the peak energy and the spectral shape (particularly the photon index $\alpha$) and is therefore at the origin of the observed spectral evolution.
The combination of the evolution of bolometric power $L_\mathrm{bol}$ and this spectral evolution allows to understand the details of the pulse shape in a given energy channel. The luminosity at a given photon energy can be written
\begin{equation}
L(E_\mathrm{obs}) = \frac{L_\mathrm{bol}}{E_\mathrm{p,obs}} \mathcal{S}\left(\frac{E_\mathrm{obs}}{E_\mathrm{p,obs}}\right)\, ,
\end{equation}
where $\mathcal{S}$ is the spectral shape, normalized by $\int_{0}^\infty \mathcal{S}(x) dx = 1$. Then  the flux in energy channel $\left[E_\mathrm{1,obs};E_\mathrm{2,obs}\right]$ is given by
\begin{equation}
F_{12}\simeq \frac{L_\mathrm{bol}}{4\pi D_\mathrm{L}^2} \times \int_{E_\mathrm{1,obs}/E_\mathrm{p,obs}}^{E_\mathrm{2,obs}/E_\mathrm{p,obs}} \mathcal{S}(x) dx\, .
\label{eq:flux12}
\end{equation}
The light curve is shaped by the two terms in \refeq{eq:flux12}. The evolution of the first term has been described above and is responsible for the general shape of the light curve, with a clear peak. The spectral correction contained in the second term is more complicated : it depends on the energy channel and 
is responsible for the time lags, evolution of the pulse shape with energy, etc. \citep{daigne:98,daigne:03,hafizi:07,boci:10}.

The temporal and spectral evolution which has just been described is valid as long as the internal shock phase is active and the observed emission is dominated by the contribution of on-axis radiation from shocked regions. After the last internal shock has disappeared, the observed emission is due to the high latitude emission and is therefore governed by the geometry of the shells and the relativistic Doppler effects, which lead to the asymptotic evolution $E_\mathrm{p}\propto 1/t_\mathrm{obs}$. The corresponding bolometric flux decreases as $F_\mathrm{bol}\propto 1/t_\mathrm{obs}^3$ leading to a spectral evolution which is much too fast to reproduce the properties of observed bursts, as pointed out by \citet{fenimore:96}. For instance the HIC would have a slope $\delta < 0.5$ instead of the typical observed value $\delta \simeq 0.5- 1$. This clearly indicates that the spectral evolution in GRB pulses is not governed by such geometrical effects but by the physics of the internal dissipative and radiative process in the outflow as described above.

%%%%%%%%%%%%%%%%%%%%%%%%%%%%%%%%%%%%%%%%%%%%%%
%%%%%%%%%%%%%%%%%%%%%%%%%%%%%%%%%%%%%%%%%%%%%%
%%%%%%%%%%%%%%%%%%%%%%%%%%%%%%%%%%%%%%%%%%%%%%
%%%
%%% Figures 4 : reference cases A,B,C : HIC and HFC
%%%
%%%%%%%%%%%%%%%%%%%%%%%%%%%%%%%%%%%%%%%%%%%%%%
%%%%%%%%%%%%%%%%%%%%%%%%%%%%%%%%%%%%%%%%%%%%%%
%%%%%%%%%%%%%%%%%%%%%%%%%%%%%%%%%%%%%%%%%%%%%%
\begin{figure*}[!t]
\begin{center}
\begin{tabular}{cc}
\centering
\includegraphics[width=0.38\textwidth]{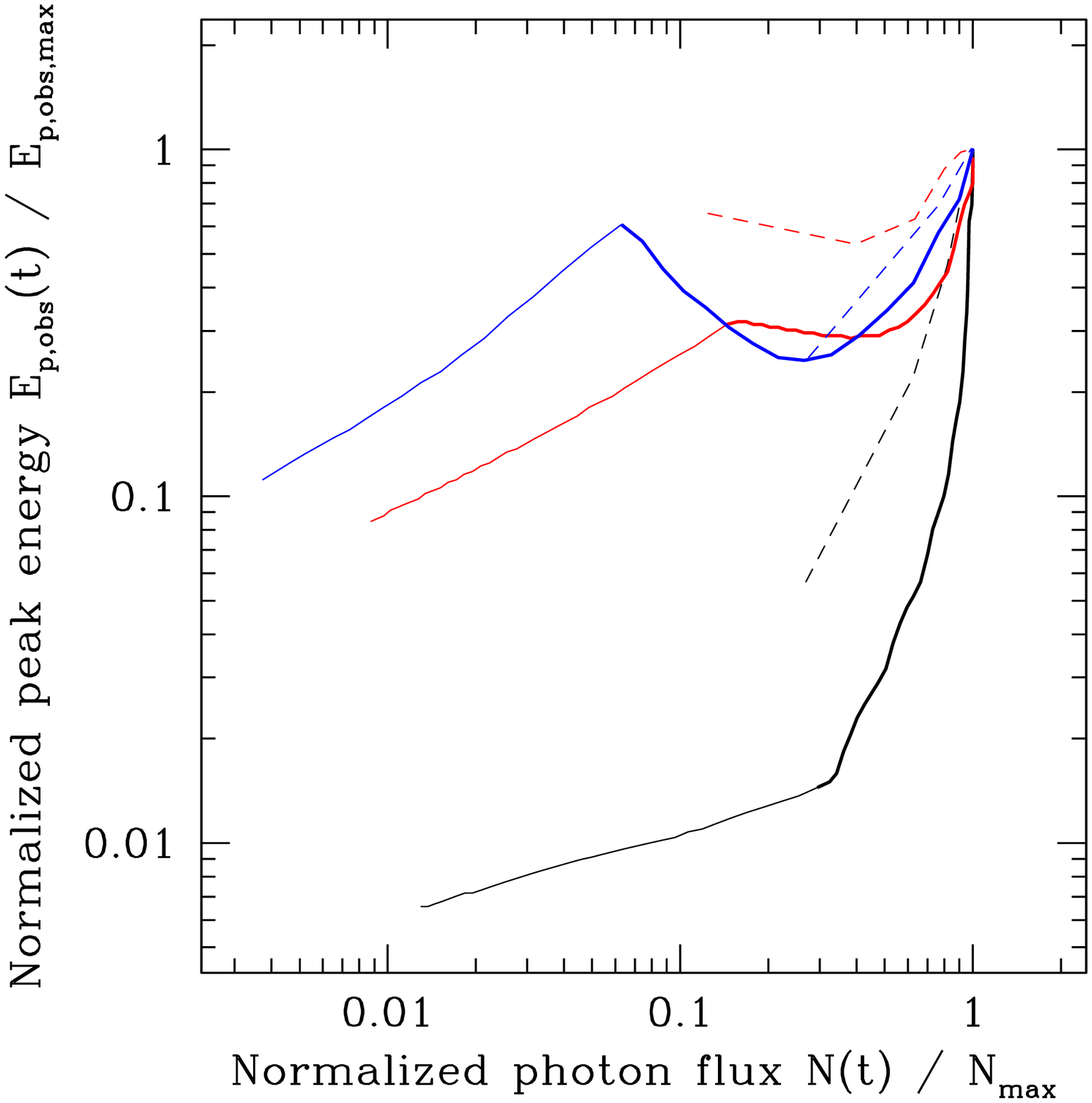}&
\includegraphics[width=0.38\textwidth]{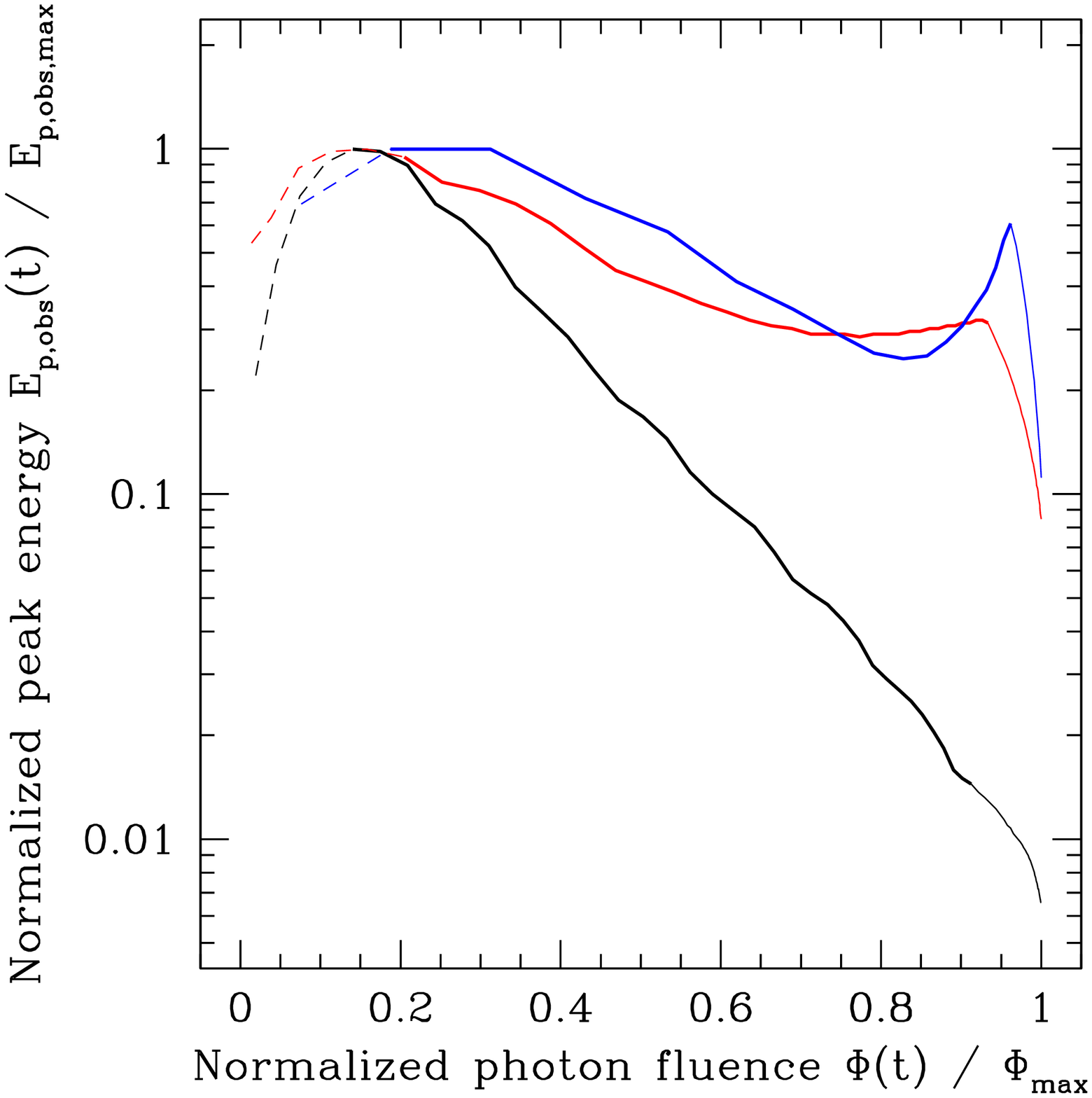}\\
\end{tabular}
\end{center}
%\vspace*{-1ex}

\caption{ \textbf{Three reference case: hardness-intensity and hardness-fluence correlations.} The peak energy energy is plotted as a function of the photon flux (left panel) and photon fluence (right panel) between 20 keV and 2 MeV for the three reference cases, A (black), B (red) and C (blue). The peak energy, photon flux and photon fluence are normalized by their respective maximum values.
The dashed lines show the behavior during the rise of the pulse, thick lines correspond to the decay phase, and the thin lines are the high latitude emission. This last stage is unlikely to be observed in complex lightcurves, except during the early steep decay in X-rays. 
}
\label{fig:ABC_HIC_HFC}
\end{figure*}
%%%%%%%%%%%%%%%%%%%%%%%%%%%%%%%%%%%%%%%%%%%%%%
%%%%%%%%%%%%%%%%%%%%%%%%%%%%%%%%%%%%%%%%%%%%%%
%%%%%%%%%%%%%%%%%%%%%%%%%%%%%%%%%%%%%%%%%%%%%%

\subsection{Comparison to observations}
We now compare more quantitatively the spectral evolution found in reference cases A, B and C 
to the available observations. \reffig{fig:ABC_4BATSEchannels} shows the normalized light curve in the four BATSE channels for each case, and \reffig{fig:ABC_Ep_alpha} shows the corresponding spectral evolution. The break at $t_\mathrm{HLE,obs}\simeq 5.4\, \mathrm{s}$ corresponds to the end of the on-axis emission and the transition to the high-latitude emission which decays faster. In a complex light curve, the probablity to observe this transition is low as it is very likely hidden by the overlapping on-axis emission of another pulse \citep[see however][]{sonbas:12}. It is only at the end of the prompt phase that the high-latitude emission can be observed (early steep decay in X-rays).
We focus here on the pulse temporal and spectral properties during the on-axis emission, i.e. for $t_\mathrm{obs}<t_\mathrm{HLE,obs}$.

There is a good qualitative agreement with 
observations (see \refsec{sec:obs}). In particular, it is found that the pulse light curve is asymetric with a fast rise and a slow decay (\refobsPulseAsymmetry); this assymetry is stronger at lower energy (\refobsPulseEnergy) and 
the width of the light curve is broader at lower energy (\refobsPulseEnergy);
the pulse light curve peaks earlier at higher energy (\refobsLags); the peak energy decreases in the decay phase of the pulse (\refobsHardToSoftEvolution). 

The more quantitative comparison is less satisfactory:

\noindent--~The shape of the pulse in channel 1 and 2 for case A  shows a double peak which does not seem to be usually observed in GRBs. 
This is due to a too strong spectral evolution in this case (see \reffig{fig:ABC_Ep_alpha}). The peak energy has a maximum $E_\mathrm{p,obs}\simeq 800\, \mathrm{keV}$ close to the first maximum of the light curve at $t_\mathrm{obs} \simeq 1.5\, \mathrm{s}$ and then decreases rapidly, reaching $E_\mathrm{p,obs}\simeq 10\, \mathrm{keV}$ at the end of the on-axis emission at $t_\mathrm{HLE,obs}$. The rapid decrease of $E_\mathrm{p,obs}$ leads to an increase of the second term in \refeq{eq:flux12} (spectral correction) for channels 2 and 1 which are successively crossed by $E_\mathrm{p,obs}$.
This compensates the decrease of the bolometric luminosity to create a second peak in the light curve for these two channels.

\noindent--~In the three cases, the pulse width does not decrease enough with energy. This is illustrated in \reffig{fig:ABC_WE_lags} (upper panel), where $W(E)$ is plotted as 
a function of the mean energy of the channel, defined by $E=\sqrt{E_\mathrm{min}E_\mathrm{max}}$ where $E_\mathrm{min}$ and $E_\mathrm{max}$ are the lower and higher energy bounds. In agreement with observations, it is found that the width follows approximatively a power-law evolution $W(E) \propto E^{-a}$.
However, the value of the index $a$, listed in \reftab{tab:allmodels}, is usually a little too small compared to 
observations
 (\refobsPulseEnergy). The best agreement is found for case A.

\noindent--~Time lags between the different channels are too large, as illustrated in \reffig{fig:ABC_WE_lags} (lower panel). 
A quantitative comparison with observations is more delicate as 
we do not
measure the lag by the maximum of correlation like in the method described by e.g. \citet{band:97}, which is usually  applied to GRB data.
We rather plot the difference between the time of the maximum of the light curve in a given channel, and the time of the maximum in channel 1. 
The observed trend is reproduced (channel 4 peaks first, channel 1 peaks the last), but, especially in case A, the typical lags are too long compared to 
observations
(\refobsLags).

%%%%%%%%%%%%%%%%%%%%%%%%%%%%%%%%%%%%%%%%%%%%%%
%%%%%%%%%%%%%%%%%%%%%%%%%%%%%%%%%%%%%%%%%%%%%%
%%%%%%%%%%%%%%%%%%%%%%%%%%%%%%%%%%%%%%%%%%%%%%
%%%
%%% Figure 5 : impact of the electron slope p - peak energy and alpha
%%%
%%%%%%%%%%%%%%%%%%%%%%%%%%%%%%%%%%%%%%%%%%%%%%
%%%%%%%%%%%%%%%%%%%%%%%%%%%%%%%%%%%%%%%%%%%%%%
%%%%%%%%%%%%%%%%%%%%%%%%%%%%%%%%%%%%%%%%%%%%%%
\begin{figure}[!t]
\includegraphics[width=0.40\textwidth]{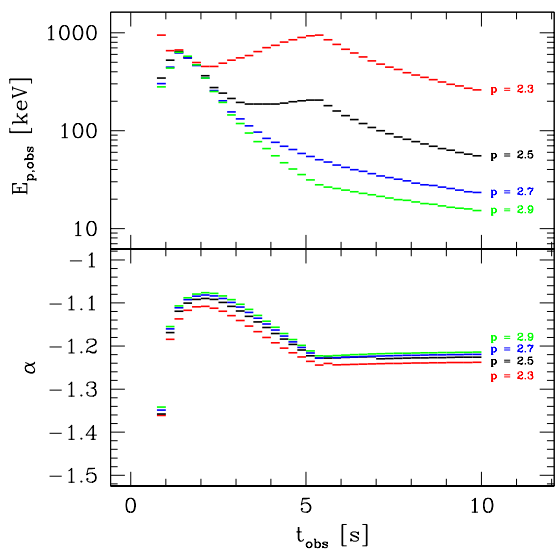}
\caption{\textbf{Impact of the electron slope $p$: spectral evolution.} 
The time evolution of the observed peak energy $E_\mathrm{p,obs}$ (top) and low-energy photon index $\alpha$ (bottom) in case B is plotted for  different values of the relativistic electron slope $p$. The evolution for $p=3.3$ and $p=3.9$ has also been computed but is not shown as the curves coincide with the case $p=2.9$.
In each case, the value of $\zeta$ is adjusted to keep the same observed peak energy of the time-integrated spectrum (see \reftab{tab:allmodels}).}
\label{fig:B_Epeak_p}
\end{figure}
%%%%%%%%%%%%%%%%%%%%%%%%%%%%%%%%%%%%%%%%%%%%%%
%%%%%%%%%%%%%%%%%%%%%%%%%%%%%%%%%%%%%%%%%%%%%%
%%%%%%%%%%%%%%%%%%%%%%%%%%%%%%%%%%%%%%%%%%%%%%

\noindent--~The HIC is qualitatively reproduced, as illustrated in  \reffig{fig:ABC_HIC_HFC} (left panel), where the peak energy $E_\mathrm{p,obs}$ is plotted as a function of the photon flux $N$ in the 20--2000 keV range in log--log scale. The peak energy increases during the rise of the light curve, reaching a maximum which precedes the maximum of the intensity; then it decreases during the pulse decay (hard to soft evolution, \refobsHardToSoftEvolution). However, the quantitative behavior is not reproduced. During the pulse decay, the peak energy should follow $E_\mathrm{p,obs} \propto N^\delta$ with $\delta \simeq 0.5-1$ or $E_\mathrm{p,obs}\propto F^\kappa$ with $\kappa\simeq 0.4-1.2$, $N$ and $F$ being the photon and energy fluxes (\refobsHIC). This is not found in our simulations. Cases B  and C do not show a simple power-law behavior during the decay phase. Case A is closer to the expected evolution, but the slopes $\delta$ and $\kappa$ are too large compared with BATSE and GBM observations.

\noindent--~The same disagreement in found for the HFC, as illustrated in \reffig{fig:ABC_HIC_HFC} (right panel).
 Again, cases B and C do not really show the expected behavior $\log{E_\mathrm{p,obs}}\propto \Phi$, whereas the agreement is better for case A, with a quasi exponentially decay for the peak energy as a function of the photon fluence (\refobsHFC). 

In the three cases, a careful analysis shows that the disagreements listed above are due to the fact that the spectral evolution, even if it reproduces qualitatively the hard-to-soft evolution (\refobsHardToSoftEvolution), is usually to strong (the peak energy, and sometimes the spectral slopes, vary too much). 
 We note that cases B and C have the strongest disagreement with the observed HIC and HFC. This is due to a peculiar spectral evolution in this case (see \reffig{fig:ABC_Ep_alpha}): the peak energy is initially decreasing during the pulse decay, as expected, but then does not evolve any more (case B: it is only slightly increasing at the end of the pulse) or starts to increase instead of decreasing (case C), which is usually not observed. 
This unexpected behaviour is analysed in \S\ref{sec:slopep}.

We conclude 
 that the three reference cases A, B and C, which are representative of the scenario where the prompt GRB emission is dominated by the synchrotron radiation from shock accelerated electrons in internal shocks, can reproduce well the qualitative spectral evolution identified in GRBs observed by BATSE and GBM and described in \refsec{sec:obs},
 but that there is no quantitative agreement, the spectral evolution being usually too strong in the three simulated bursts compared to observations. This apparent disagreement may help us to shed light on some uncertainties in the considered scenario. From \refeq{eq:lbol} and \refeqs{eq:Epeak}{eq:wm}, there are two groups of factors that can impact our results:
 the assumptions for the microphysics in the shocked region, and the assumptions for the initial conditions in the outflow that impact the dynamics of the internal shock phase.
 We now investigate in \refsec{sec:EffectMicrophysics} and \refsec{sec:EffectDynamics} how these factors
affect the predicted spectral evolution and if
the observed  evolution can be better reproduced.

\section{Impact of the uncertainties on the microphysics}
\label{sec:EffectMicrophysics}

%%%%%%%%%%%%%%%%%%%%%%%%%%%%%%%%%%%%%%%%%%%%%%
%%%%%%%%%%%%%%%%%%%%%%%%%%%%%%%%%%%%%%%%%%%%%%
%%%%%%%%%%%%%%%%%%%%%%%%%%%%%%%%%%%%%%%%%%%%%%
%%%
%%% Figure 6 : impact of the electron slope p - HIC
%%%
%%%%%%%%%%%%%%%%%%%%%%%%%%%%%%%%%%%%%%%%%%%%%%
%%%%%%%%%%%%%%%%%%%%%%%%%%%%%%%%%%%%%%%%%%%%%%
%%%%%%%%%%%%%%%%%%%%%%%%%%%%%%%%%%%%%%%%%%%%%%
\begin{figure}[!t]
\includegraphics[width=0.42\textwidth]{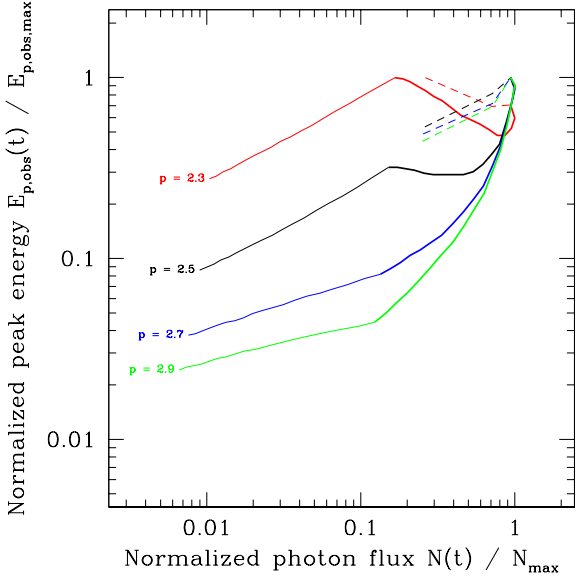}
\caption{\textbf{Impact of the electron slope $p$: hardness-intensity correlation.}
The HIC diagram in case B is plotted for  different values of the relativistic electron slope $p$.}
\label{fig:B_HIC_p}
\end{figure}
%%%%%%%%%%%%%%%%%%%%%%%%%%%%%%%%%%%%%%%%%%%%%%
%%%%%%%%%%%%%%%%%%%%%%%%%%%%%%%%%%%%%%%%%%%%%%
%%%%%%%%%%%%%%%%%%%%%%%%%%%%%%%%%%%%%%%%%%%%%%

Due to a lack of better, physically motivated, prescriptions, the microphysics in the shocked region (magnetic field amplification, electron acceleration) is simply parametrized by $\epsilon_\mathrm{B}$, $\epsilon_\mathrm{e}$, $\zeta$ and $p$ in our model of the internal shock phase. Such an over-simplistic description may be at the origin of some disagreements between the simulated and observed spectral evolution pointed out in the previous section. We investigate here this possibility and focus on the two most relevant cases, i.e. A and B.

\subsection{Steeper electron slopes ?}
\label{sec:slopep}
\begin{figure}[!t]
\includegraphics[width=0.43\textwidth]{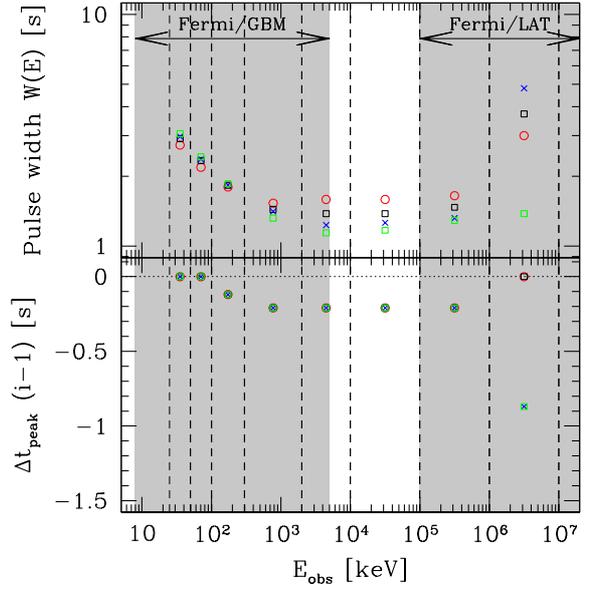}
\caption{\textbf{Impact of the electron slope $p$: pulse width and time lags.}
Same as in \reffig{fig:ABC_WE_lags} for Case B with four different values of the electron slope $p$. Color code is the same as in \reffig{fig:B_Epeak_p}. In the lower panel, the time lags are the same for the four cases except for the highest energy channel where the $p=2.3$ and $p=2.5$ cases show an important lag whereas the $p=2.7$ and $p=2.9$ cases do not show any lag.}
\label{fig:B_p_WE_lags}
\end{figure}

Unfortunately, the current physical understanding of shock acceleration in the mildly relativistic regime does not allow to predict the value of the slope $p$ of the electron distribution.
A 'standard' value $p=2.5$ is adopted in the three reference cases of \citet{daigne:11}. 
We investigate here how changing $p$ affects the predicted temporal and spectral properties of the simulated pulse. The most dramatic change is for case B. For low values of $p$, case B shows a very peculiar spectral evolution in the decay phase of the pulse, the peak energy starting to increase after an initial decrease.
The more standard simple hard to soft evolution is recovered above  a threshold $p\simeq 2.7$, as shown in \reffig{fig:B_Epeak_p} where the time evolution of $E_\mathrm{p,obs}$ and $\alpha$ is plotted for case B for different values of $p$.
Moreover, we find that increasing $p$ improves the quantitative comparison between the predicted and observed spectral evolution in 
 cases A and B: see for instance the values of the index $a$
 and the slopes $\delta$ and $\kappa$
  in \reftab{tab:allmodels} and the corresponding \reffig{fig:B_p_WE_lags} (pulse width, time lags) and
  \reffig{fig:B_HIC_p} (HIC).

The peculiar evolution of the peak energy for $p\la 2.7$ can be understood from the theoretical spectra computed by \citet{daigne:11} in this spectral regime where the synchrotron spectrum is affected by inverse Compton scattering in Klein Nishina regime (see Fig.~2 in \citet{daigne:11}). For large values of the Compton parameter $Y_\mathrm{Th}$ 
which favors the scatterings, not only the low-energy slope of the synchrotron spectrum is affected, but also the peak, which moves towards higher energy, with the spectrum around the peak becoming very flat.  The synchrotron peak energy in the comoving frame is not any more simply proportional to $B \Gamma_\mathrm{m}^2$ and the standard spectral evolution governed by \refeq{eq:Epeak} is lost. For higher $p$, we find numerically that the peak is not shifted any more, even for very large values of $Y_\mathrm{Th}$. This is also confirmed by the semi-analytical calculation of the Klein-Nishina effects on optically thin synchrotron and synchrotron self-Compton spectrum made by  \citet{nakar:09}: see the discussion of their case IIc, which is relevant for our reference case B\footnote{See \citet{daigne:11} for a correspondence between the notations of \citet{nakar:09} and \citet{daigne:11} and a 
 comparison.}. They find a threshold at $p=3$, slightly larger than $p\simeq 2.7$. This difference is probably due to the additional approximations which are necessary to allow for the analytical treatment, compared to the full numerical resolution we use here.

%%%%%%%%%%%%%%%%%%%%%%%%%%%%%%%%%%%%%%%%%%%%%%
%%%%%%%%%%%%%%%%%%%%%%%%%%%%%%%%%%%%%%%%%%%%%%
%%%%%%%%%%%%%%%%%%%%%%%%%%%%%%%%%%%%%%%%%%%%%%
%%%
%%% Figure 7 : impact of the electron slope p - pulse width and time lags
%%%
%%%%%%%%%%%%%%%%%%%%%%%%%%%%%%%%%%%%%%%%%%%%%%
%%%%%%%%%%%%%%%%%%%%%%%%%%%%%%%%%%%%%%%%%%%%%%
%%%%%%%%%%%%%%%%%%%%%%%%%%%%%%%%%%%%%%%%%%%%%%
%%%%%%%%%%%%%%%%%%%%%%%%%%%%%%%%%%%%%%%%%%%%%%
%%%%%%%%%%%%%%%%%%%%%%%%%%%%%%%%%%%%%%%%%%%%%%
%%%%%%%%%%%%%%%%%%%%%%%%%%%%%%%%%%%%%%%%%%%%%%

%%%%%%%%%%%%%%%%%%%%%%%%%%%%%%%%%%%%%%%%%%%%%%
%%%%%%%%%%%%%%%%%%%%%%%%%%%%%%%%%%%%%%%%%%%%%%
%%%%%%%%%%%%%%%%%%%%%%%%%%%%%%%%%%%%%%%%%%%%%%
%%%
%%% Figure 8 : varying zeta : time-evolving spectrum
%%%
%%%%%%%%%%%%%%%%%%%%%%%%%%%%%%%%%%%%%%%%%%%%%%
%%%%%%%%%%%%%%%%%%%%%%%%%%%%%%%%%%%%%%%%%%%%%%
%%%%%%%%%%%%%%%%%%%%%%%%%%%%%%%%%%%%%%%%%%%%%%
\begin{figure}[!t]
\includegraphics[width=0.48\textwidth]{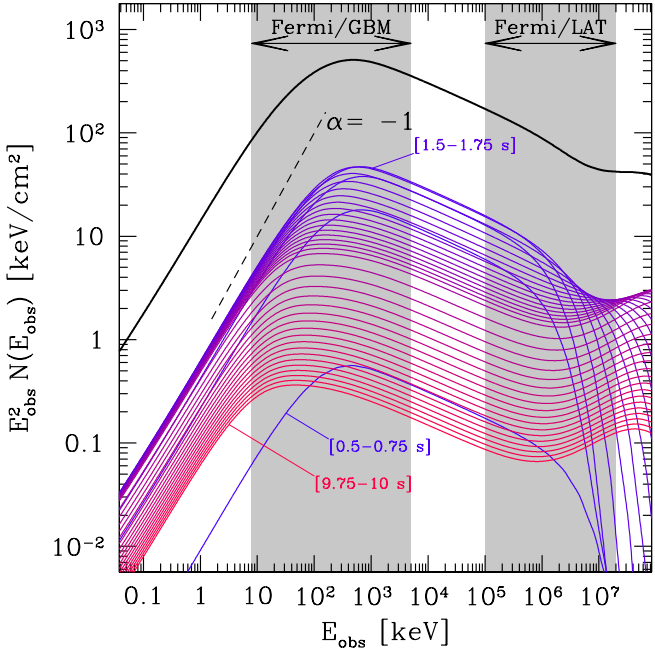}
\caption{\textbf{Impact of a varying accelerated electron fraction $\zeta$: spectral evolution.}
Time resolved spectra are plotted for case B assuming 
a varying parameter $\zeta$ during the propagation of internal shocks (see text).
The corresponding light curves are shown in the right panel of  \reffig{fig:AB_zeta_4BATSEchannels}. 
Each curve corresponds to a time interval of $0.25$ s, starting with the bluest curve (0.5--0.75 s) and finishing with the reddest one (9.75--10 s). The time bin corresponding to the pulse maximum is indicated (1.5--1.75 s). The time-integrated spectrum is also plotted as a thick black line. A thin dashed line of photon index $\alpha=-1$ is indicated for comparison. }
\label{fig:B_zeta_Spectrum}
\end{figure}
%%%%%%%%%%%%%%%%%%%%%%%%%%%%%%%%%%%%%%%%%%%%%%
%%%%%%%%%%%%%%%%%%%%%%%%%%%%%%%%%%%%%%%%%%%%%%
%%%%%%%%%%%%%%%%%%%%%%%%%%%%%%%%%%%%%%%%%%%%%%

%%%%%%%%%%%%%%%%%%%%%%%%%%%%%%%%%%%%%%%%%%%%%%
%%%%%%%%%%%%%%%%%%%%%%%%%%%%%%%%%%%%%%%%%%%%%%
%%%%%%%%%%%%%%%%%%%%%%%%%%%%%%%%%%%%%%%%%%%%%%
%%%
%%% Figure 9 : varying zeta : HIC
%%%
%%%%%%%%%%%%%%%%%%%%%%%%%%%%%%%%%%%%%%%%%%%%%%
%%%%%%%%%%%%%%%%%%%%%%%%%%%%%%%%%%%%%%%%%%%%%%
%%%%%%%%%%%%%%%%%%%%%%%%%%%%%%%%%%%%%%%%%%%%%%
\begin{figure*}[!t]
\begin{center}
\begin{tabular}{cc}
\centering
\includegraphics[width=0.41\textwidth]{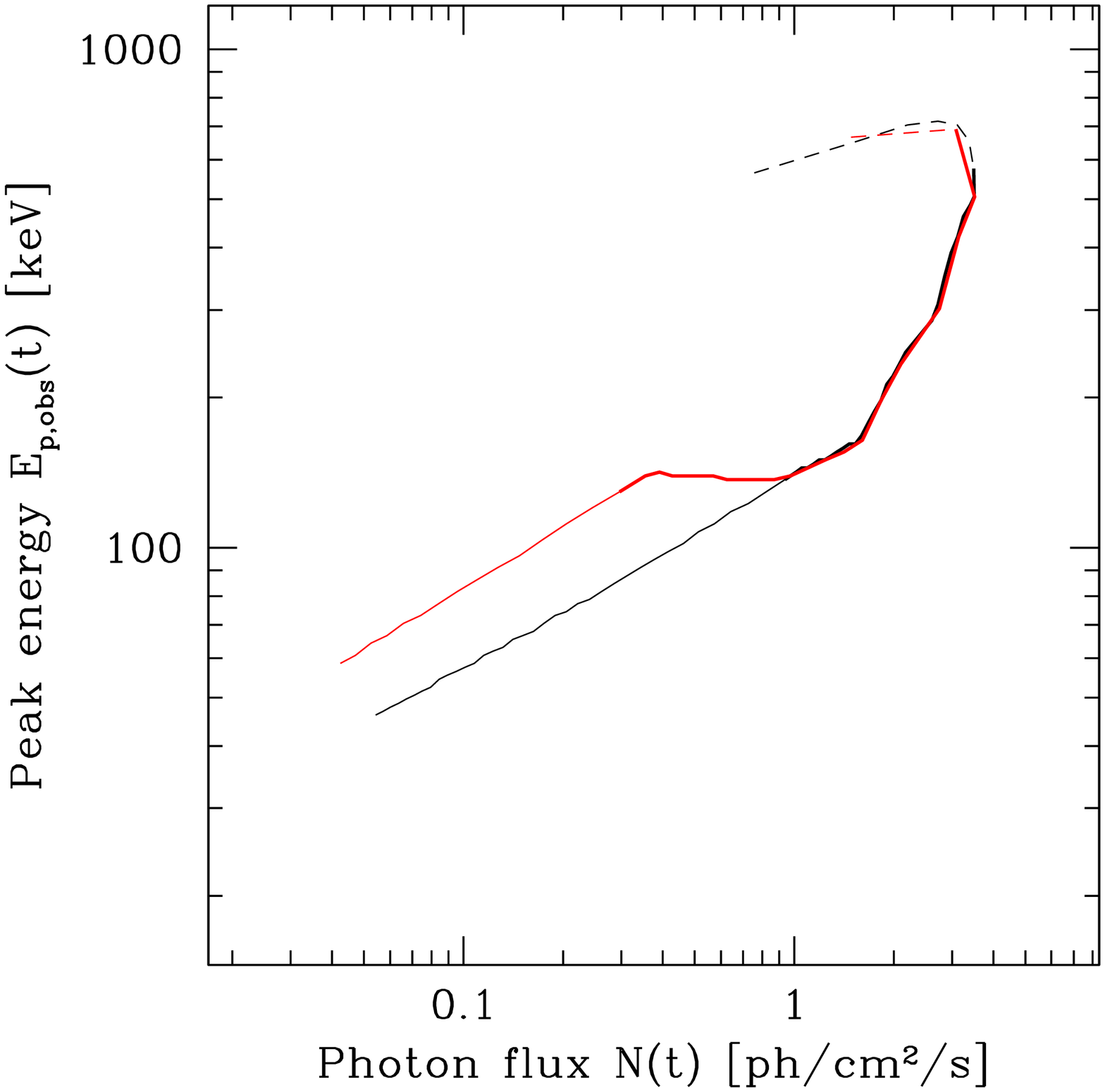} &
\includegraphics[width=0.41\textwidth]{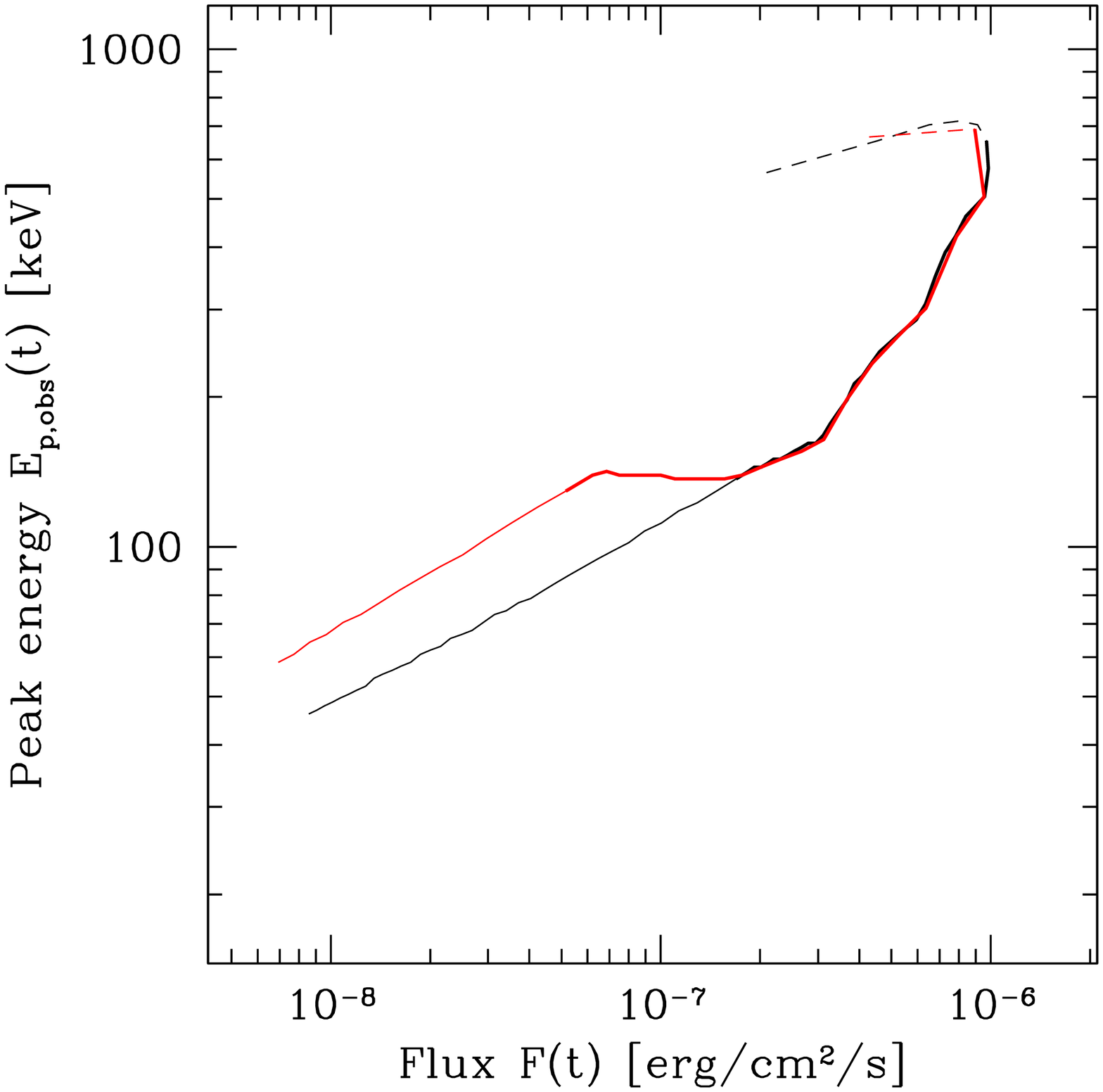}
\\
\end{tabular}
\end{center}
%\vspace*{-3ex}

\caption{\textbf{Impact of a varying accelerated electron fraction $\zeta$: hardness-intensity correlation.}
Same as in \reffig{fig:ABC_HIC_HFC} for case B with a varying parameter $\zeta$ during the propagation of internal shocks (see text), either with the same ejection duration as in the reference case B (black) or with an extended high-Lorentz factor tail in the ejecta (red) : see \S\ref{sec:InitGamma}. }
\label{fig:B_zeta_HIC}
\end{figure*}
%%%%%%%%%%%%%%%%%%%%%%%%%%%%%%%%%%%%%%%%%%%%%%
%%%%%%%%%%%%%%%%%%%%%%%%%%%%%%%%%%%%%%%%%%%%%%
%%%%%%%%%%%%%%%%%%%%%%%%%%%%%%%%%%%%%%%%%%%%%%

%%%%%%%%%%%%%%%%%%%%%%%%%%%%%%%%%%%%%%%%%%%%%%
%%%%%%%%%%%%%%%%%%%%%%%%%%%%%%%%%%%%%%%%%%%%%%
%%%%%%%%%%%%%%%%%%%%%%%%%%%%%%%%%%%%%%%%%%%%%%
%%%
%%% Figure 10 : varying zeta : pulse shape
%%%
%%%%%%%%%%%%%%%%%%%%%%%%%%%%%%%%%%%%%%%%%%%%%%
%%%%%%%%%%%%%%%%%%%%%%%%%%%%%%%%%%%%%%%%%%%%%%
%%%%%%%%%%%%%%%%%%%%%%%%%%%%%%%%%%%%%%%%%%%%%%
\begin{figure*}[!t]
\begin{center}
\begin{tabular}{cc}
\centering
\includegraphics[width=0.41\textwidth]{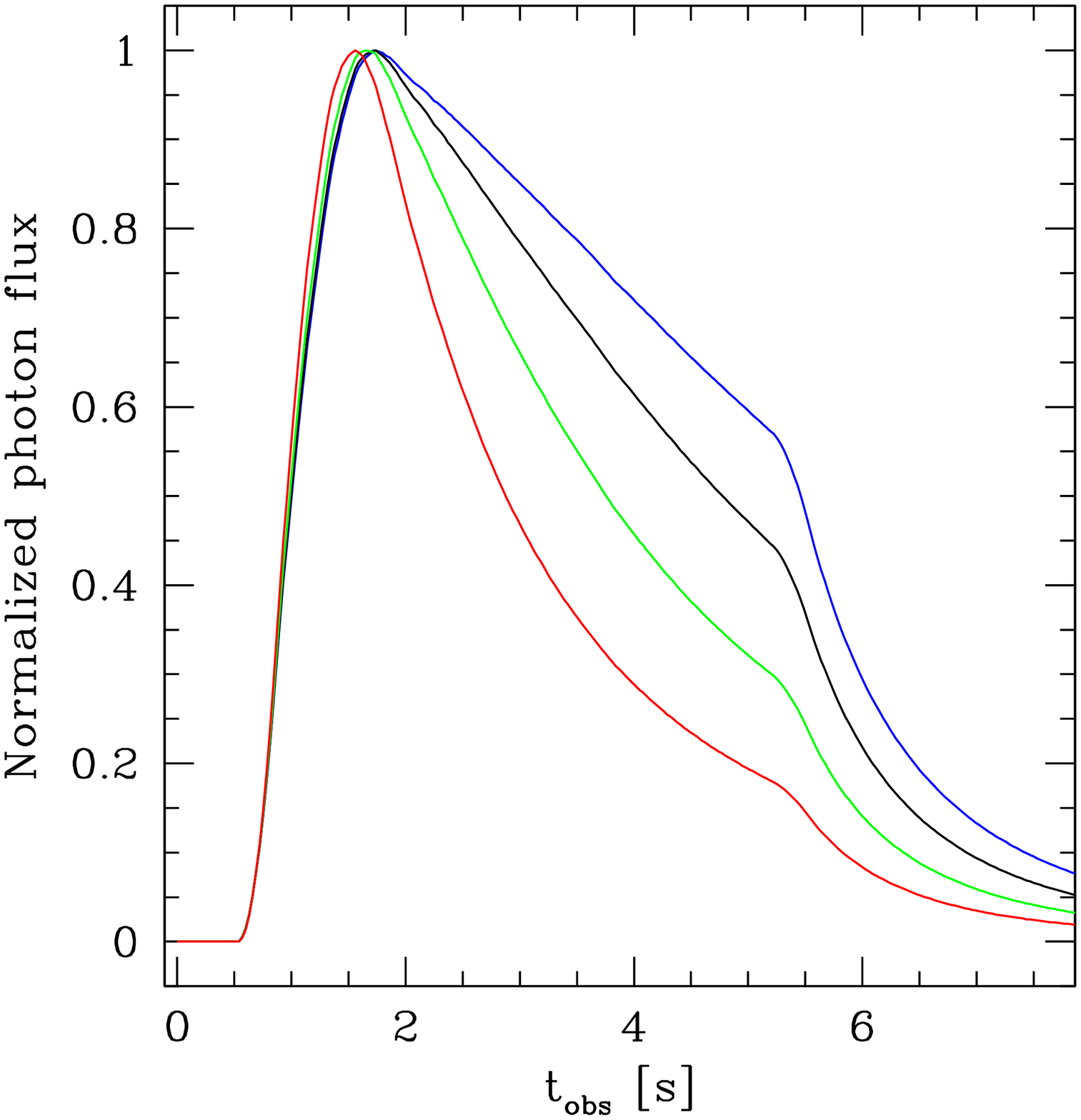}&
\includegraphics[width=0.41\textwidth]{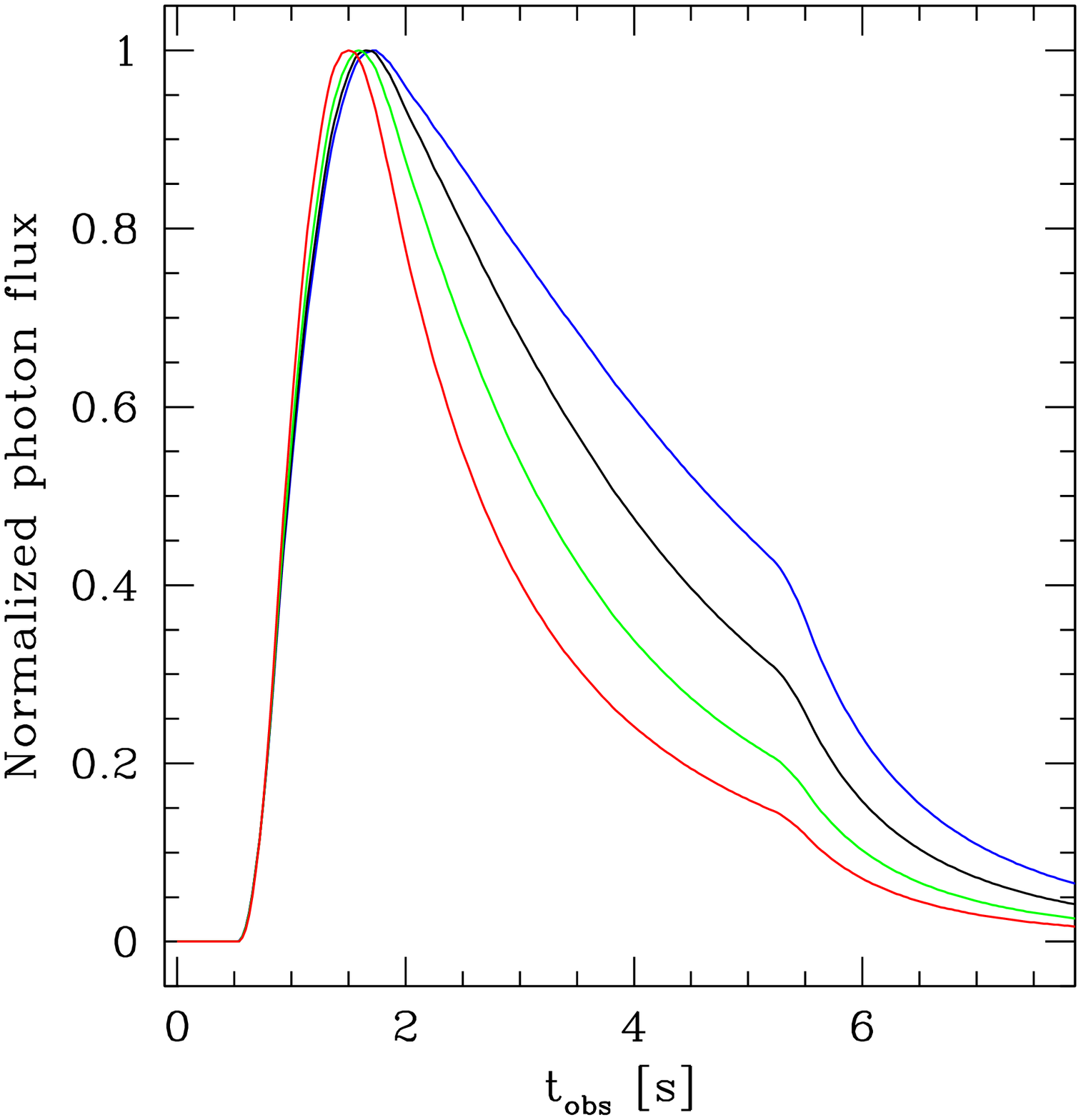}\\
\end{tabular}
\end{center}
%\vspace*{-3ex}

\caption{\textbf{Impact of a varying accelerated electron fraction $\zeta$: pulse shape.} Same as in \reffig{fig:ABC_4BATSEchannels} for case A (left panel) and B (right panel) assuming a varying parameter $\zeta$ during the propagation of internal shocks (see text). Conventions are the same as in \reffig{fig:ABC_4BATSEchannels}. The corresponding time-evolving spectrum for case B is shown in \reffig{fig:B_zeta_Spectrum}.}
\label{fig:AB_zeta_4BATSEchannels}
\end{figure*}
%%%%%%%%%%%%%%%%%%%%%%%%%%%%%%%%%%%%%%%%%%%%%%
%%%%%%%%%%%%%%%%%%%%%%%%%%%%%%%%%%%%%%%%%%%%%%
%%%%%%%%%%%%%%%%%%%%%%%%%%%%%%%%%%%%%%%%%%%%%%

%%%%%%%%%%%%%%%%%%%%%%%%%%%%%%%%%%%%%%%%%%%%%%
%%%%%%%%%%%%%%%%%%%%%%%%%%%%%%%%%%%%%%%%%%%%%%
%%%%%%%%%%%%%%%%%%%%%%%%%%%%%%%%%%%%%%%%%%%%%%
%%%
%%% Figure 11 : varying zeta : pulse width and time lags
%%%
%%%%%%%%%%%%%%%%%%%%%%%%%%%%%%%%%%%%%%%%%%%%%%
%%%%%%%%%%%%%%%%%%%%%%%%%%%%%%%%%%%%%%%%%%%%%%
%%%%%%%%%%%%%%%%%%%%%%%%%%%%%%%%%%%%%%%%%%%%%%
\begin{figure*}[!t]
\begin{center}
\begin{tabular}{cc}
\centering
\includegraphics[width=0.40\textwidth]{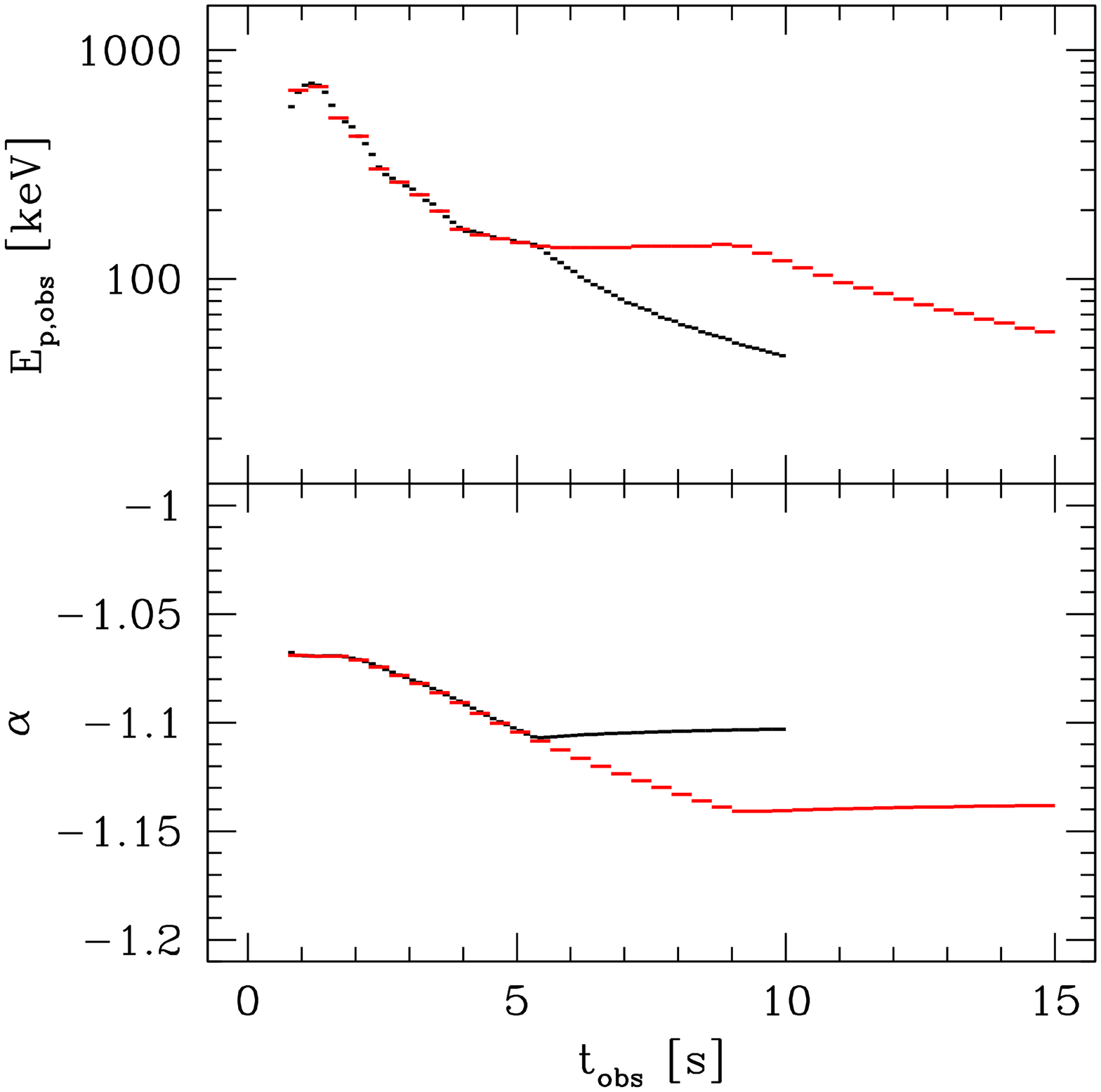}&
\includegraphics[width=0.40\textwidth]{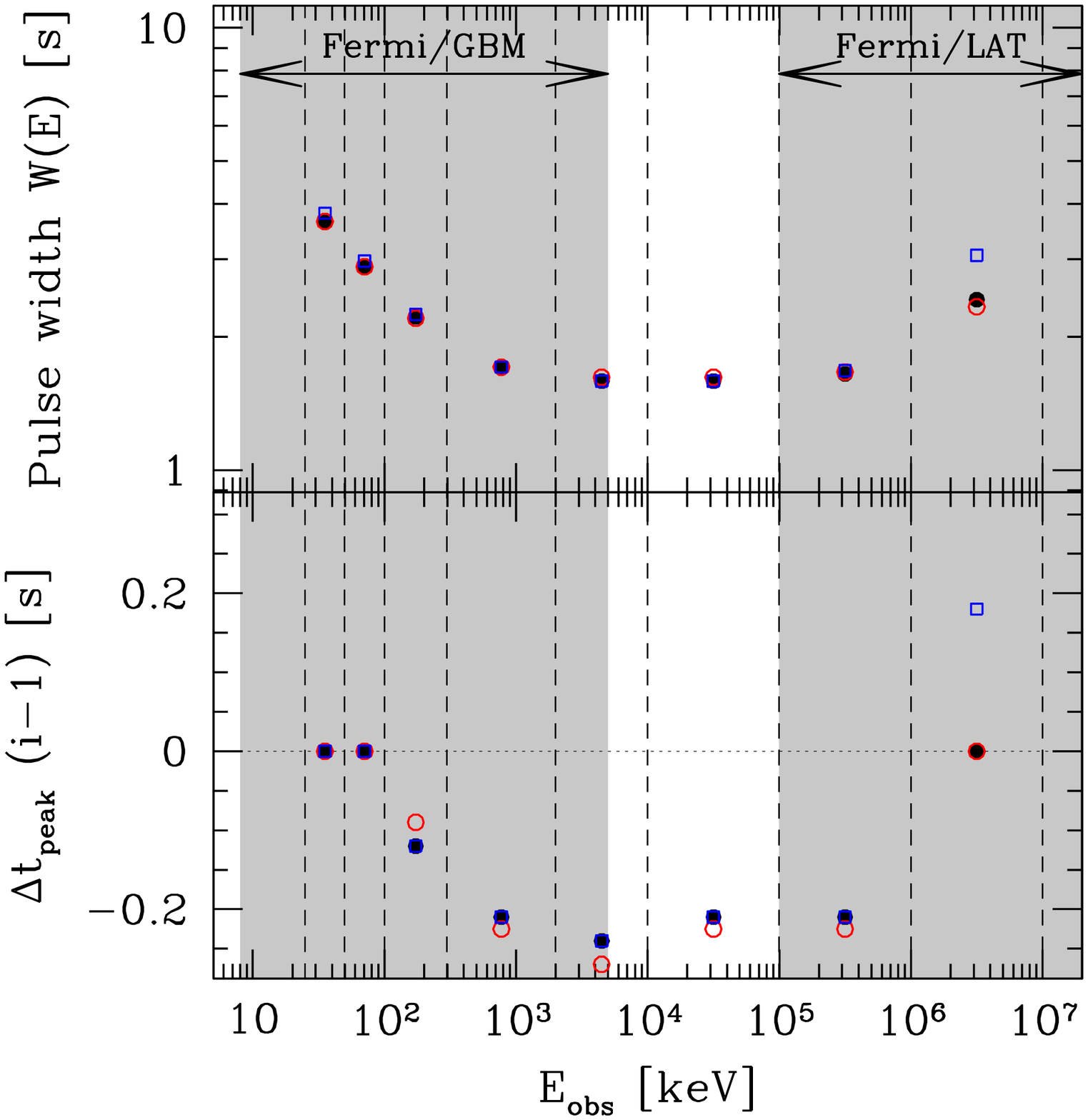}\\
\end{tabular}
\end{center}
\caption{\textbf{Impact of a varying accelerated electron fraction $\zeta$: spectral evolution -- pulse width and time lags.}
Curves are plotted for case B with a varying parameter $\zeta$ during  internal shocks (see text), either with the same ejection duration as in the reference case B (black) or with an extended high-Lorentz factor tail in the ejecta (red) : see \S\ref{sec:InitGamma}. 
For the first case (black, 2 s duration for the ejection), the corresponding spectra and light curves are plotted  in \reffig{fig:B_zeta_Spectrum} and in the right panel of \reffig{fig:AB_zeta_4BATSEchannels}.
\textit{Left panel:} the time evolution of the peak energy and the low-energy photon index.
The dashed vertical line indicates the start of the high latitude emission. \textit{Right panel:} evolution of the pulse width and time lags, as in \reffig{fig:ABC_WE_lags}.
}
\label{fig:B_zeta_Epeak_WE_lags}
\end{figure*}
%%%%%%%%%%%%%%%%%%%%%%%%%%%%%%%%%%%%%%%%%%%%%%
%%%%%%%%%%%%%%%%%%%%%%%%%%%%%%%%%%%%%%%%%%%%%%
%%%%%%%%%%%%%%%%%%%%%%%%%%%%%%%%%%%%%%%%%%%%%%

\subsection{Varying microphysics parameters ?}
\label{sec:zetavar}
A strong but common assumption 
is
to assume constant microphysics parameters in a GRB. 
The fact that different values of the parameters are needed to fit observations from different GRBs (see for instance the case of GRB afterglows as in the study by \citealt{panaitescu:01}) indicates on the other hand that there are no universal values. Then, it is highly probable that $\epsilon_\mathrm{B}$, $\epsilon_\mathrm{e}$, $\zeta$ and $p$ depend on the shock conditions and evolve during 
a GRB.
In \citet{daigne:03}, where only synchrotron radiation was included in a simple way,
it has been shown that assuming such variations of the microphysics parameters could greatly improve the comparison between the predicted and the observed spectral evolution 
in a pulse.
As there are no physically motivated prescriptions for such variations, we cannot fully explore this possibility. We only illustrate the effect in cases A and B, assuming a simple variation law for one
parameter, i.e.  
 $\zeta \propto \epsilon_*$
as suggested by \cite{bykov:96}, so that a larger fraction of electrons is accelerated when the shock is more violent.  We have normalized $\zeta$ to have a similar peak energy of the time-integrated spectrum as in the reference cases (see \reftab{tab:allmodels}).
From \refeq{eq:Epeak}, it appears clearly that this prescription will reduce the variations of $E_\mathrm{p,obs}$ during the pulse, as it is now only 
proportional to $\epsilon_*^{1/2}$ rather than $\epsilon_*^{5/2}$. 
This 
 is confirmed by our detailed radiative calculation,  as illustrated for case B in \reffig{fig:B_zeta_Spectrum} (spectrum)
and in 
\reffig{fig:B_zeta_Epeak_WE_lags}
 (left panel: spectral evolution).
  The peak energy decreases only by a factor $\sim 4$
 during the pulse decay and 
 $\alpha$ 
 remains close to $-1.1$ for most of the evolution.
In agreement with the results of \citet{daigne:98,daigne:03}, 
the pulse shape is dramatically improved in both cases (\reffig{fig:AB_zeta_4BATSEchannels}), and especially in case A where the double peak in the reference case has disappeared 
(left panel). 
In case B, there is in addition a general improvement of most relations, as illustrated in \reffig{fig:B_zeta_Epeak_WE_lags}, bottom-right panel (time lags) and top-right panel (pulse width),  \reffig{fig:B_zeta_HIC} (HIC), and from \reftab{tab:allmodels}.
  In summary, this modified case B with a varying electron fraction $\zeta$ has an overall good agreement with BATSE and GBM observations. Especially, the predicted spectral evolution reproduces now qualitatively and quantitatively  the observational constraints
  \refobsPulseAsymmetry\,
  to \refobsHFC\, 
  described in \refsec{sec:obs}.

The assumption $\zeta\propto \epsilon_*$ is suggested by \citet{bykov:96}, but one may expect variations of other microphysics parameters as well. We expect that any modification of the microphysics leading to a reduced dependence of the peak energy to the shock conditions will produce a similar improvement as described here.

We conclude 
 that the disagreement between the observed spectral evolution in GRBs and the predictions of the simplest version of the internal shock model illustrated by our reference cases A, B and C can be largely due to over-simplifying assumptions regarding the microphysics in the emitting shocked regions. Our current knowledge of the physics of mildly relativistic shocks does not allow yet 
to improve this description but we have illustrated 
 that both a qualitative and quantitative agreement can be achieved, for instance if the fraction of accelerated electron is varying with the shock strength.

\section{Impact of the uncertainties on the dynamics}
\label{sec:EffectDynamics}

\begin{figure*}[!t]
\begin{center}
\begin{tabular}{ccc}
\includegraphics[width=0.32\textwidth]{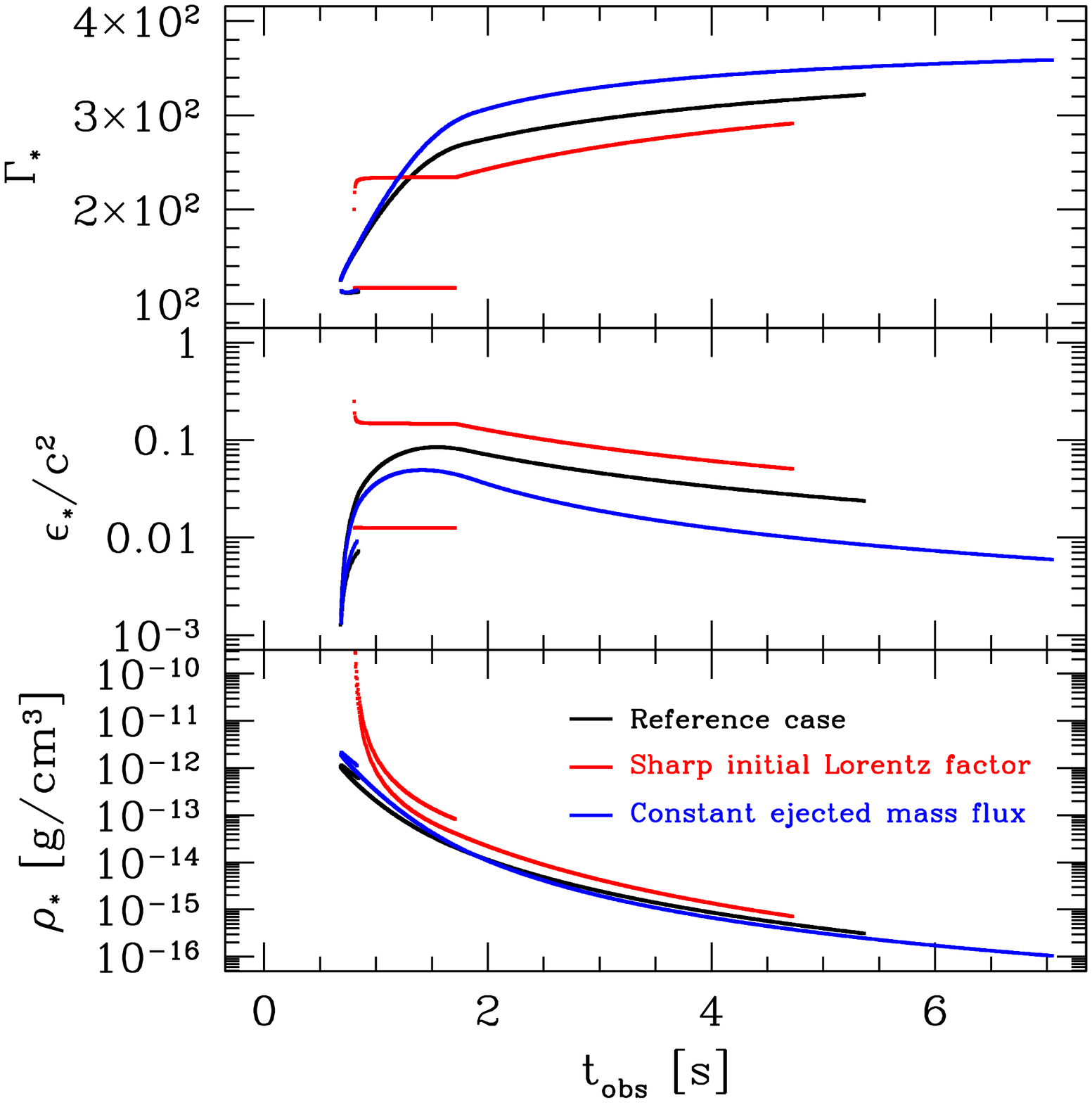} &
\includegraphics[width=0.32\textwidth]{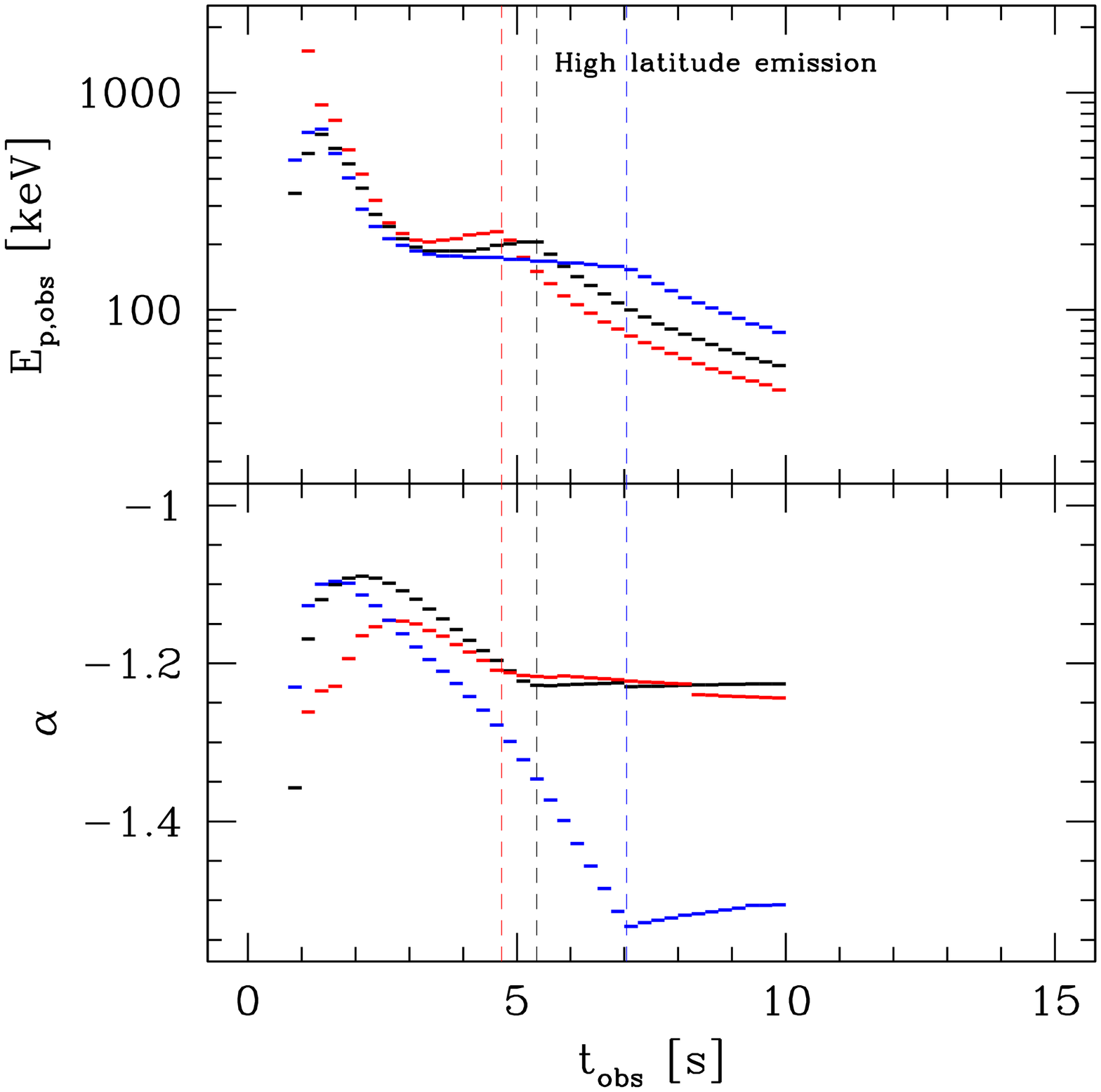} &
\includegraphics[width=0.32\textwidth]{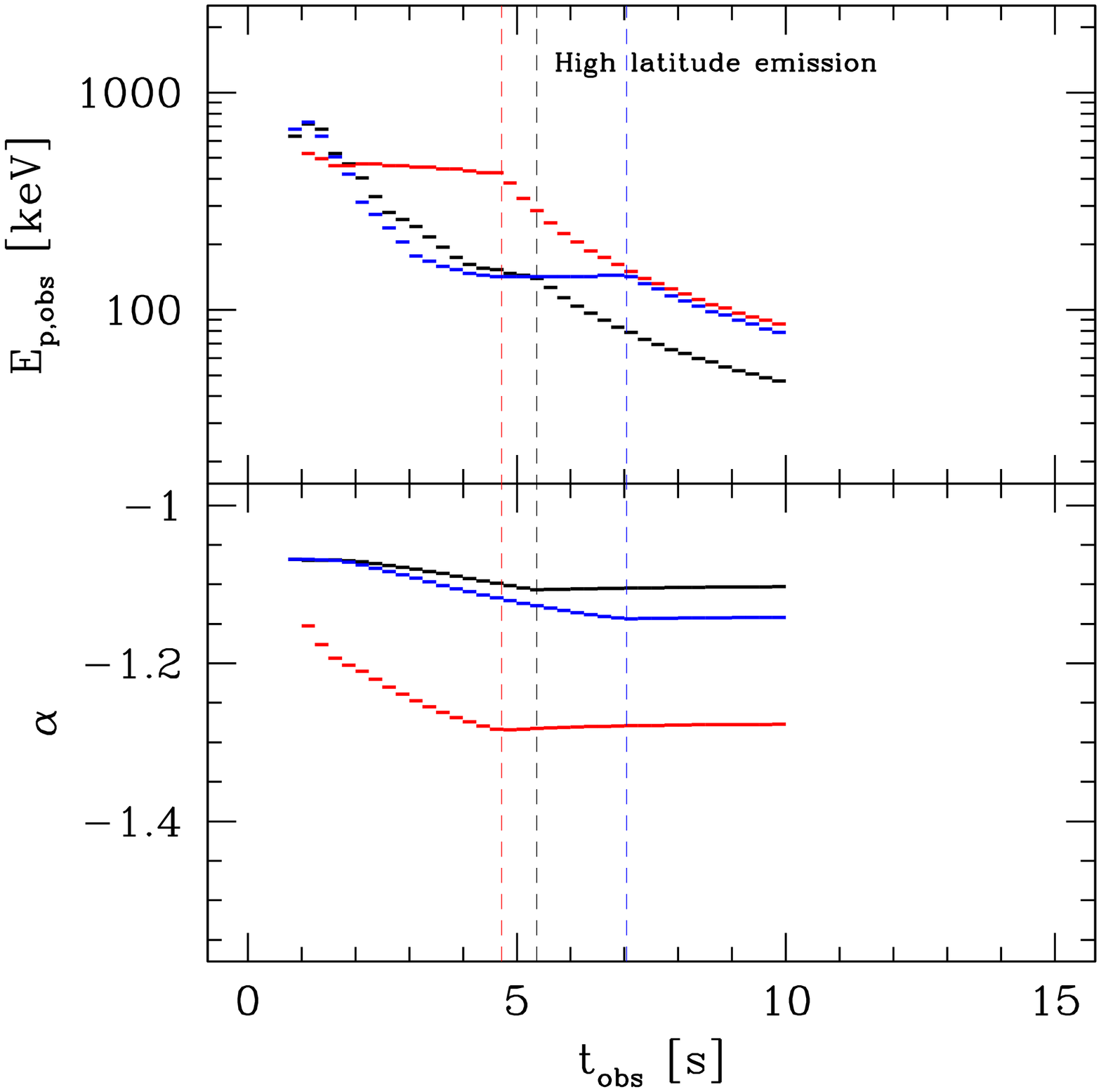}\\
\end{tabular}
\end{center}
\caption{\textbf{Evolution of the physical conditions in the shocked region.}  
\textit{left panel:}
The evolution of the Lorentz factor $\Gamma_*$ (top), the specific energy density $\epsilon_*$ (middle) and the mass density $\rho_*$ (bottom) is plotted for the reference case B (case A would show exactly the same evolution), case B with a sharp distribution of the initial Lorentz factor (see text) and case B with a constant ejected mass flux $\dot{M}$ rather than a constant injected kinetic power $\dot{E}$ (see text). For each case, two curves are seen, corresponding to the propagation of a  'forward' and a 'reverse' internal shocks;
\textit{Middle panel:} corresponding spectral evolution (same color code) assuming a constant fraction $\zeta$ of accelerated electrons;
\textit{Right panel:} corresponding spectral evolution (same color code) assuming a varying fraction $\zeta$ of accelerated electrons (see text).
}
\label{fig:B_Dynamics}
\end{figure*}

The physics of the central engine, 
and of the acceleration of the relativistic outflow, is not well  
understood. This does not allow to predict the initial conditions in the jet before the internal shock phase. 
 Typically, a pulse is due to the collision between two regions  
 with different Lorentz factors (in the reference cases, the 'slow' region correspond to
 $\Gamma(t_\mathrm{ej})=\Gamma_\mathrm{min}\to\Gamma_\mathrm{max}$, and the 'rapid' region to 
 $\Gamma=\Gamma_\mathrm{max}$, see \refeq{eq:SmoothGamma}). Two internal shocks form, a 'forward' and a 'reverse' shock (see \reffig{fig:B_Dynamics}). In the reference cases, the emission is entirely dominated by the 'reverse' internal shock. 
This assumption has
the advantage of simplicity to simulate a single pulse burst (as a building block for more complex light curves) but is
 not physically motivated. 
We now investigate how these assumptions can affect the dynamics of the internal shock phase and therefore the spectral evolution in pulses.

%%%%%%%%%%%%%%%%%%%%%%%%%%%%%%%%%%%%%%%%%%%%%%
%%%%%%%%%%%%%%%%%%%%%%%%%%%%%%%%%%%%%%%%%%%%%%
%%%%%%%%%%%%%%%%%%%%%%%%%%%%%%%%%%%%%%%%%%%%%%
%%%
%%% Figure 12 : physical conditions in the shocked regions
%%%
%%%%%%%%%%%%%%%%%%%%%%%%%%%%%%%%%%%%%%%%%%%%%%
%%%%%%%%%%%%%%%%%%%%%%%%%%%%%%%%%%%%%%%%%%%%%%
%%%%%%%%%%%%%%%%%%%%%%%%%%%%%%%%%%%%%%%%%%%%%%
%%%%%%%%%%%%%%%%%%%%%%%%%%%%%%%%%%%%%%%%%%%%%%
%%%%%%%%%%%%%%%%%%%%%%%%%%%%%%%%%%%%%%%%%%%%%%
%%%%%%%%%%%%%%%%%%%%%%%%%%%%%%%%%%%%%%%%%%%%%%

%%%%%%%%%%%%%%%%%%%%%%%%%%%%%%%%%%%%%%%%%%%%%%
%%%%%%%%%%%%%%%%%%%%%%%%%%%%%%%%%%%%%%%%%%%%%%
%%%%%%%%%%%%%%%%%%%%%%%%%%%%%%%%%%%%%%%%%%%%%%
%%%
%%% Figure 13 : impact of the initial Lorentz factor
%%%
%%%%%%%%%%%%%%%%%%%%%%%%%%%%%%%%%%%%%%%%%%%%%%
%%%%%%%%%%%%%%%%%%%%%%%%%%%%%%%%%%%%%%%%%%%%%%
%%%%%%%%%%%%%%%%%%%%%%%%%%%%%%%%%%%%%%%%%%%%%%
\begin{figure*}[!t]
\begin{center}
\begin{tabular}{ccc}
\centering
\includegraphics[width=0.32\textwidth]{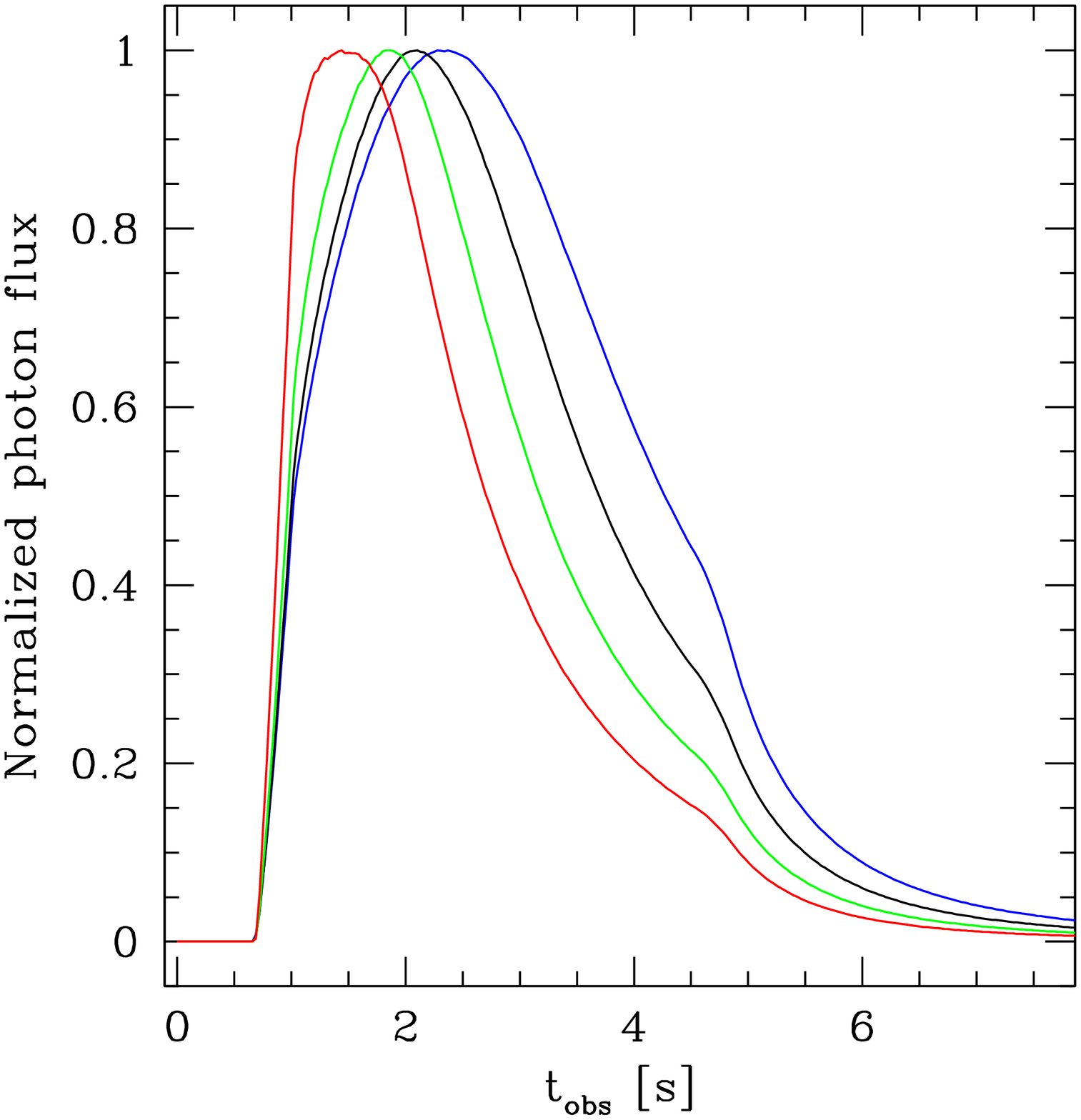}&
\includegraphics[width=0.32\textwidth]{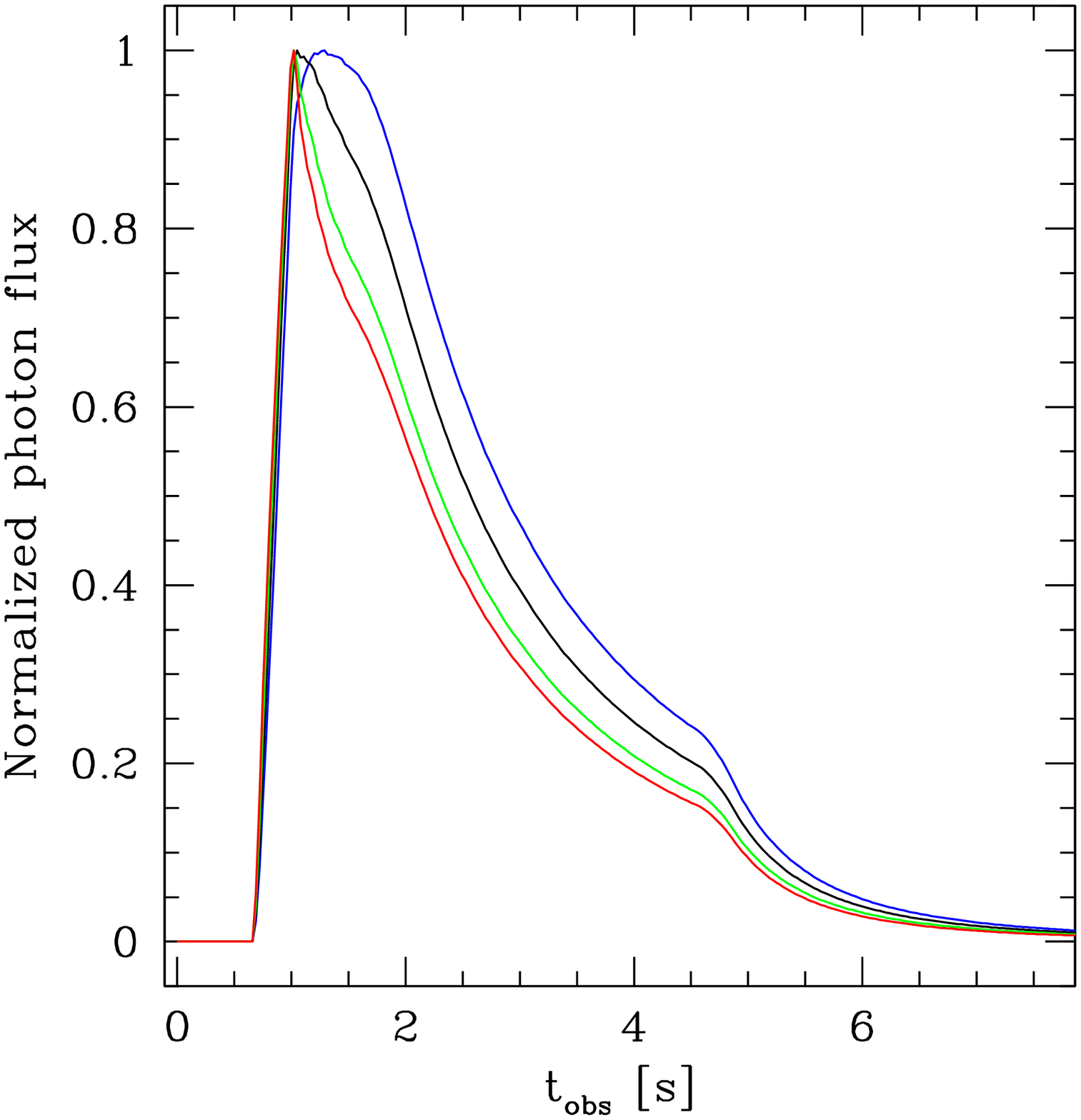}&
\includegraphics[width=0.32\textwidth]{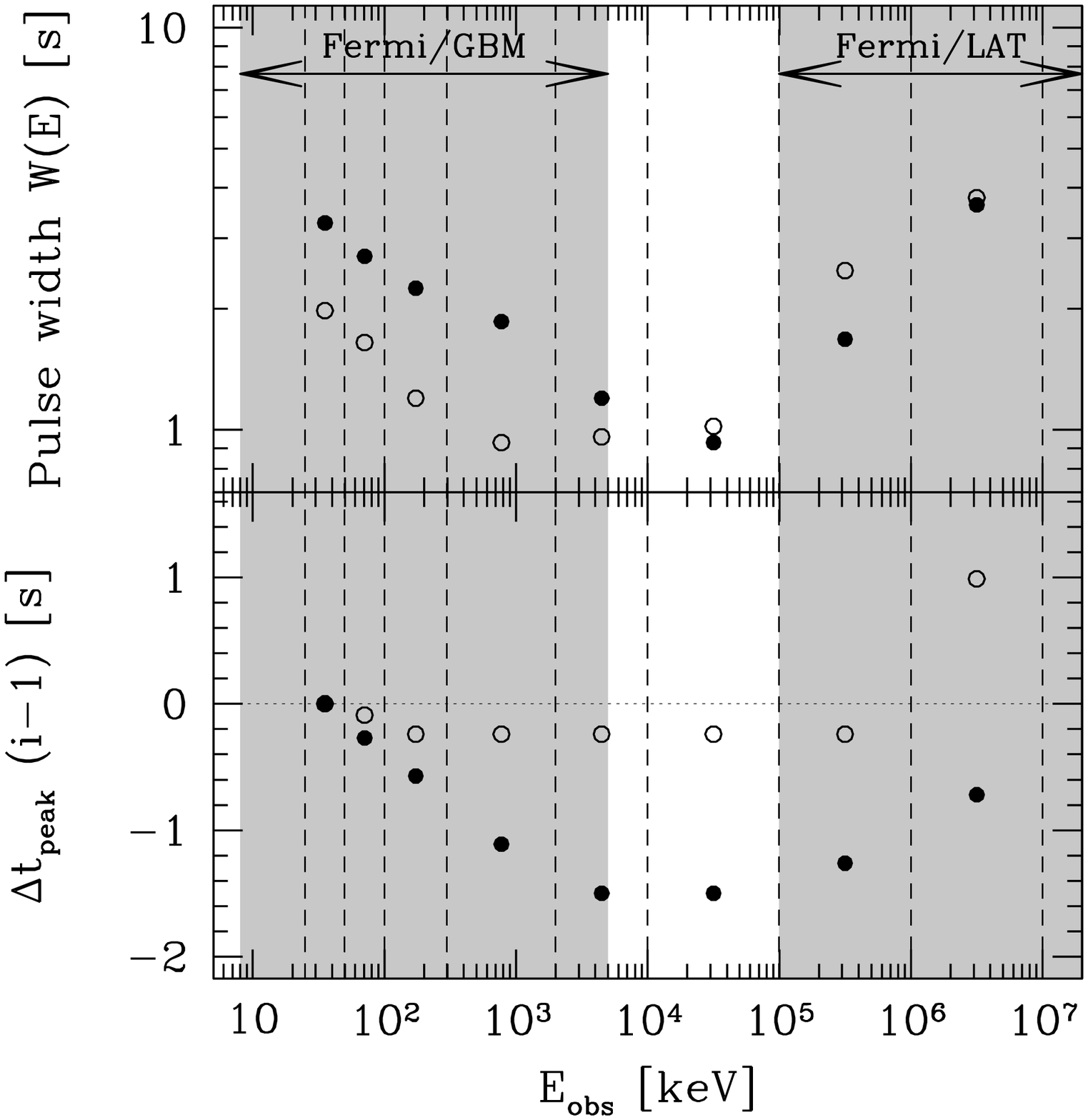}\\
\end{tabular}
\end{center}
\caption{\textbf{Impact of the initial distribution of the Lorentz factor in the outflow.} 
\textit{Left and middle panels:} normalized light curves corresponding to the 4 BATSE energy channels for case B computed with the sharp initial distribution of the Lorentz factor given by \refeq{eq:SteepGamma} and a constant (left) or varying (middle) accelerated electron fraction (see text). Conventions are the same as in \reffig{fig:ABC_4BATSEchannels};
\textit{Right panel:} pulse width and pulse maximum position as a function of energy for the same cases (filled circles: constant $\zeta$, open circles: varying $\zeta$).  The shaded regions and lines have the same meaning as in \reffig{fig:ABC_WE_lags}.}
\label{fig:B_Gamma_4BATSEchannels_WE_lags}
\end{figure*}
%%%%%%%%%%%%%%%%%%%%%%%%%%%%%%%%%%%%%%%%%%%%%%
%%%%%%%%%%%%%%%%%%%%%%%%%%%%%%%%%%%%%%%%%%%%%%
%%%%%%%%%%%%%%%%%%%%%%%%%%%%%%%%%%%%%%%%%%%%%%
 
%%%%%%%%%%%%%%%%%%%%%%%%%%%%%%%%%%%%%%%%%%%%%%
%%%%%%%%%%%%%%%%%%%%%%%%%%%%%%%%%%%%%%%%%%%%%%
%%%%%%%%%%%%%%%%%%%%%%%%%%%%%%%%%%%%%%%%%%%%%%
%%%
%%% Figure 14 : impact of the injected kinetic power
%%%
%%%%%%%%%%%%%%%%%%%%%%%%%%%%%%%%%%%%%%%%%%%%%%
%%%%%%%%%%%%%%%%%%%%%%%%%%%%%%%%%%%%%%%%%%%%%%
%%%%%%%%%%%%%%%%%%%%%%%%%%%%%%%%%%%%%%%%%%%%%%
\begin{figure*}[!t]
\begin{center}
\begin{tabular}{ccc}
\centering
\includegraphics[width=0.32\textwidth]{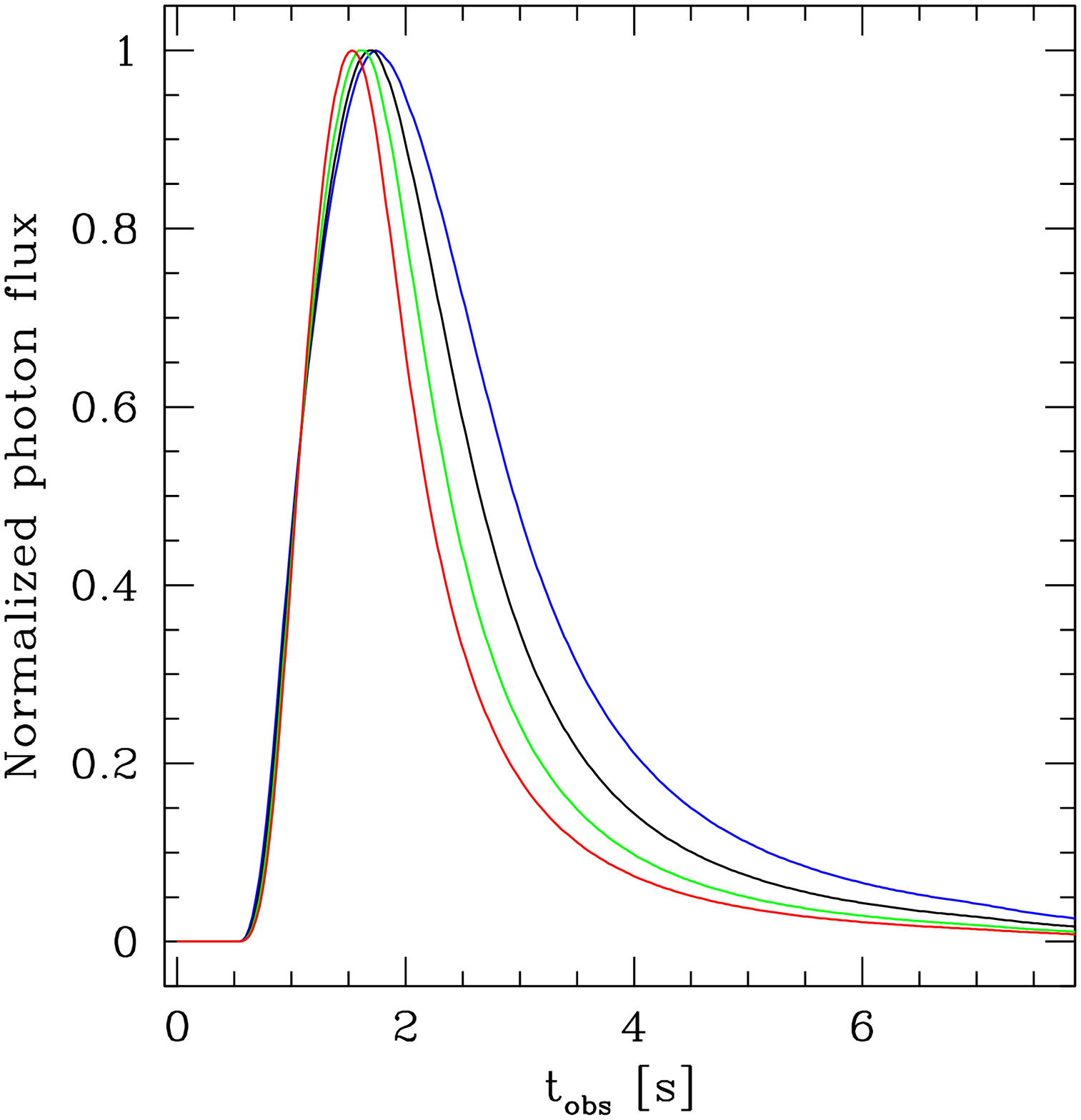}&
\includegraphics[width=0.32\textwidth]{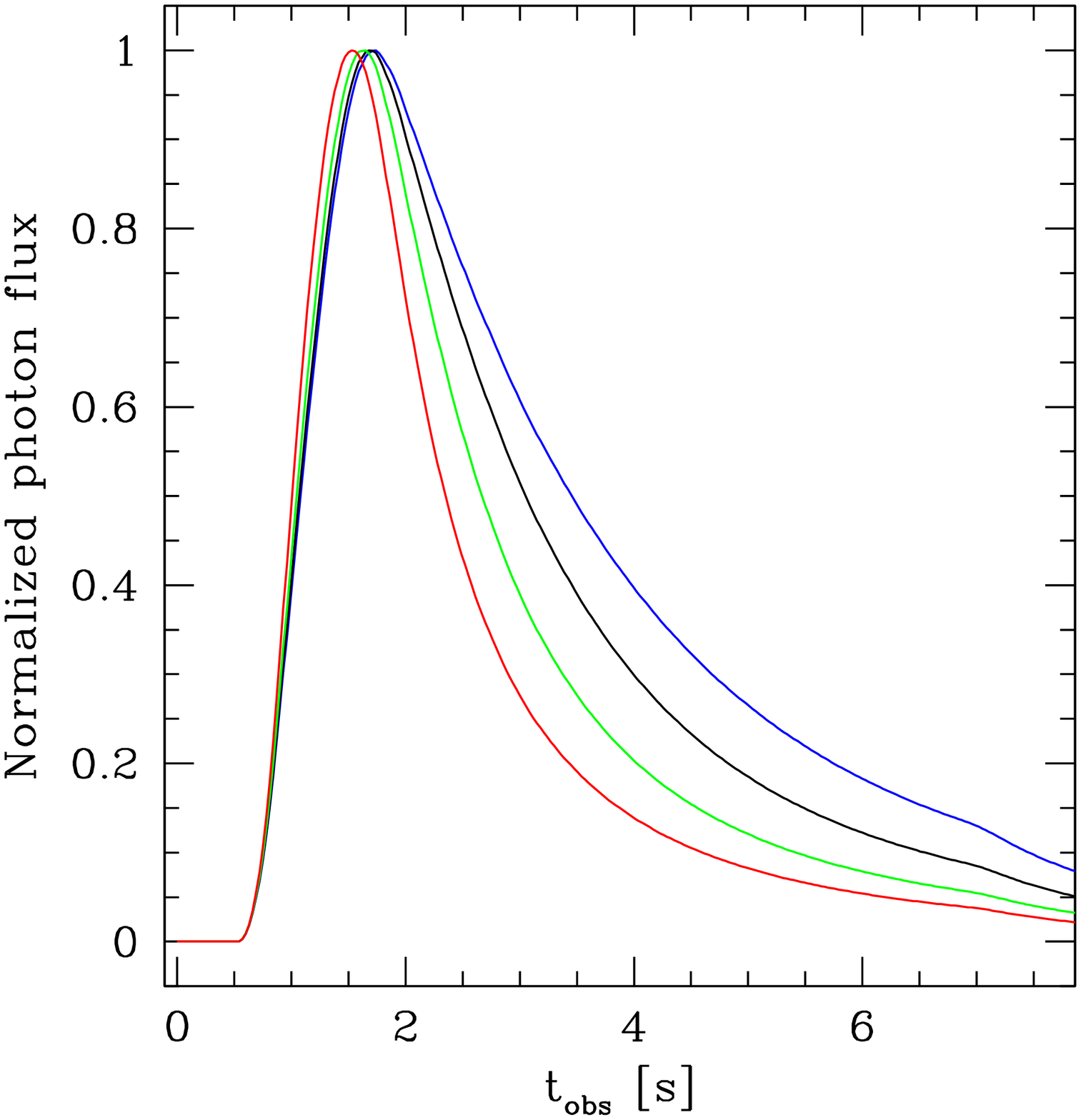}&
\includegraphics[width=0.32\textwidth]{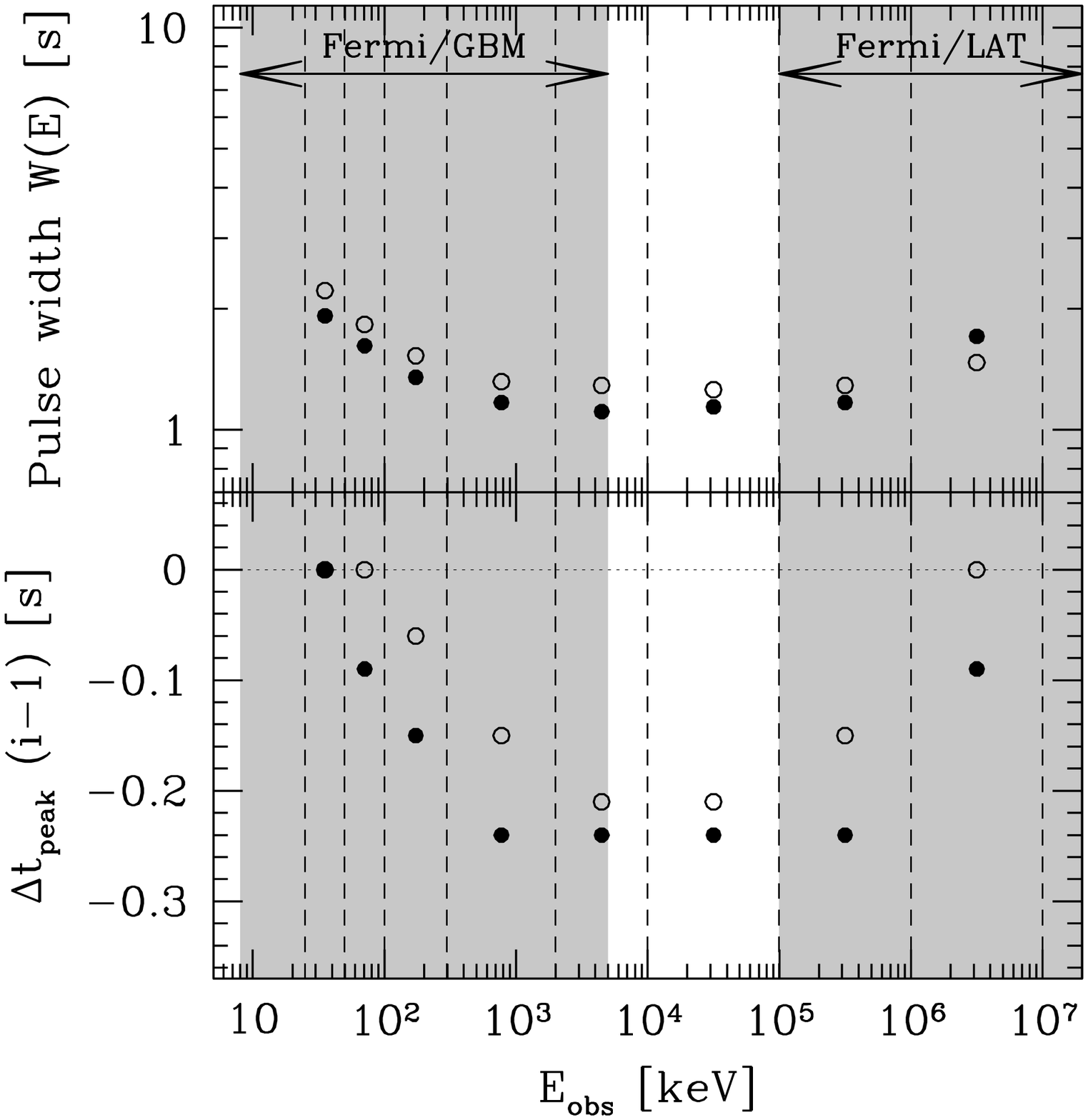}\\
\end{tabular}
\end{center}
\caption{\textbf{Impact of the shape of the injected kinetic power  in the outflow.}  
\textit{Left and middle panels:} normalized light curves corresponding to the 4 BATSE energy channels for case B computed by assuming a constant ejected mass flux $\dot{M}$ rather than a constant injected kinetic power $\dot{E}$ (see text) and a constant (left) or varying (middle) accelerated electron fraction (see text). Conventions are the same as in \reffig{fig:ABC_4BATSEchannels};
\textit{Right panel:} pulse width and pulse maximum position as a function of energy for the same cases (filled circles: constant $\zeta$, open circles: varying $\zeta$).  The shaded regions and lines have the same meaning as in \reffig{fig:ABC_WE_lags}.}
\label{fig:B_Mdot_4BATSEchannels_WE_lags}
\end{figure*}
%%%%%%%%%%%%%%%%%%%%%%%%%%%%%%%%%%%%%%%%%%%%%%
%%%%%%%%%%%%%%%%%%%%%%%%%%%%%%%%%%%%%%%%%%%%%%
%%%%%%%%%%%%%%%%%%%%%%%%%%%%%%%%%%%%%%%%%%%%%%

\subsection{Effect of the initial Lorentz factor}
\label{sec:InitGamma}
In the reference cases, the Lorentz factor during the relativistic ejection increases continuously from $\Gamma_\mathrm{min}$ to $\Gamma_\mathrm{max}$.
The shape of the transition can impact the radius where the internal shocks form and the initial strength of the shocks. It will therefore affect the pulse shape and the spectral evolution, especially in the early phase. To investigate this effect, we have simulated case B assuming a much steeper transition: 
\begin{equation}
\Gamma(t_\mathrm{ej}) = \left\lbrace\begin{array}{cl}
\Gamma_\mathrm{min} & \mathrm{for}\, 0\le t_\mathrm{ej} \le 0.2 t_\mathrm{w}\\ 
\Gamma_\mathrm{max} & \mathrm{for}\,  0.2 t_\mathrm{w} \le t_\mathrm{ej} \le t_\mathrm{w}
\end{array}
\right.\, ,
\label{eq:SteepGamma}
\end{equation}
 where the value
 $0.2 t_\mathrm{w}$
  has be chosen 
to have the same mean Lorentz factor than in the reference cases.
 For such initial conditions, the internal shock phase starts earlier and the shocks are immediately stronger than in the reference cases as shown in \reffig{fig:B_Dynamics}. The corresponding efficiency is increased and therefore, we have adjusted the injected kinetic power to keep the same radiated energy as in the reference cases (see \reftab{tab:allmodels}).
As expected, the fact that the shock is initially stronger affects the rising part of the pulse: see \reffig{fig:B_Gamma_4BATSEchannels_WE_lags} (light curves, pulse width and time lags) and \reftab{tab:allmodels}.
The main change is the increase of the time lags, which become too large compared to observations.
However, we checked that, as for the reference case, a varying fraction of accelerated electrons $\zeta\propto \epsilon_*$ solves this problem 
(\reffig{fig:B_Gamma_4BATSEchannels_WE_lags}, middle panel). 
We conclude that the main effect of the shape of the initial distribution of the Lorentz factor
is on the rising part of the pulse, especially at high energy as discussed in \refsec{sec:LAT}.

Another effect is also related to the initial distribution of the Lorentz factor. As can be seen in \reffig{fig:ABC_4BATSEchannels}, the decaying part of the light curve is interrupted at $t_\mathrm{HLE,obs}$ with a break towards a steeper decline. Such breaks are usually not observed, which can be easily understood as complex GRB light curves show the superimposition of many pulses, which makes the observation of the very end of the decay of a pulse difficult. We note however that $t_\mathrm{HLE,obs}$ is directly related to the radius where the propagation of the internal shocks responsible for the pulse ends \citep{hascoet:12}. In the reference case, it can easily occur at later times
by simply increasing the duration of the phase of the relativistic ejection where $\Gamma=400$, as illustrated in \reffigs{fig:B_zeta_Epeak_WE_lags}{fig:B_zeta_HIC} where it assumed that the $\Gamma=400$ phase in the ejection lasts for 2.2 s instead of 1.2 s. The pulse is  exactly the same except for the break in the decay  at $t_\mathrm{HLE,obs}\simeq 9$ s instead of 5.4 s.
It slightly affects the HIC and the HFC (\reffig{fig:B_zeta_HIC}), but at late times where the flux is too low to be considered in observed HIC diagrams.
  
We do not discuss here two other factors also related to the initial distribution of the Lorentz factor: (i) the contrast $\Gamma_\mathrm{max}/\Gamma_\mathrm{min}$ has a direct impact on the collision radius and on the strength of the shock. 
Very high contrasts may appear unrealistic for central engine models and very low contrasts lead to too low internal shock efficiencies.
We have favored here a intermediate value
 leading to a total efficiency of a few percents (ratio of the radiated energy over the kinetic energy): 
 see also \citet{bosnjak:09} and figure 8.d therein; (ii) the mean value $\bar{\Gamma}$ of the initial Lorentz factor in the ejecta. The main effects of this parameter are on the peak energy ($E_\mathrm{p,obs}$ decreases if $\bar{\Gamma}$ increases, all other parameters being constant, \citealt{barraud:05,bosnjak:09}) and at high energy ($\gamma\gamma$ annihilation, see Sect.\ref{sec:LAT}).

\subsection{Effect of the injected kinetic power}
The reference cases are computed using the simple assumption that the injected kinetic power $\dot{E}$ is constant during the relativistic ejection, which corresponds to an ejected mass flux that evolves as $\dot{M}\propto 1/\Gamma$. Other assumptions are of course possible and can again affect the dynamics and the predicted spectral evolution. To investigate this possibility, we have simulated case B assuming a constant ejected mass flux $\dot{M}$, i.e. 
$\dot{E}\propto \Gamma$. We fixed the value of $\dot{M}$ so that the total radiated energy 
is the same in both cases and we adjusted $\zeta$ to have similar peak energies. A clear drawback of the $\dot{M}=\mathrm{cst}$ assumption, as already explained in \citet{kobayashi:97,daigne:98} is a smaller internal shock efficiency,
which leads to increase $\dot{E}$ to have 
 the same GRB fluence (see \reftab{tab:allmodels}).
As seen in \reffig{fig:B_Dynamics}, the impact on the dynamics is weaker than 
in the case studied in \S\ref{sec:InitGamma}.
The qualitative comparison with observations (see  \reffig{fig:B_Mdot_4BATSEchannels_WE_lags} and \reftab{tab:allmodels}) shows that the pulse shape is improved, especially the values of the time lags (\refobsLags) and the evolution of the pulse width with energy (\refobsPulseEnergy). The HIC (\refobsHIC) is also slightly improved.
There is also a weak impact on the high-energy emission in the LAT range that is discussed in \refsec{sec:LAT}.

\subsection{Effect of the duration of the relativistic ejection}
We have not commented yet the evolution of the pulse properties with duration. Observations show that pulses of short duration are more symmetric (\refobsPulseAsymmetry), have very short, or zero, time-lags (\refobsLags), and are harder (\refobsHardnessDuration), for a large part due to higher peak energies as shown by the analysis of three bright GBM short GRBs \citep{guiriec:10}. The internal shock model reproduces qualitatively well these observations \citep{daigne:98}. 
Short GRBs have similar luminosities compared to long GRBs \citep{nakar:07}, then it is reasonable to assume similar kinetic power $\dot{E}$. Short GRBs emit MeV photons like long GRBs \citep[see e.g.][]{guiriec:10}, and even GeV photons in the case of GRB 090510 \citep{ackermann:10}. Then, one would also expect similar Lorentz factors. The main difference seems to be
simply limited to
 the shorter duration, and more generally a compression of all variability time scales \citep{guiriec:10,bhat:12}. If all input parameters are kept constant except for the timescales, a simple -- two-shell collision-- model of the internal shock phase \citep[see e.g.][]{barraud:05}, shows that the radius and comoving mass density evolve as $R\propto t_\mathrm{var}$ and $\rho_*\propto t_\mathrm{var}^{-2}$, where $t_\mathrm{var}$ is the variability timescale,  and that  $\Gamma_*$ and $\epsilon_*$ are unchanged. Then \refeq{eq:lbol} shows that $L_\mathrm{bol}$ is not affected but \refeq{eq:Epeak} indicate that $E_\mathrm{p,obs}\propto t_\mathrm{var}^{-1}$ : shorter pulses are naturally expected to have higher peak energies. An increase of $E_\mathrm{p,obs}$ has a direct impact on the second factor in \refeq{eq:flux12} (spectral correction) and then affect many observed features, such as the time lags, the pulse width, etc., as it has been illustrated in all the examples shown in this paper. A secondary effect can also play a role. The parameter $w_\mathrm{m}$ also evolves with the variability timescale, as $w_\mathrm{m}\propto t_\mathrm{var}^{-1}$, so the importance of Klein-Nishina corrections should increase when the duration decreases, which can affect the general spectral shape and especially the low-energy photon index: see Figure 8.d in \citet{bosnjak:09}.
 
%%%%%%%%%%%%%%%%%%%%%%%%%%%%%%%%%%%%%%%%%%%%%%
%%%%%%%%%%%%%%%%%%%%%%%%%%%%%%%%%%%%%%%%%%%%%%
%%%%%%%%%%%%%%%%%%%%%%%%%%%%%%%%%%%%%%%%%%%%%%
%%%
%%% Figure 15 : impact of the duration
%%%
%%%%%%%%%%%%%%%%%%%%%%%%%%%%%%%%%%%%%%%%%%%%%%
%%%%%%%%%%%%%%%%%%%%%%%%%%%%%%%%%%%%%%%%%%%%%%
%%%%%%%%%%%%%%%%%%%%%%%%%%%%%%%%%%%%%%%%%%%%%% 
\begin{figure}[!t]
\centering
\includegraphics[width=0.5\textwidth]{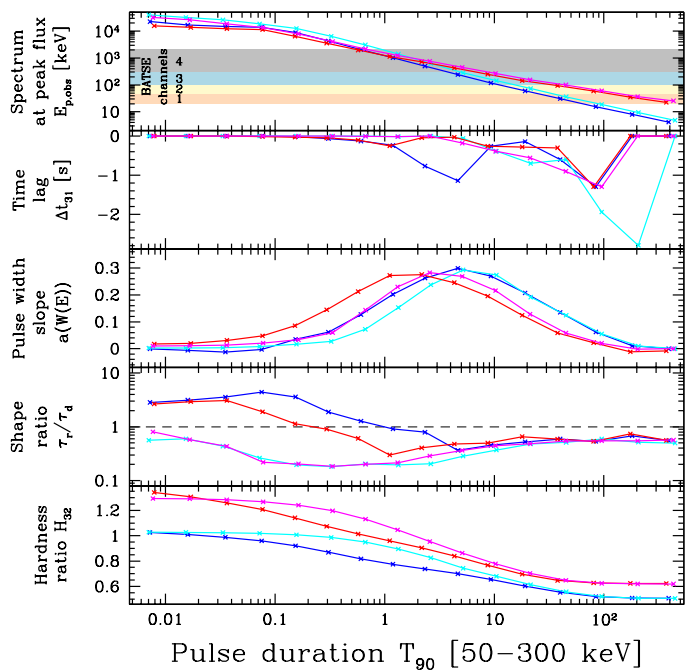}
\caption{\textbf{Effect of the duration of the ejection.} The peak energy at the maximum of the pulse (top panel), the difference between the time of maximum in BATSE channel 3 and 1 (second panel), the index of the power-law giving the evolution of the pulse width with energy $a=-\mathrm{d}\ln{W}/\mathrm{d}\ln{E}$ (third panel), the ratio of the rise and decay time of the pulse in BATSE channel 2+3 (fourth panel) and the hardness ratio $H_{32}$ (ratio of the photon fluence in BATSE channel 3 over the photon fluence in BATSE channel 2) are plotted as a function of the duration $T_{90}$ in BATSE channel 2+3 for case A with a constant (blue) or varying (cyan) $\zeta$ , and case B  with a constant (red) or varying (magenta) $\zeta$, with $p=2.7$ in all cases. The other model parameters are constant and equal the values given in \reftab{tab:allmodels}, except for the duration of the ejection $t_\mathrm{w}$ which is varied from $2$ ms to $200$ s.
}
%\vspace*{-3ex}
\label{fig:duration}
\end{figure}
%%%%%%%%%%%%%%%%%%%%%%%%%%%%%%%%%%%%%%%%%%%%%%
%%%%%%%%%%%%%%%%%%%%%%%%%%%%%%%%%%%%%%%%%%%%%%
%%%%%%%%%%%%%%%%%%%%%%%%%%%%%%%%%%%%%%%%%%%%%% 

To test in more details the predicted pulse evolution with duration, we have computed a series of synthetic pulses  keeping the same parameters as in the reference cases A et B ($\Gamma_\mathrm{min}$, $\Gamma_\mathrm{max}$, $\dot{E}$ and microphysics parameters), except for the duration $t_\mathrm{w}$ which varies from 2 ms to 200 s. We have also performed the same series of simulations for cases A and B with a varying fraction of accelerated electrons $\zeta$, as discussed in \S\ref{sec:zetavar} and for each series, we have considered both the case with $p=2.5$ and with $p=2.7$. The results for $p=2.7$ are shown in \reffig{fig:duration}, where we plot the evolution of different features of the pulse as a function of the duration $T_{90}$ measured between 50 and 300 keV (BATSE channels 2+3). 
As predicted, and in agreement with observations, we find that
(i) shorter pulses have higher peak energies; 
(ii) short pulses have negligible time lags. Indeed
 the peak energy is well above BATSE channel 4 and all light curves in the low gamma-ray range (BATSE) correspond to the same part of the synchrotron spectrum
  (the spectral correction in \refeq{eq:flux12} is constant with time for all channels).
  A similar behavior is observed at very long duration as the peak energy is below the spectral range of BATSE channel 1 (such an evolution is more difficult to test with observations as very long pulses are rare);
(iii)
for the same reason, short pulses have
the same width in all BATSE channels, $a\to 0$; 
(iv) for the same reason, short pulses have more symmetric shapes (i.e. the pulse decay time becomes comparable to the rise time).
We notice that for constant $\zeta$, the effect is too strong, with the shortest pulses having a rise time time longer than the decay time, which is sometimes observed but remains rare. The case with a varying $\zeta$ shows a much better agreement with observations;
(v) short pulses have a larger hardness ratio $H_{32}$.
This indicator was used to identify the hardness-duration relation in \citet{kouveliotou:93}. It depends mainly on $E_\mathrm{p,obs}$ and therefore the observed trend is reproduced. At very short durations, as $E_\mathrm{p,obs}$ is above the spectral range of BATSE channel 3, the hardness ratio becomes constant, and has a value that depends only on $\alpha$, i.e. $H_{32}\to \left(300^{1-\alpha}-100^{1-\alpha}\right)/\left(100^{1-\alpha}-50^{1-\alpha}\right)$. It is therefore distinct for case A ($\alpha\simeq -1.5$ and $H_{32}\to 1.02$) and case B ($\alpha\simeq -1.2$ to $1.1$ and $H_{32}\to 1.33$ to $1.45$). 

To make a more realistic test of the predictions of the internal shock model as a function of duration, one should consider multi-pulses light curves, which is out of the scope of this paper. However, our results show
 clearly that it
  predicts correctly the observed hardness-duration relation \refobsHardnessDuration, and that the other properties (symmetry of short pulses, \refobsPulseAsymmetry\, and \refobsPulseEnergy; vanishing lags for short GRBs, \refobsLags) are also explained as a consequence of higher peak energies 
 at short duration.

\section{High energy signatures ($>$ 100 MeV)}
\label{sec:LAT}

We have described in the previous section how the different assumptions in the internal shock model (dynamics or microphysics) can affect the temporal and spectral properties of pulses in the soft gamma-ray range. Is it possible to distinguish among the different possibilities from the high energy emission above 100 MeV? The few bursts detected by {\it Fermi}-LAT show several interesting features \citep{LATcatalog:13}: 
(i) LAT GRBs 
are among the brightest ones detected by the GBM, with the exception of few cases (e.g. GRB 081024 or GRB 090531). The measured E$_\mathrm{iso}$ in the  subsample of bursts with a measured redshift shows that the LAT bursts are intrinsically brighter. The ratio of fluences in the high ($>$ 100 MeV) and low ($<$ 1 MeV) energy channels is $\la$ 20\%. 
The highest energy photons ($>$ 10 GeV) are coming from the highest fluence GRBs (080916C, 090510, 090902B); (ii)
several LAT
 GRBs require an extra power-law component in addition to the Band model in the high energy portion of the spectrum. It
 can have a significant contribution to the total energy budget (10 - 30\%) and 
becomes prominent at energies $E_\mathrm{obs}\ga$ 100 MeV. The slope of the
 power-law lies within the range $-2$ to $-1.6$; (iii)
the emission above 100 MeV systematically starts later with respect to the GBM light curve. The ratio of the time delay over the total duration in the GBM is larger for longer bursts. In long GRBs, e.g.  GRB 080916C, the typical delay is of the order of a few seconds, while it is less than one second in short GRBs, e.g. respectively $\sim 0.5$ s and $0.05$ s in GRB 090510 or GRB 081024B; (iv) the
 emission
 in the LAT 
is long lasting compared to the GBM.
 It decays smoothly with time and can be fitted with a power-law, $F_\nu \propto t^{-\alpha}$ with $\alpha$ close to $1$ in most of the cases. A break in the decay of the extended emission
 is detected in GRB 090510, GRB 090902B, and GRB 090926A, with a transition from $\alpha\simeq 2.2$ to $0.9$.

The  LAT long-lasting emission  indicates that the high-energy emission in GRBs has, at least at late times, an external origin, i.e. is due to the deceleration of the ejecta by the external medium \citep{kumar:09,gao:09,kumar:10,ghisellini:10,wang:10,wang:13}. On the other hand, several arguments suggest that there is also a contribution of internal origin at early times: (i) a break in the temporal decay of the LAT emission is observed in at least three bright GRBs close to the end of the prompt emission in the GBM and suggest a change in the dominant mechanism \citep{LATcatalog:13}; (ii) the LAT emission at early times is known to be variable. For instance, variability on the 100 ms (resp. 20 ms) timescale is found in GRB 090902B (resp. 090510). This is difficult to reconcile with an external origin.
As emphasized by \citet{beloborodov:13b}, there is an additional theoretical argument against the scenario where the whole LAT emission (prompt and long lasting) would be associated to the external shock (rise and decay):
when using large time integration bins, 
the LAT flux starts to decay well before the end of the prompt emission in GBM, whereas the self-similar stage of the blast wave cannot be reached at such early times (typically, not earlier than the duration of the prompt emission).

In the following we examine which of the properties of the early high-energy emission in GRBs can be accommodated within the internal shock model
and if the LAT observations may offer a way to distinguish among the different scenarios studied in the two previous sections.
 We do not consider the possible contribution of shock-accelerated protons to the emission, as various studies have shown that it requires extreme parameters to be dominant in the LAT range, due to a weak efficiency \citep{asano:09,asano:12}. We do not include additional processes that may be important, such as the scattering of photospheric photons by shock accelerated electrons in internal shocks \citep{toma:11} or the scatterings of prompt photons in the pair-enriched shocked external medium at early stages of the deceleration \citep{beloborodov:13b}.

\subsection{High energy emission in reference cases A and B}
\begin{figure}
\begin{center}
\includegraphics[width=0.44\textwidth]{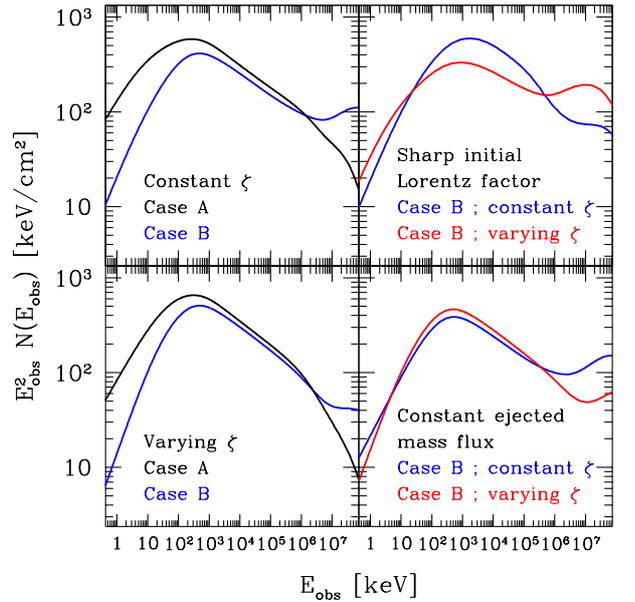}
\end{center}
\caption{\textbf{High-energy emission: spectra.} 
The time-integrated spectrum (0 -- 15 s)  is plotted from the keV to the GeV range for the same cases as in \reffig{fig:HELC}. \textit{Top left panel:} reference cases A and B; \textit{Bottom left panel:} cases A and B with a varying accelerated electron fraction $\zeta$; \textit{Top right panel:} case B with a sharp initial distribution of the Lorentz factor (see text) and a constant or a varying accelerated electron fraction $\zeta$;  \textit{Bottom right panel:} case B with a constant ejected mass flux (see text) and a constant or a varying accelerated electron fraction $\zeta$.
}
\label{fig:HESP}
\end{figure}

As illustrated in  \reffig{fig:HESP}, cases A and B, with a constant or a varying $\zeta$, have very different high-energy spectra as the efficiency of inverse Compton scatterings strongly depend on $\epsilon_\mathrm{B}/\epsilon_\mathrm{e}$.
The inverse Compton component is negligible in case A, whereas it creates a well defined additional component at high-energy in case B. This additional component is stronger when $\zeta$ is constant. As shown in \S\ref{sec:zetavar}, the peak energy of the synchrotron component is decaying faster in this case, so that Klein-Nishina corrections become more and more negligible in the pulse decay. On the other hand, the assumption $\zeta\propto \epsilon_*$ maintains a higher value of the peak energy during the decay, and then a less efficient inverse Compton emission. It is interesting to note that the additional component in the GeV range is very flat in the $\nu F_\nu$ spectrum (see \reffig{fig:HESP} bottom-left panel) and would probably be fitted by a power-law with a photon index close to $-2$, as observed in several LAT bursts \citep{LATcatalog:13}. 
%%%%%%%%%%%%%%%%%%%%%%%%%%%%%%%%%%%%%%%%%%%%%%
%%%%%%%%%%%%%%%%%%%%%%%%%%%%%%%%%%%%%%%%%%%%%%
%%%%%%%%%%%%%%%%%%%%%%%%%%%%%%%%%%%%%%%%%%%%%%
%%%
%%% Figure 16 : HE spectrum
%%%
%%%%%%%%%%%%%%%%%%%%%%%%%%%%%%%%%%%%%%%%%%%%%%
%%%%%%%%%%%%%%%%%%%%%%%%%%%%%%%%%%%%%%%%%%%%%%
%%%%%%%%%%%%%%%%%%%%%%%%%%%%%%%%%%%%%%%%%%%%%% 
%%%%%%%%%%%%%%%%%%%%%%%%%%%%%%%%%%%%%%%%%%%%%%
%%%%%%%%%%%%%%%%%%%%%%%%%%%%%%%%%%%%%%%%%%%%%%
%%%%%%%%%%%%%%%%%%%%%%%%%%%%%%%%%%%%%%%%%%%%%% 

\begin{figure}
\begin{center}
\includegraphics[width=0.44\textwidth]{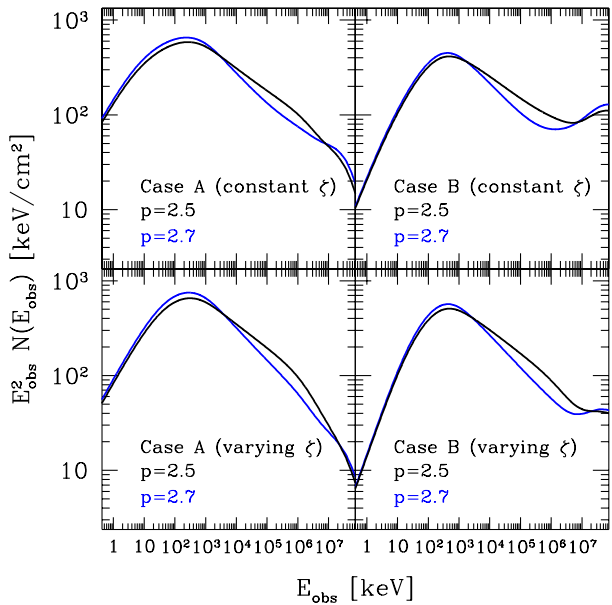}
\end{center}
\caption{\textbf{High-energy emission: effect of the electron slope $p$.} 
The time-integrated spectrum (0 -- 15 s)  is plotted from the keV to the GeV range for reference case A (left panels) and B (right panels), either assuming a constant (top panels) or a varying (bottom panels) fraction $\zeta$ of accelerated electrons, with %an electron slope 
$p=2.5$ (black) or $2.7$ (blue). 
}
\label{fig:HESPslopep}
\end{figure}

These examples  cannot be directly compared to \textit{Fermi}-LAT bursts as they 
radiate $\sim 10^{52}$ erg, 
 whereas LAT bursts are much brighter \citep{LATcatalog:13}. In addition, we did not try to adjust the model parameters to improve the peak energy  of the additional component. 
In our examples, the additional component typically appears above 1-10 GeV, whereas it is already detected at lower energy in LAT bursts. The shape of the additional component and its peak energy
are determined in a complex manner by the relative efficiency of the synchrotron and inverse Compton emission, the slope $p$ of the shock-accelerated electrons, and the $\gamma\gamma$ annihilation. 
This is illustrated in \reffig{fig:HESPslopep} where the spectrum in cases A and B is plotted for two different values of $p$, which
directly impacts the photon index $\beta$ of the high-energy part of the dominant (synchrotron) component. Increasing $p$ and $\beta$ allows to observe the emergence of the additional component at lower
energy and affects its
measured slope.

\begin{figure*}[!t]
\begin{center}
\begin{tabular}{cc}
\includegraphics[width=0.44\textwidth]{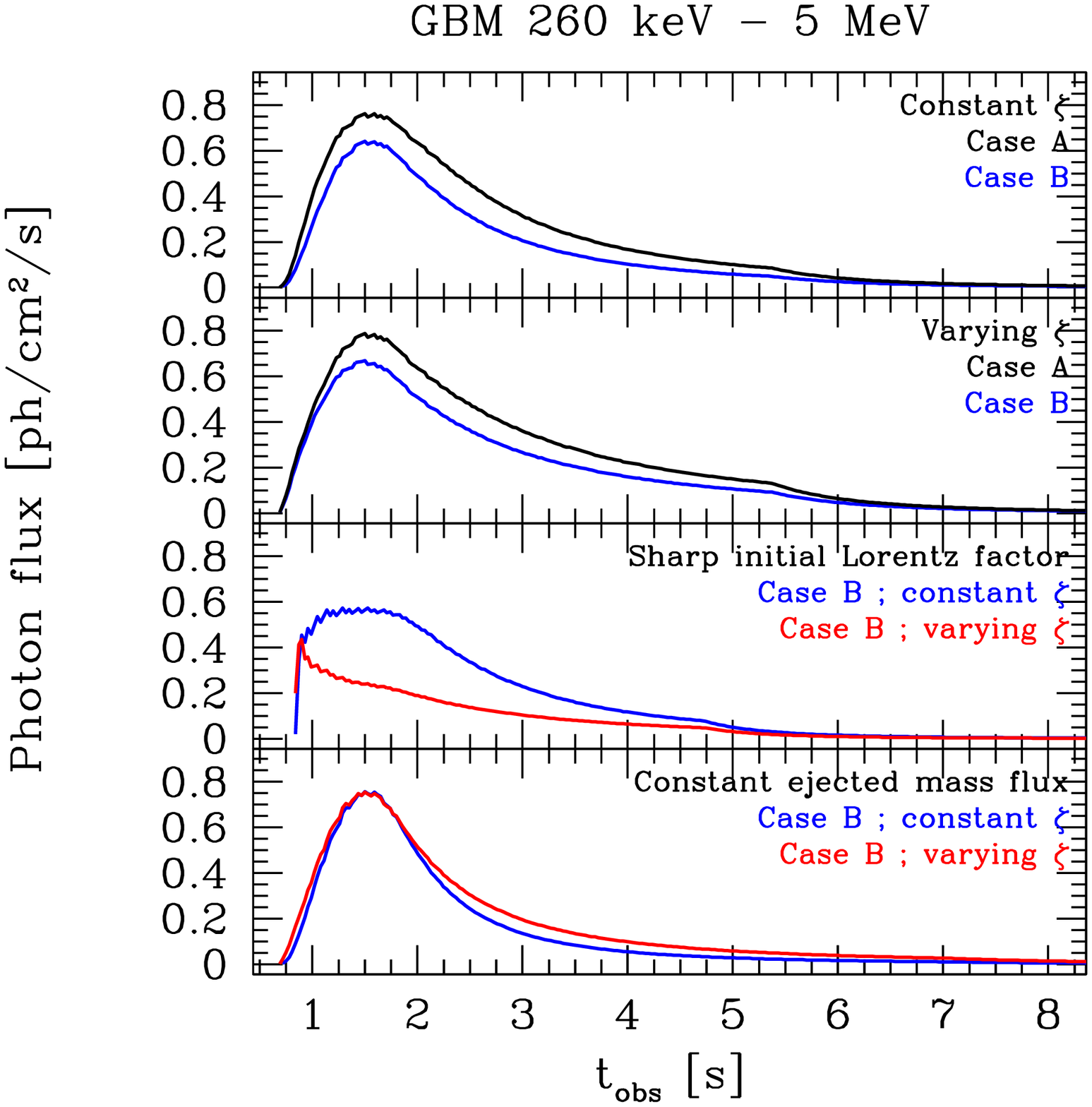} &
\includegraphics[width=0.44\textwidth]{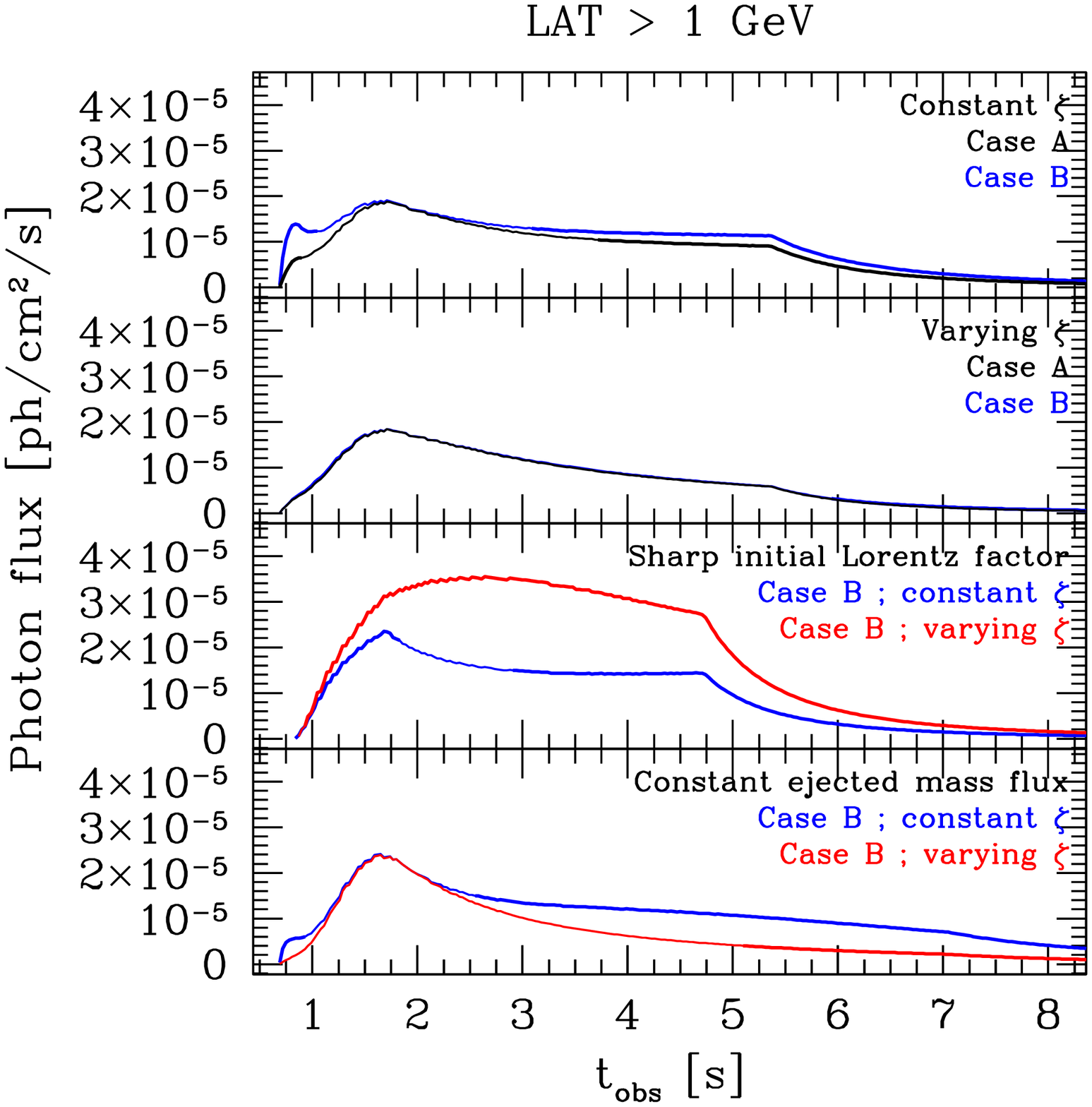}\\
\end{tabular}
\end{center}
\caption{\textbf{High-energy emission: light curves.} 
The light curves in the soft gamma-ray range (260 keV -- 5 MeV, left figure) and in the high-energy gamma-ray range ($>1$ GeV, right figure) are plotted for different cases discussed in the paper. For the high-energy light curves, a thin solid line indicates that the synchrotron emission is dominant above 1 GeV whereas a thick solid line indicates that the inverse Compton emission is dominant. \textit{Top panel:} reference cases A and B; \textit{Second panel:} cases A and B with a varying accelerated electron fraction $\zeta$; \textit{Third panel:} case B with a sharp initial distribution of the Lorentz factor (see text) and a constant or a varying accelerated electron fraction $\zeta$;  \textit{Bottom panel:} case B with a constant ejected mass flux (see text) and a constant or a varying accelerated electron fraction $\zeta$.
}
\label{fig:HELC}
\end{figure*}

 Due to the high peak energies of the inverse Compton component in our reference cases, the light curves above 1 GeV are mainly governed by the synchrotron radiation and peak approximatively at the same time as the soft gamma-ray component, with only a very small delay (see \reffig{fig:HELC}), contrary to the observed delayed onset of the GeV emission 
   \citep{LATcatalog:13}. To increase this delay, one should either increase the $\gamma\gamma$ annihilation in the early phase by decreasing $\bar{\Gamma}$ as illustrated 
    in \citet{hascoet:12}, or adjust the parameters so that the inverse Compton emission peaks at lower energy \citep[see e.g.][]{asano:12}, or both. 
 Nevertheless, in case B with a constant $\zeta$, where the inverse Compton emission is the most efficient, the additional component starts to be visible in the light curve during the pulse decay (see \reffig{fig:HELC}, right, top panel). One also sees a small high-energy precursor 
that appears because
  the shock is initially weak, with a low peak energy and a high inverse Compton efficiency \citep{bosnjak:09}.
  This precursor, never observed in LAT GRBs, can be suppressed either by changing the assumptions for the microphysics  
  or the dynamics (\reffig{fig:HELC}).

%%%%%%%%%%%%%%%%%%%%%%%%%%%%%%%%%%%%%%%%%%%%%%
%%%%%%%%%%%%%%%%%%%%%%%%%%%%%%%%%%%%%%%%%%%%%%
%%%%%%%%%%%%%%%%%%%%%%%%%%%%%%%%%%%%%%%%%%%%%%
%%%
%%% Figure 17 : HE spectrum, effect of slope p
%%%
%%%%%%%%%%%%%%%%%%%%%%%%%%%%%%%%%%%%%%%%%%%%%%
%%%%%%%%%%%%%%%%%%%%%%%%%%%%%%%%%%%%%%%%%%%%%%
%%%%%%%%%%%%%%%%%%%%%%%%%%%%%%%%%%%%%%%%%%%%%% 
%%%%%%%%%%%%%%%%%%%%%%%%%%%%%%%%%%%%%%%%%%%%%%
%%%%%%%%%%%%%%%%%%%%%%%%%%%%%%%%%%%%%%%%%%%%%%
%%%%%%%%%%%%%%%%%%%%%%%%%%%%%%%%%%%%%%%%%%%%%% 

%%%%%%%%%%%%%%%%%%%%%%%%%%%%%%%%%%%%%%%%%%%%%%
%%%%%%%%%%%%%%%%%%%%%%%%%%%%%%%%%%%%%%%%%%%%%%
%%%%%%%%%%%%%%%%%%%%%%%%%%%%%%%%%%%%%%%%%%%%%%
%%%
%%% Figure 18 : HE light curves
%%%
%%%%%%%%%%%%%%%%%%%%%%%%%%%%%%%%%%%%%%%%%%%%%%
%%%%%%%%%%%%%%%%%%%%%%%%%%%%%%%%%%%%%%%%%%%%%%
%%%%%%%%%%%%%%%%%%%%%%%%%%%%%%%%%%%%%%%%%%%%%% 
%%%%%%%%%%%%%%%%%%%%%%%%%%%%%%%%%%%%%%%%%%%%%%
%%%%%%%%%%%%%%%%%%%%%%%%%%%%%%%%%%%%%%%%%%%%%%
%%%%%%%%%%%%%%%%%%%%%%%%%%%%%%%%%%%%%%%%%%%%%% 

\subsection{Impact of the assumptions on the dynamics}

The assumptions 
for the dynamics have a large impact on the high energy emission. 
In the case 
$\dot{M}=\mathrm{cst}$ (rather than $\dot{E}=\mathrm{cst}$), the inverse Compton emission is more efficient during the pulse decay
(but not during the rise, as in reference case B). This is due to a more rapid decrease of the peak energy during the 
 decay (see \reffig{fig:B_Dynamics}) and therefore a more rapid decrease of the Klein Nishina corrections. 
  This improves the light curves (\reffig{fig:HELC} bottom panels), which show a more intense tail due to inverse Compton emission, and a larger delay between the peaks of the LAT and GBM light curves, however
still too small to explain the observed delayed onset. The additional component in the spectrum is already detected between 1 GeV and 10 GeV (\reffig{fig:HESP}).

The case where the initial distribution of the Lorentz factor has a sharp transition from $\Gamma_\mathrm{min}$ to $\Gamma_\mathrm{max}$ has the strongest impact. In this case, the shocks are immediately violent so that  the weak precursor observed in the LAT in other cases does not appear (see \reffig{fig:HELC}): the peak energy of the 'forward' internal shock is indeed immediately very high and the corresponding inverse Compton emission is suppressed by Klein-Nishina corrections. This is only at late times that 
the inverse Compton emission becomes bright when more 
scatterings occur in Thomson regime. 
However, an important difference in this case 
is the fact that the emission of the 'forward' internal shock lasts longer and is not negligible (see 
\reffig{fig:B_Dynamics}). It is even dominant in the LAT for the first seconds. Due to a lower peak energy, the corresponding additional high energy component is well seen in the LAT, either with a constant accelerated electron fraction $\zeta$, or even more with a varying $\zeta$, which is the only simulated case in all the examples presented in this paper where the choice of parameters immediately leads to a peak energy of the IC component at 10 GeV (see \reffig{fig:HESP}).  For this reason, this case is the brightest in the LAT range and illustrates well that the high energy emission from internal shocks is not only sensitive to the details of the assumptions regarding the microphysics but also to the dynamics. 

Interestingly, we note that in most of the scenarios discussed in \refsec{sec:LAT}, the plots showing the time lag with respect to the low-energy channel as a function of the energy (Figs.~\ref{fig:ABC_WE_lags}, \ref{fig:B_p_WE_lags}, \ref{fig:B_zeta_Epeak_WE_lags}, \ref{fig:B_Gamma_4BATSEchannels_WE_lags}, \ref{fig:B_Mdot_4BATSEchannels_WE_lags}) shows a U-shape, the light curves initially peaking earlier when the energy is increasing, with a reversal of this trend above $\sim 10-100$ MeV. Such a behavior is found in GBM+LAT data, as studied by \citet{foley:11,foley:12}.

We conclude 
that the high energy emission from internal shocks is highly sensitive to the details of the assumptions regarding both the microphysics and the dynamics, and can therefore provide valuable diagnostics to distinguish among the various scenarios discussed in this paper. However a direct comparison of our results with observations reveals to be delicate as 
 LAT GRBs are among the brightest, with isotropic energies much larger than the average 'typical' value considered here. As this paper is mainly focussed on the temporal and spectral properties in the soft gamma-ray range, we leave to a forthcoming study a more detailed comparison to \textit{Fermi} data, which will be based on simulated bursts with more extreme parameters, especially regarding the total injected energy and the Lorentz factor.

\section{Discussion}
\label{sec:discussion}
The spectro-temporal evolution in the internal shock model 
 is governed by the hydrodynamics :
the physical conditions in the shocked regions vary on 
the hydrodynamical timescale associated to the propagation of the internal shocks.
This evolution affects in a complex manner the respective efficiency of the radiative processes  (synchrotron radiation, inverse Compton scatterings)  as well as the peak energy and spectral shape of each component. 
The model parameters 
 can be divided in two groups: assumptions for the microphysics and for the dynamics. Both can strongly affect
 the spectro-temporal evolution in GRBs.

\subsection{Impact of the microphysics parameters.} 
The dissipation of the energy in the shocked region is parameterized by ($\epsilon_\mathrm{e}, \zeta, p$) describing the energy injection in the relativistic electrons distribution, and $\epsilon_\mathrm{B}$ describing the amplification of the magnetic field. The values of these microphysics parameters are broadly constrained by the observations. 
As GRBs are extremely bright, a high $\epsilon_e$ is required to avoid an energy crisis. As \textit{Fermi}-LAT observations are not compatible with a SSC spectrum \citep{piran:09}, the soft gamma-rays must be directly produced by synchrotron radiation, which requires a low fraction of accelerated electrons $\zeta \la 10^{-3}-10^{-2}$ \citep{daigne:98,bosnjak:09,daigne:11,beniamini:13}.  
The fact that the observed low energy index $\alpha$ 
 is usually 
larger than the standard synchrotron fast cooling value $-3/2$
 favors low values of $\epsilon_\mathrm{B}$, typically 
$\epsilon_B \la 10^{-3}-10^{-4}$ \citep{daigne:11,barniolduran:12}, if such
photon indices
 are mainly due to the effect of inverse Compton scattering in the Klein Nishina regime. The fact that many bursts show also a steep high-energy photon index $\beta$ or are even well fitted with a power-law + an exponential cutoff \citep{kaneko:06,goldstein:12} implies that the electron slope $p$
 can be  larger than the usually considered value $p\simeq 2.2-2.5$.

We found that the spectro-temporal evolution predicted by the internal shock model is qualitatively in agreement with the observations. This is illustrated by the two reference cases A and B defined in \citet{daigne:11} and corresponding to $\epsilon_\mathrm{e}=1/3$, a low $\zeta$, $p=2.5$ and
either a high or a low $\epsilon_\mathrm{B}$, leading respectively to a standard $\alpha\simeq -1.5$ or a large 
 $\alpha\simeq -1.1$ low-energy photon index for the synchrotron spectrum: the pulse shape is asymmetric, having a faster rise than the decay, the pulse width is larger in lower energy channels, and the pulse maximum is reached at earlier times for higher energy channels. The typical hard-to-soft evolution is reproduced during the pulse decay. 

We investigated if a quantitative agreement can  be achieved, since for the reference cases A \& B the observed spectrum evolves too rapidly. We studied the effect of the electron distribution slope (Fig 6), and found that  $p \simeq$ 2.7 (slightly steeper than the common assumption
$p$ = 2.5) improved the evolution of the spectral peak energy, especially in case B (low magnetic field, large $\alpha\la -1$).  
Such steeper values are also in better agreement with observed high-energy photon indices $\beta$. However, the evolution of the peak energy remains usually too rapid compared to observations. This problem had already been identified by \citet{daigne:98,daigne:03} based on a much simpler treatment of the radiative processes. They also suggested that this may be related to the common assumption of constant microphysics parameters during the evolution of the shocks, which may appear unrealistic.
To investigate the impact of these assumptions, they considered  
a simple prescription for varying microphysics parameters -- in the absence of still missing physically motivated prescriptions based on shock acceleration theory -- where the fraction of accelerated electrons is evolving with the shock Lorentz factor, such as $\zeta \propto \epsilon_\mathrm{*}$.
We simulated the spectro-temporal evolution predicted by the internal shock model under such an assumption. This indeed leads 
to a much better quantitative agreement: the evolution of the peak energy is slower, and, as it governs most of the other properties, the general agreement is much better for the hardness intensity correlation, the evolution of the pulse shape and time of pulse maximum 
with energy channels, etc. (see \refsec{sec:EffectMicrophysics} and \reffigs{fig:B_zeta_HIC}{fig:B_zeta_Epeak_WE_lags}).

\subsection{How realistic are our assumptions for the microphysics ?} 
There are no theoretical arguments why microphysics parameters should be universal in mildly relativistic shocks \citep[see e.g.][]{bykov:12}. Even in the ultra-relativistic regime, GRB afterglows
already
 show the opposite, as a broad distribution of parameters is necessary to fit the observations \citep[see e.g.][]{panaitescu:01,cenko:10}. 
In absence of a well established shock theory, 
we have tested here variations following
 the prescription $\zeta\propto\epsilon_\mathrm{*}$, which is suggested by the work of \citet{bykov:96}.
 Our result that varying microphysics parameters improves the quantitative agreement between the predictions of the internal shock model and the observed spectro-temporal evolution observed in GRBs is therefore encouraging. 
On the other hand, some of the typical values of the microphysics parameters in the simulations presented in this paper may appear unrealistic, compared to recent progress in shock acceleration modelling, especially results from large Particle-In-Cell (PIC) simulations. Steep $p\simeq 2.7$ and large $\epsilon_\mathrm{e}\simeq 0.1-0.3$ may be achieved, but on the other hand, low accelerated electron fraction $\zeta\simeq 10^{-4}-10^{-2}$ and low magnetic field energy fraction $\epsilon_\mathrm{B}\simeq 10^{-3}$ may appear in contradiction with shock simulations, as mentioned for instance by \citet{barniolduran:12,beloborodov:13,beniamini:13}. This calls for several comments :

\noindent(i)~current PIC simulations are limited to ultra-relativistic shocks and do not describe yet the parameter space of mildly relativistic shocks such as in internal shocks, i.e. with typical shock
 Lorentz factors $\gamma_\mathrm{sh}\la 2$. A direct comparison is therefore difficult.
For 
$\gamma_\mathrm{sh}=15$, PIC simulations show that acceleration does not occur for magnetized ($\sigma\ga 10^{-3}$) perpendicular shocks, but is observed either for weakly magnetized or quasi-parallel "subluminal" shocks  with typically $\epsilon_\mathrm{e}\sim 0.1$, $\zeta\sim 10^{-2}$ and $p\simeq 2.3-2.4$ \citep{martins:09,sironi:11}; 

\noindent(ii)~these PIC simulations predict a low value of $\zeta\sim 10^{-2}$. Theoretical investigations of the energy transfer from protons to electrons in mildly relativistic shocks also predict that only a fraction of electrons are accelerated, with $\zeta$ as low as $10^{-3}$ \citep{bykov:96}. Therefore the values of the accelerated electron fraction $\zeta$ in the simulations discussed here are not in strong contradictions with shock acceleration modelling, but are usually too small. The low values of $\zeta$ in our simulations are necessary to reach high peak energies for the synchrotron component. However, a detailed comparison between the ballistic ("solid shells" model) approach used here for the dynamics of internal shocks with a more precise calculation based on a 1D Lagragian special relativistic hydrocode shows that the agreement between the two calculations is usually very good, except for 
the mass density $\rho_\mathrm{*}$ and specific internal energy density $\epsilon_*$ in the shocked region, which are underestimated by the simple model \citep{daigne:00}. For similar Lorentz factor 
$\Gamma_\mathrm{*}$ 
 microphysics parameters $\epsilon_\mathrm{e}$,  $\zeta$, $p$, and $\epsilon_\mathrm{B}$,  larger
values of $\rho_*$ and $\epsilon_*$ lead
to a larger peak energy. As $\rho_\mathrm{*}$ is typically underestimated by at least a factor $\sim 10^3$ and $\epsilon_*$ by a factor $\sim 4$
in the initial phase of the shock propagation which dominates the pulse emission
 (see Fig.~5 in \citealt{daigne:00}), the values of $\zeta$ deduced from the simple dynamical model may be underestimated by a factor $\ga 10^{3/4}\times 4^{5/4}\simeq 30$ as $E_\mathrm{p,obs}\propto \rho_\mathrm{*}^{0.5}\zeta^{-2}$ (see \refeq{eq:Epeak}). Taking into account this effect, the cases listed in \reftab{tab:allmodels} would correspond to 
 the range $\zeta\sim \left(1-10\right)\, \%$, in better agreement with theoretical predictions. In addition, we have simulated very smooth single pulse bursts for simplicity and to better identify the spectro-temporal evolution but more variable outflows would lead to more efficient collisions with higher values of the dissipated specific internal energy $\epsilon_\mathrm{*}$, allowing to reach high peak energies for larger values of $\zeta$;

\noindent(iii)~shock acceleration is accompanied by the amplification of the magnetic field in the shocked region. Both processes cannot be dissociated. Therefore, the low values of $\epsilon_\mathrm{B}$ considered in reference case B and the derived cases may appear unrealistic, as discussed in \citet{barniolduran:12}. However, one should remember that $\epsilon_\mathrm{B}$ should be understood here as fixing the typical strength of the magnetic field seen by radiative electrons, i.e. on a length scale fixed by the electron radiative timescale. This length scale is much larger than the plasma scale and $\epsilon_\mathrm{B}$ is therefore not only determined by the amplification at the shock, but also by the evolution of the magnetic field on larger scales. Recently, \citet{lemoine:13} demonstrated that if the large magnetic field generated in a thin microturbulent layer at the shock front decays over some hundreds of skin depths -- as suggested by recent simulations \citep{keshet:09} -- the effective $\epsilon_\mathrm{B}$ deduced from observations may be much lower than the value predicted by PIC simulations at the shock \citep[see also][]{derishev:07,kumar:12,uhm:13}. In addition, the low value of $\epsilon_\mathrm{B}$ used in case B is required to favor inverse Compton scatterings in Klein-Nishina regime and to increase the low-energy slope of the synchrotron spectrum (see \citet{daigne:11} for a discussion of the detailed conditions). As discussed below (\S\ref{sec:extensions}), the precise shape of the observed spectrum on the low-energy part of the gamma-ray range is debated \citep[see e.g.][]{guiriec:11}. If it happens that the occurence of 
large photon indices ($\alpha\ga -1$) is over-estimated by current spectral  analyses, the constraint on $\epsilon_\mathrm{B}$ would be relaxed and higher values could be considered.

\subsection{Impact of the dynamical parameters.} The dynamics of the relativistic outflow is determined by the initial conditions described by the variation of the bulk Lorentz factor of the flow $\Gamma(t_\mathrm{ej})$, the kinetic power $\dot{E}(t_\mathrm{ej})$, and the duration of the relativistic ejection $t_\mathrm{w}$. As the spectral evolution is mainly governed by the details of the  propagation of the internal shock waves, any change in the initial Lorentz factor or kinetic power directly affects the light curve shape (see \reffigs{fig:B_Gamma_4BATSEchannels_WE_lags}{fig:B_Mdot_4BATSEchannels_WE_lags}). Typically, we find that steeper variations of the Lorentz factor lead to internal shocks which are immediately efficient, with a peak energy which is already high at early times in the pulse. The biggest impact is an improvement of the light curve at high energy (GeV range) compared to \textit{Fermi} observations, as illustrated in \reffig{fig:HELC}. Changing the assumptions on the injected kinetic power also impacts the results, mostly at high energy, and can affect the overall efficiency of the internal shock phase.

Unfortunately, the current understanding of GRB central engines and of the relativistic ejection phase does not allow a detailed prediction of the input parameters $\Gamma(t_\mathrm{ej})$ and $\dot{E}(t_\mathrm{ej})$. 
We can investigate
which assumptions favor the best agreement with observations, but we can not conclude if these assumptions are realistic and -- when different assumptions lead to a similar agreement -- which assumption should be preferred.

We have also tested an interesting property of the internal shock model : the dependence of the temporal and spectral properties on the duration of a pulse. For that purpose, we have simulated a series of pulses keeping all parameters constant except for the total duration of the relativistic ejection, $t_\mathrm{w}$. It is very encouraging to observe that, despite its simplicity (in reality, variations of $t_\mathrm{w}$ are probably accompanied by variations of other input parameters),  the model reproduces well the observations : short pulses become more symmetric, have smaller or zero time lags, have a higher hardness ratio (see \reffig{fig:duration}). This is mainly due  to the fact that the peak energy is higher for a shorter variability timescale, a clear prediction of the internal shock model. At very short duration, most of the pulse light curve in the soft gamma-ray range occurs in the same portion of the spectrum (below the peak energy), which explains why the lags vanish and the hardness ratio tends to be constant.
All these properties of short pulses have been observed in real GRBs since the BATSE era, and have been confirmed by \textit{Fermi} \citep{guiriec:11,guiriec:13}, which in addition has shown the dominant effect of higher peak energies  to explain this evolution in short pulses.

\subsection{Possible extensions of this work}
\label{sec:extensions}
There are several potential additional effects that are not taken into account in this work but should be examined in the future.

\noindent--~We have shown in \refsec{sec:LAT} that the different sub-scenarios of the internal shock model may differ by their predictions for the high-energy gamma-ray emission. However, a special modeling effort is necessary to compare these predictions to observations, as \textit{Fermi}-LAT bursts are among the brightest GRBs ever detected, whereas the pulses simulated here
have average properties.

\noindent--~Recent \textit{Fermi/}GBM observations have shown a disagreement in the soft gamma-ray range between the observed spectrum and the phenomenological Band function \citep{band:93} usually used for spectral fits  \citep{guiriec:11,axelsson:12,mcglynn:12,guiriec:13}. This leads to reconsider the distribution of the low-energy photon index $\alpha$, as it is usually found in these bright GBM bursts that adding a new spectral component at low energy to better reproduce the spectral shape leads to smaller values of $\alpha$. As discussed above, this relaxes the constraint on one microphysics parameter, $\epsilon_\mathrm{B}$. A promising interpretation for these new observations is that an extra component associated to the photospheric emission is detected, in agreement with theoretical predictions \citep{guiriec:11,guiriec:13}. 
A possible diagnostic to distinguish among the different scenarios discussed here would be to simultaneously simulate the photospheric and internal shock emission, as the predicted spectral evolution for these two components has not the same dependence on the properties of the relativistic outflow  \citep{hascoet:13};

\noindent--~To be able to explore a large range of the parameter space, some simplifications have been made in the present calculations. There are several possible improvements which may be investigated in the future : (i) what is the contribution to the emission of the thermal electrons which are not shock-accelerated \citep[see e.g][]{giannios:09}. This depends of course on the fraction $\epsilon_\mathrm{e}^\mathrm{th}$ of the dissipated energy which remains in the fraction $1-\zeta$ of electrons that are not accelerated. We have checked that for $\epsilon_\mathrm{e}^\mathrm{th}/\epsilon_\mathrm{e}\la 0.1$, the additional component due to the emission of the thermal electrons does not affect the gamma-ray spectrum, and therefore does not change the results of the present paper. On the other hand, it may contribute in certain conditions to the prompt optical emission.
We note that the absence of a clear signature of thermal electrons in afterglow observations may indicate that the ratio $\epsilon_\mathrm{e}^\mathrm{th}/\epsilon_\mathrm{e}$ is not very large in relativistic shocks; (ii) what is the contribution to the emission of the secondary leptons produced by $\gamma\gamma$ annihilation. As shown by \citet{asano:11}, this could have an important contribution to the observed extra power-law component identified by \textit{Fermi/}LAT. In addition, a more precise calculation of the $\gamma\gamma$ annihilation may help in better reproducing the delayed onset of the GeV light curve, also identified by \textit{Fermi} \citep{hascoet:12}; 
(iii)  what is the effect of the injection timescale of the accelerated particles~? A slow injection may  improve the spectral shape at low energy, as investigated recently by \citet{asano:09b}; (iv) 
what is the effect of a decaying magnetic field behind the shock front~? Such an evolution is expected from shock acceleration modelling and may  improve the shape of the synchrotron spectrum (increasing low-energy photon index) without implying as low values of $\epsilon_\mathrm{B}$ as what is considered in this paper \citep[see e.g.][]{derishev:07,wang:13}.

\section{Conclusions}
\label{sec:conclusions}

Motivated by the results from the \textit{Fermi} satellite which significantly extends the spectral coverage of the GRB phenomenon and improves particularly the spectral analysis of the prompt emission, we investigated in this paper the origin of the observed spectral evolution in GRBs.
We presented the results of a set of numerical simulations of the GRB prompt emission in the framework of the internal shock model. We made a detailed comparison of the model predictions with the observed temporal and spectral GRB properties in the soft gamma-ray range. We focussed on the simplest case of a single pulse burst associated to the synchrotron radiation from shock-accelerated electrons in the internal shocks formed after the collision
 between a 'fast' and a 'slow' region in an ultra-relativistic ejecta.
We considered three reference case with 
a duration of $2-3$ s, 
an isotropic radiated energy of respectively $1.9\times 10^{52}$, $1.3\times 10^{52}$ and $1.3\times 10^{51}\, \mathrm{erg}$, 
a peak energy of 730, 640 and 160 keV, and a low-energy photon index of -1.5, -1.1 and -0.7.

We show that many observed properties or common trends -- namely (i) the pulse asymmetry, (ii) the energy dependent pulse asymmetry (evolution of the pulse width with energy channel), (iii) the time lags between the light curves in different energy channels, (iv) the hard-to-soft evolution within pulses, (v) the hardness-intensity correlation, (vi) the hardness-fluence correlation -- can be accounted for and are governed by the details of the spectral evolution, i.e. the evolution of the peak-energy and the spectral slopes.

We showed that there is a qualitative agreement between the model results for our three reference cases and the large set of observations listed above.
With a comprehensive set of simulations, 
we demonstrated that a quantitative agreement can be achieved under some constraints on the model parameters. We distinguished between the effects of the microphysics  (details of the energy distribution in shocked regions) and the dynamical parameters (initial conditions in the outflow). We found that the agreement with the observed spectral evolution can be significantly improved if (i) the distribution of shock-accelerated electrons is steeper than what is usually assumed, with a slope $p\ga 2.7$; (ii) the microphysics parameters vary with the shock conditions in a manner that reduces the dependency of the peak energy on the shock conditions. It is illustrated here by
the case where the fraction of accelerated electrons increases for stronger shocks;
 (iii) the initial variations of the Lorentz factor in the outflow are steeper. An additional advantage of this assumption
  is the increase of the efficiency of internal shocks; (iv) the relativistic ejection proceeds with a constant mass flux rather than a constant kinetic energy flux. A drawback of this last possibility is a reduced efficiency of the shocks. As the microphysics parameters are not well constrained by the current stage of shock acceleration modelling in the mildly relativistic regime relevant for internal shocks, and as the initial conditions in the outflow are also poorly constrained due to many uncertainties regarding the mechanism responsible for the relativistic ejection by the central engine, we cannot conclude if one of these four possibilities may be expected or should be preferred.

We also specifically investigated the impact of the duration of the relativistic ejection, as many of the properties listed above are known to evolve with  pulse duration. The internal shock model naturally predicts a larger peak energy for short pulses, and possibly a harder photon index due to a deeper Klein-Nishina regime for inverse Compton scatterings. We showed that -- in agreement with observations -- this leads to a hardness-duration correlation and to the following consequences: pulses become more symmetric, with almost no evolution of the pulse width with energy, and with very short or zero lags. The prompt emission from short GRBs could then be due to the same mechanism as in long GRBs, but for different  model parameters due to the fact that all timescales are contracted, probably because of a different central engine.

Finally, we investigated the signature at high-energy (\textit{Fermi}-LAT range). In this domain, the observed flux is made of the high-energy tail of the synchrotron component and a new component produced by inverse Compton scattering. 
A direct comparison with \textit{Fermi}-LAT results is not possible as  
LAT bursts are among the brightest whereas we have simulated here average pulses. However, we note  a qualitative agreement with data:
due to the evolving efficiency of the scatterings -- they usually occur in the Klein-Nishina regime at early times and enter the Thomson regime during the pulse decay -- the resulting emission at high-energy can differ significantly from the keV-MeV range; specifically, the rise of the light curve is delayed and the emission lasts longer.  This leads to a U-shape curve when plotting time lags with respect to the low-energy channel as a function of energy, in agreement with GBM+LAT observations. However, we do not have a quantitative agreement : the onset of the high-energy light curve is not delayed enough. Interestingly, some of the effects listed above 
--
a steeper electron slope, a varying electron acceleration fraction, and especially steeper variations of the initial Lorentz factor
--
have also a positive impact on the properties of the high-energy emission. The time-integrated spectrum at high-energy depends strongly on the efficiency of the inverse Compton scatterings. In some cases, it is found to be very close of the extrapolation of the MeV component, possibly with a cutoff at high-energy; in other cases, it clearly shows an additional component, which can either be rising (photon index greater than $-2$) or flat (photon index close to $-2$).
As there are significant differences between the various scenarios discussed in the paper, this motivates a specific comparison to \textit{Fermi}-LAT bursts which will hopefully provide diagnostics to distinguish among the various theoretical possibilities.

This study illustrates the capacity of the internal shock model to reproduce most of the observed properties of the GRB prompt emission related to the spectral evolution, both for long and short bursts. Our conclusions are limited by many uncertainties in the ingredients of the model, namely the details of the microphysics in mildly relativistic shocks and the initial conditions in the GRB relativistic outflows. However, in a more optimistic view, we showed that this poorly understood physics may have a detectable imprint in GRB data, which should allow for some progress in the future.

\begin{acknowledgements} 
The authors thank R. Mochkovitch for many valuable discussions on this work, and a careful reading of the manuscript.
F.D. and Z.B. acknowledge the French Space Agency (CNES) for financial support.  
\end{acknowledgements}

\bibliographystyle{aa}
\bibliography{grbspectralevolution}

\end{document}